\documentclass[acmsmall,screen,nonacm]{acmart}

\AtBeginDocument{%
  }

\usepackage{wrapfig}
\usepackage[ruled,vlined,linesnumbered]{algorithm2e}
\SetKwInOut{Input}{Input}
\SetKwInOut{Output}{Output}
\SetKwBlock{Loop}{repeat}{end}
\SetKwProg{Proc}{procedure}{ begin}{end}
\usepackage{listings}
\usepackage[most]{tcolorbox}
\usepackage[x11names]{xcolor}
\usepackage{soulpos}
\usepackage{tikz}
\usepackage{mathtools}
\usepackage{tikz-3dplot}
\usepackage{multirow}
\usetikzlibrary{arrows.meta, positioning, fit, 3d, shapes.geometric}
\usepackage{graphicx}
\usepackage{subcaption}
\usepackage{enumitem}
\usepackage{pifont}

\newcommand{\set}[1]{\left\{#1\right\}}
\newcommand{\seq}[1]{\left<#1\right>}
\DeclareMathOperator\dom{dom}
\DeclareMathOperator\dist{Dist}
\newtcolorbox{todobox}{
  colback=yellow,
  coltext=Red2,
  boxrule=0pt,
  left=2pt, right=2pt, top=1pt, bottom=1pt,
  sharp corners,
  enhanced, breakable
}
\newcommand{\ib}{\vartriangleleft}
\newcommand{\nat}{\mathbb{N}_0}
\newcommand{\coord}[2]{#1 \xrightarrow{#2} \nat}
\newcommand{\coordsp}[2]{\left[ \coord{#1}{#2} \right]}
\newcommand{\subcoord}[3]{#1 \xrightarrow[#2]{#3} \nat}
\newcommand{\subcoordsp}[3]{\left[ \subcoord{#1}{#2}{#3} \right]}
\newcommand{\ucoord}[2]{#1 \xrightharpoonup{#2} \nat}
\newcommand{\ucoordsp}[2]{\left[ \ucoord{#1}{#2} \right]}
\newcommand{\usubcoord}[3]{#1 \xrightharpoonup[#2]{#3} \nat}
\newcommand{\usubcoordsp}[3]{\left[ \usubcoord{#1}{#2}{#3} \right]}

\newcommand{\err}[1]{\colorbox{red!30!white}{\texttt{#1}}}
\newcommand{\ambig}[1]{\colorbox{yellow!90!white}{\texttt{#1}}}
\newcommand{\ok}[1]{\colorbox{green!50!white}{\texttt{#1}}}
\lstdefinestyle{rustish}{
    basicstyle=\ttfamily\footnotesize,
    aboveskip=3pt,
    belowskip=3pt,
    stringstyle=\color{gray},
    commentstyle=\color{gray},
    breakatwhitespace=false,
    breaklines=true,
    captionpos=b,
    keepspaces=true,
    numbers=left,
    columns=fullflexible,
    showspaces=false,
    showstringspaces=false,
    showtabs=false,
    tabsize=2,
    numberstyle=\tiny,
    numbersep=8pt,
    xleftmargin=12pt,
}
\newcommand{\dimP}{\textcolor{gray}{\texttt{p}}}
\newcommand{\dimF}{\textcolor{Green4}{\texttt{f}}}
\newcommand{\dimT}{\textcolor{Turquoise4}{\texttt{t}}}
\newcommand{\dimR}{\textcolor{Gold4}{\texttt{r}}}
\newcommand{\dimS}{\textcolor{red}{\texttt{s}}}
\newcommand{\dimG}{\textcolor{violet}{\texttt{g}}}
\newcommand{\cmark}{\text{\ding{51}}}
\newcommand{\xmark}{\text{\ding{55}}}
\newtheorem*{lemma*}{Lemma}
\newtheorem*{proposition*}{Proposition}
\newtheorem*{corollary*}{Corollary}
\newtheorem*{theorem*}{Theorem}

\begin{document}

\title{Operon: Incremental Construction of Ragged Data via Named Dimensions}

\author{Sungbin Moon}
\email{sb.moon@asteromorph.com}
\affiliation{
  \institution{Asteromorph}
  \city{Seoul}
  \country{Republic of Korea}
}

\author{Jiho Park}
\email{jh.park@asteromorph.com}
\affiliation{
  \institution{Asteromorph}
  \city{Seoul}
  \country{Republic of Korea}
}

\author{Suyoung Hwang}
\email{sy.hwang@asteromorph.com}
\affiliation{
  \institution{Asteromorph}
  \city{Seoul}
  \country{Republic of Korea}
}

\author{Donghyun Koh}
\email{dh.koh@asteromorph.com}
\affiliation{
  \institution{Asteromorph}
  \city{Seoul}
  \country{Republic of Korea}
}

\author{Seunghyun Moon}
\email{sh.moon@asteromorph.com}
\affiliation{
  \institution{Asteromorph}
  \city{Seoul}
  \country{Republic of Korea}
}

\author{Minhyeong Lee}
\authornote{Correspondence to Minhyeong Lee.}
\email{mh.lee@asteromorph.com}
\affiliation{
  \institution{Asteromorph}
  \city{Seoul}
  \country{Republic of Korea}
}

\begin{abstract}

Modern data processing workflows frequently encounter ragged data: multidimensional collections whose inner lengths vary and are revealed only at runtime.
Ragged data is prevalent in domains throughout science and computing, yet existing workflow engines treat it as nested containers.
This representation leaves shape semantics ambiguous, obscures task dependencies when upstream coordinates are not yet enumerated, and imposes coarse synchronization barriers that stall independent work across tasks.
We argue that ragged data is fundamentally a flat function over a coordinate space, and that making this view explicit resolves all three difficulties at once.

We realize this insight in Operon, a statically verifiable workflow engine built around a formalism of named dimensions with explicit dependency relations.
Users declare pipelines in the Operon DSL by annotating data flow with dimension names, where the static checker verifies that every task does not include faulty iterations regardless of runtime lengths.
At runtime, the system models in-progress arrays as partial functions over downward-closed coordinate domains, which provides a natural representation for partially-unknown data states, an exact notion of per-task indegree before all upstream coordinates are enumerated, and a sound basis for fault recovery without programmer intervention.\footnote{Proofs and additional details are available in the appendix.}

\end{abstract}



\maketitle

\section{Introduction}
\label{sec:introduction}


Modern data processing increasingly relies on workflows over \emph{ragged data}~\cite{karpitschka2015droplets,chambers2010flumejava}, or multidimensional collections of data with non-uniform lengths.
Ragged data arises in many data processing tasks, such as natural language processing~\cite{pouransari2024dataset,krell2023efficient}, scientific measurements~\cite{heybrock2020scipp,peterson2015event,rubel2019nwb}, computer vision~\cite{gao2025survey,li2025open}, and autonomous agentic systems~\cite{piotrowski2025will}.
However, current frameworks remain tied to the nested container (or array-of-arrays) interpretation of ragged data.
While nested containers suffice for data at rest, they leave it an auxiliary concern to handle workflow settings where structure is revealed incrementally at runtime.

Ragged data is fundamentally a function over a coordinate space.
We realize this insight with \textbf{Operon}, a statically verifiable workflow engine that indexes data and task instances by a shared and evolving coordinate space.
Operon provides a domain-specific language (DSL) where users declare data flow and annotate iteration axes using named dimensions.
For all well-formed programs, Operon generates an incremental execution model based on a formalization of dimensional dependencies, enforcing well-behavedness guarantees---namely \emph{completeness}, \emph{confluence}, \emph{work-conservation}, and \emph{fault-tolerance} (all four defined in Section~\ref{subsec:problem_statement})---by construction.

\paragraph{Challenges}

Ragged data poses three interlocking challenges in a workflow setting.
First, its evolving structure creates \emph{shape ambiguity}.
The flat type signatures of array operations cannot disambiguate intent: an operation over nested lists might zip its arguments elementwise, or aggregate them across an axis.
As a result, faulty iteration over nonexistent entries surfaces only as a runtime error~\cite{henriksen2021towards,paszke2021getting}.
Avoiding this requires static evidence, independent of runtime values, that certain lengths match.
Naming dimensional lengths supplies such evidence for regular tensors, but the idea does not readily generalize to ragged data, where a length is not fixed even within a single array.

\begin{wrapfigure}{r}{0.5\columnwidth}
  \begin{lstlisting}[numbers=left,xleftmargin=1.5em,basicstyle=\footnotesize\ttfamily]
for i (*@\(\in [0, N)\)@*) do
  A[i] (*@\(\leftarrow\)@*) alpha(i)
for i (*@\(\in [0, N)\)@*) do
  B[i] (*@\(\leftarrow\)@*) beta(i)
for i (*@\(\in [0, N)\)@*) do
  for j (*@\(\in\)@*) A[i].coords do
    for k (*@\(\in\)@*) B[i].coords do
      C[i][j][k] (*@\(\leftarrow\)@*) gamma(A[i][j], B[i][k])
\end{lstlisting}
  \caption{
    A simple ragged data pipeline.
    We abbreviate an execution such as \texttt{gamma(A[i][j], B[i][k])} by its iterating variables alone, as \texttt{gamma(i,j,k)}, throughout the paper.
    \texttt{gamma(i,j,k)} can start as soon as \texttt{alpha(i)} and \texttt{beta(i)} complete, without waiting on other values of \texttt{i}.
  }
  \label{fig:dyn-sched}
\end{wrapfigure}

Second, \emph{dynamic scheduling} is unavoidable.
Because the number of tasks and their dependent data is not known ahead of time, a static DAG-based scheduler is ruled out~\cite{wu2000runtime}.
Nevertheless, an ideal engine would still launch each task as soon as its inputs are ready, running as many in parallel as possible.
Figure~\ref{fig:dyn-sched} illustrates the difficulty: \texttt{alpha} and \texttt{beta} are immediately parallelizable across \texttt{i}, but \texttt{gamma} can only run once both finish and reveal the \texttt{j} and \texttt{k} coordinates it depends on.

Third, \emph{partial data} must be represented explicitly.
While a workflow runs, its target arrays and execution state are only partially known.
However, the scheduler must still name dependencies before the instances they refer to exist.
For example, a downstream \texttt{delta(i,k)} awaits every \texttt{gamma(i,j,k)} that shares its \texttt{i} and \texttt{k}.
A nested container \texttt{C[i][j][k]} offers no such reference: it files each \texttt{k} under a \texttt{j} bucket, even though \texttt{k} is fixed by \texttt{i} alone.
This leaves no natural way to express the \emph{quota} or \emph{indegree} of a partially-realized task, and it complicates fault-tolerance, since recovery must reconstruct the execution structure from partial data.

\paragraph{Our approach}

\begin{table}[t]
  \centering\small
  \caption{
    Shape ambiguity in Operon.
    Each row's \textbf{Pseudocode} highlights an array operation whose safety depends on the shapes of its data so that the same operation can be correct in one pipeline and incorrect in another.
    The two \textbf{Operon DSL} columns place that operation in one context of each kind: a set of declarations Operon's static checker accepts, and one it rejects.
    The checker decides each case without inspecting any runtime values.
  }
  \begin{tabular}{c|cc}
\toprule[0.1em]
\multicolumn{1}{c}{\textbf{Pseudocode}} & \multicolumn{2}{c}{\textbf{Operon DSL examples} (Accepted / Rejected)} \\ \midrule[0.1em]
\noindent\begin{minipage}{.305\textwidth}
\textit{Reducing across outer dimension.}
\begin{lstlisting}[numbers=none,xleftmargin=0pt]
for j (*@\(\in\)@*) foo[0].coords do
  bar[j] (*@\(\leftarrow\)@*) f((*@\ambig{foo[..][j]}@*))
\end{lstlisting}
\end{minipage}
&
\noindent\begin{minipage}{.305\textwidth}
\begin{lstlisting}[numbers=none,xleftmargin=0pt]
Baz<i> = baz();
Qux<j> = qux();
Foo = foo(Baz, Qux) for i, j;
Bar = f((*@\ok{Foo<i>}@*)) for j;
\end{lstlisting}
\end{minipage}
&
\noindent\begin{minipage}{.305\textwidth}
\begin{lstlisting}[numbers=none,xleftmargin=0pt]
Baz<i> = baz();
Foo<j> = foo(Baz) for i;
Bar = f((*@\err{Foo<i>}@*)) for j;
\end{lstlisting}
\end{minipage}

\\ \midrule

\noindent\begin{minipage}{.305\textwidth}
\textit{Zipping.}
\begin{lstlisting}[numbers=none,xleftmargin=0pt]
for (i, j) (*@\(\in\)@*) foo.coords do
  qux[i][j] (*@\(\leftarrow\)@*) g(
    foo[i][j], (*@\ambig{bar[i][j]}@*)
  )
\end{lstlisting}
\end{minipage}
&
\noindent\begin{minipage}{.305\textwidth}
\begin{lstlisting}[numbers=none,xleftmargin=0pt]
Baz<i> = baz();
Foo<j> = foo(Baz) for i;
Bar = bar(Foo) for i, j;
Qux = g(Foo, (*@\ok{Bar}@*)) for i, j;
\end{lstlisting}
\end{minipage}
&
\noindent\begin{minipage}{.305\textwidth}
\begin{lstlisting}[numbers=none,xleftmargin=0pt]
Baz<i> = baz();
Foo<j> = foo(Baz) for i;
Bar<k> = bar(Baz) for i;
Qux = g(Foo, (*@\err{Bar}@*)) for i, j;
\end{lstlisting}
\end{minipage}

\\ \midrule

\noindent\begin{minipage}{.305\textwidth}
\textit{Broadcasting.}
\begin{lstlisting}[numbers=none,xleftmargin=0pt]
for (i, j) (*@\(\in\)@*) foo.coords do
  qux[i][j] (*@\(\leftarrow\)@*) h(
    foo[i][j], (*@\ambig{bar[j]}@*)
  )
\end{lstlisting}
\end{minipage}
&
\noindent\begin{minipage}{.305\textwidth}
\begin{lstlisting}[numbers=none,xleftmargin=0pt]
Baz<i> = baz();
Bar<j> = bar();
Foo = foo(Baz, Bar) for i, j;
Qux = h(Foo, (*@\ok{Bar}@*)) for i, j;
\end{lstlisting}
\end{minipage}
&
\noindent\begin{minipage}{.305\textwidth}
\begin{lstlisting}[numbers=none,xleftmargin=0pt]
Baz<i> = baz();
Foo<j> = foo(Baz) for i;
Bar<k> = bar();
Qux = h(Foo, (*@\err{Bar}@*)) for i, j;
\end{lstlisting}
\end{minipage}
\\ \bottomrule[0.1em]
\end{tabular}

  \label{tab:shape_ambiguity}
\end{table}

We address these challenges by treating ragged data as a \emph{flat function} from a shared coordinate space to values, rather than as nested containers.
The organizing primitive is the \emph{named dimension}: each axis of repetition is assigned a symbolic name, and structural relationships between axes are declared explicitly in the Operon DSL.
Because these annotations identify, for every array and every task instance, precisely which coordinates it inhabits, the DSL can resolve shape ambiguity without inspecting runtime lengths.
A static checker validates that every data flow edge in the pipeline connects compatible coordinate spaces, ruling out the class of indexing errors that depend-typed systems can only catch at runtime; Table~\ref{tab:shape_ambiguity} gives concrete DSL examples of shape ambiguities accepted and rejected under this checker.

The core formalism is an \emph{order-theoretic model of dimensional dependencies}.
Each named dimension may depend on one or more enclosing dimensions, giving rise to a partial order over coordinates.
Within this order, we model ragged arrays as partial functions whose domains are downward-closed sets (sets containing all dimension ancestors of each of their members).
At any point during execution, the scheduler can inspect this partial domain to determine which task instances are immediately runnable, which are waiting on upstream coordinates to materialize, and which cannot yet be enumerated because an enclosing dimension is incomplete.
In this design, the scheduler never requires a global synchronization barrier: as in Figure~\ref{fig:dyn-sched}, a \texttt{gamma} instance is dispatched as soon as the corresponding \texttt{alpha} and \texttt{beta} finish, regardless of other values of \texttt{i}.

For every well-formed Operon program, this model generates an incremental execution plan whose well-behavedness properties, which we will enumerate in Section~\ref{subsec:problem_statement}, hold by construction.
Tasks are assumed to be pure (possibly stochastic) functions, giving effectively-once semantics.
Each coordinate is executed at most once under normal operation, and re-executed only when a prior attempt failed before its result was atomically persisted.

\paragraph{Contributions}

Our main contributions are:
\begin{itemize}
  \item \emph{Named-dimension formalism for ragged data}
        (Section~\ref{sec:ragged_and_named_dimensions}).
        We introduce an order-theoretic framework in which named dimensions and their dependencies are first-class objects, and ragged arrays are modeled as partial functions over the resulting coordinate space.

  \item \emph{Structured model for partial data and incremental execution}
        (Section~\ref{sec:incremental_resolutions}).
        We develop explicit representations for the partially-known data states that arise while a ragged pipeline runs, characterize which transformations are admissible at each state, and prove that the resulting incremental execution model is confluent under parallel and stochastic evaluation.

  \item \emph{Operon DSL and runtime system}
        (Section~\ref{sec:operon}).
        We present the syntax and static verification rules for Operon pipelines and describe how the runtime system translates a verified program into a work-conserving incremental scheduler.
\end{itemize}


\section{Preliminaries}
\label{sec:preliminaries}

\subsection{Problem Statement}
\label{subsec:problem_statement}

We first delimit the class of pipelines we target, and then develop the guarantees we ask of any system that declares and executes them.
We also discuss the design questions and tradeoffs to motivate our approach.

\paragraph{Pipelines}

We base our construction on the single program, multiple data (SPMD) model where each \emph{task} may comprise multiple computation nodes.
In Figure~\ref{fig:dyn-sched}, for example, each execution of \texttt{alpha(i)}, \texttt{beta(i)}, or \texttt{gamma(i,j,k)} would represent a unit of computation, while each function \texttt{alpha}, \texttt{beta}, or \texttt{gamma} would serve as a task.
A pipeline is then understood as the composition of such tasks, where each task may acyclically depend on the results of other tasks.

From a coarse view, tasks are data transformations that must produce the full array of outputs after all executions are complete.
For instance, assuming sequential execution, the \texttt{gamma} task is equivalent to a function that takes two 2D ragged arrays \texttt{A} and \texttt{B} to produce a 3D ragged array \texttt{C}.
It is thus natural to maintain that each data-parallel execution runs exactly once per input combination, i.e., each task is total over the iteration space.

Tasks need not be deterministic: our target workloads include inherently stochastic operations, most prominently inference calls to large language and vision-language models.
An execution is thus best understood as \emph{independently sampling} its output entries from a distribution, and it is only necessary that this distribution depend on the input combination alone.
This is what later lets us speak of a well-defined distribution over terminal states despite the randomness of individual executions.

We recapitulate this definition of pipelines with representative properties for future reference.

\begin{definition}[Ragged data pipelines]\label{def:ragged_data_pipelines}
    A \emph{ragged data pipeline} is a finite set of \emph{tasks}.
    Each task reads from zero or more \emph{input sources} and writes to a single \emph{output source}, where a source is a multidimensional collection of \emph{entries}.
    A pipeline satisfies the following conditions.
    \begin{itemize}
        \item \emph{Ragged.} Each execution of a task emits finite output entries whose number may differ across executions.
        \item \emph{Acyclic.} The dependency relation induced by tasks reading from one another's output sources is acyclic.
        \item \emph{Total.} For each task, exactly one execution is performed per \emph{input combination} (a tuple with one entry from each of the task's input sources).
        \item \emph{Independent.} The distribution of an execution's output entries depends only on the entries in its input combination.
    \end{itemize}
\end{definition}

\paragraph{Design questions and tradeoffs.}

We encounter several interconnected questions in designing a system that declares and executes such pipelines. We resolve each for the coarse-grained problem space where ragged data is most prevalent.
We draw out these insights into a set of requirements that a sensible system should satisfy.

Note that these questions are not entirely novel: Chapel~\cite{chamberlain2007parallel} notably identifies analogous tradeoffs in parallel language design under the headings of a \emph{global view of computation}, \emph{general composable parallelism}, and \emph{separation of algorithm and implementation}.
Section~\ref{sec:related_work} discusses the precise connections and divergences.

\textbf{Pointwise persistence versus performance.}
Fine-grained, in-memory scheduling keeps per-execution overhead low.
However, it makes a lost computation node expensive to recover, complicates resumption after a failure, and bounds a run by aggregate memory.
We instead persist every entry pointwise as it is produced.
Though unthinkable from a high-performance-computing standpoint, this choice hands us \emph{fault-tolerance}, which becomes arguably more valuable as task executions coarsen and lost progress becomes costly.

This tradeoff only makes sense when the following assumptions are met: (1) tasks are \emph{coarse-grained}, where individual executions take roughly milliseconds or longer, and (2) \emph{durable backing storage} is always available.
We give up the scheduling performance of fine-grained tasks in return for the ability to reconstruct exact progress after a crash.
Section~\ref{sec:evaluation} further discusses this tradeoff and quantifies the overhead in our target regime.

\textbf{Governing task-parallelism.}
A ragged pipeline exposes two axes of parallelism that a scheduler must exploit at once.
\emph{Task-parallelism} runs independent tasks concurrently, as \texttt{alpha} and \texttt{beta} may.
\emph{Data-parallelism} runs the executions within a single task concurrently, across the iteration space it spans.
Mixing the two is hard (Figure~\ref{fig:dyn-sched}): the tasks of a ragged pipeline are heterogeneous, some bound by external API rate limits, others by CPU or GPU capacity, others by a strictly sequential dependency.
A single global concurrency limit cannot express these independent bounds without throttling the pipeline to its most constrained stage, so we attach a concurrency limit to each task.
This gives a per-task analog of \emph{work-conservation}.
Just as a work-conserving scheduler never idles while there is work it could run, we launch each instance as early as its own inputs and limit allow, so that none sits idle while its task has spare capacity.
Conventional workflow engines fall short of this.
Airflow~\cite{apache2025airflow} and Luigi~\cite{spotify2025luigi} require the iteration to be hand-mapped, and both impose stage-level synchronization barriers: a downstream instance launches only after an entire upstream array is populated.
Prefect~\cite{prefecthq2025prefect} generates instances dynamically, but delegates their scheduling to a generic thread pool that starves under scale (Section~\ref{sec:evaluation}).

Realizing such early launching calls for an event mechanism that dispatches an instance the moment its inputs are committed.
Events serve this role cleanly when the participating task nodes are fixed in advance~\cite{treichler2014realm}.
Ragged data breaks that premise.
Each execution emits an array of unknown length, so neither the iteration space nor the set of downstream task nodes it induces is fixed until the upstream executions run.
Treating each node discovered at runtime as its own event handler compounds the problem: it opens an unbounded number of concurrent channels, whereas global concurrency is a contended resource (Figure~\ref{fig:experiment2}).
Our design instead makes \emph{tasks} the participants of event channels---one \emph{individual scheduler} per task---rather than the individual nodes.
Nodes not yet fleshed out are tracked internally as \emph{tickets}.
This holds the event paths under static control while still admitting the internal dynamism that ragged data demands.

\textbf{The anatomy of a declaration.}
Data flow should be expressed \emph{declaratively}, as what each task consumes and produces, rather than \emph{imperatively}, as explicit loops over indices.
This raises two questions for ragged data: how to make dimensional intent legible in a user's declarations, and how to decide which \emph{sets} of declarations fit together into a coherent pipeline.
The first is largely answered by existing named-dimension notation.
Futhark's size parameters~\cite{henriksen2021towards}, for instance, name each length in a signature so that repeated names pin shapes statically:
\begin{center}
    \begin{tabular}{c}
        \begin{lstlisting}[numbers=none,xleftmargin=0pt,basicstyle=\normalsize\ttfamily]
def matmul [i][j][k] 'T (A: [i][j]T, B: [j][k]T) : [i][k]T
\end{lstlisting}
    \end{tabular}
\end{center}
This holds only for regular arrays, however, where each named length is a single fixed value.
Ragged data breaks that assumption, since a length can vary even within one array.
Deciding pipeline coherence is not covered by existing literature, and therefore is the central theoretical contribution of this paper.
Before any data exists, we must certify that every data-flow edge connects compatible coordinate spaces, so that no execution can read a nonexistent entry.
The guarantee we establish is that a declaration passing this check maps to a runtime satisfying the remaining constraints.
We discuss the typing rules that realize this guarantee further in Section~\ref{sec:operon}.

\textbf{Hiding pointwise access.}
Arbitrary dynamic access into a data array can only be validated once the surrounding data exists, which forces shape errors to runtime.
We therefore disallow random access from outside the system.
Every task is instead \emph{total} over its declared domain: it runs exactly once per input combination.
The runtime then knows the full shape of what each task will read and write without inspecting a single value.
Every required instance runs exactly once without ever touching a nonexistent input, and the run reaches a fully populated terminal state.
What we retain in exchange is dynamic \emph{generation}.
A task may still emit arrays of unpredictable length, though never read at unpredictable positions.
Trading dynamic access for dynamic generation is what sets our approach apart from systems built on arbitrary indexing; Section~\ref{sec:related_work} develops the contrast.

\textbf{Correctness of scheduling.}
The design so far grants the scheduler wide freedom in deciding the order of execution and degree of parallelism.
To obtain a coherent semantic model despite this freedom, we require the scheduler to satisfy two properties.
The first is \emph{confluence}: the order in which independent executions interleave must not perturb the outcome.
While the \emph{total} and \emph{independent} conditions of Definition~\ref{def:ragged_data_pipelines} make this a reasonable target, attaining that meaning is not automatic.
The set of executions is not known in advance but discovered incrementally, as each upstream output fixes a downstream iteration space, so different interleavings pass through different intermediate states and could in principle converge to different distributions.
Showing that they cannot is what we prove in Section~\ref{sec:incremental_resolutions}.
The second property, \emph{work-conservation}, was motivated above; what remains is to say when an instance's inputs count as ready.
They become \emph{available} once every upstream entry in the input combination has been committed to durable storage, so readiness is tracked per entry rather than per array.
In Figure~\ref{fig:dyn-sched}, a \texttt{gamma(i,j,k)} instance may be dispatched as soon as \texttt{alpha(i)} and \texttt{beta(i)} complete, without waiting on any other value of \texttt{i}.

We conclude with the design constraints we promised.

\begin{definition}[Design constraints]\label{def:design_constraints}
    A system for declaring and executing ragged data pipelines satisfies the \emph{design constraints} if it meets the following criteria.
    \begin{enumerate}
        \item \textbf{Statically verifiable.} Any pipeline declaration that passes the system's static checks maps to a runtime that satisfies the remaining four constraints.
        \item \textbf{Complete.} The runtime executes all required task instances exactly once, reaching a fully populated terminal state without accessing nonexistent inputs.
        \item \textbf{Confluent.} The runtime terminates with the same probability distribution over terminal states regardless of scheduling order, assuming terminating and non-erroring tasks.
        \item \textbf{Work-conserving.} Each task instance is launched as soon as its inputs become available, subject to a per-task concurrency limit.
        \item \textbf{Fault-tolerant.} The system persists sufficient metadata to reconstruct the maximal valid state after failures, where resumption yields effectively-once task execution.
    \end{enumerate}
\end{definition}

\subsection{Motivating Example}
\label{subsec:motivating_example}

Let us consider a concrete workflow to motivate our work.
The SciCap+ dataset~\cite{yang2024scicap+} defines scientific figure captioning as the prediction of a caption \(C\) given a figure \(I\) and an associated knowledge context \(K\) extracted from the paper.
We construct such a dataset by starting from raw paper PDFs, extracting figures and paragraphs using existing tools such as PDFFigure 2.0~\cite{clark2016pdffigures}, and then employing a vision-language model (VLM) agent to score paragraph relevance~\cite{yang2024automatic}.
A representative paper yields around 10 figures, 8 sections, and 1--5 paragraphs per section.
Across a batch of \(N\) papers, this amounts to hundreds to thousands of VLM evaluations.
The exact count is discovered only as the pipeline runs.

\paragraph{Reading the DSL}

Operon expresses this pipeline as a sequence of task declarations.
Each declaration has the form
\begin{center}
    \texttt{Output\(<\)new\_dim\(>\) = task(Input\textsubscript{1},~\ldots) for dim\textsubscript{1},~\ldots;}
\end{center}
and carries three pieces of information.

\textbf{What is produced.}
The left-hand side names the output data collection.
If the task introduces a new ragged axis, the angle brackets give that axis a name.
\texttt{CaptionedFig<f>} says each invocation emits a variable-length list of figures, whose indices are named \texttt{f}.
A left-hand side without angle brackets (e.g., \texttt{ParsedPaper}) produces exactly one value per invocation.

\textbf{What is consumed.}
Writing \texttt{ParsedPaper} (no angle brackets) passes the single entry at the current coordinate.
Writing \texttt{Paragraph<s,g>} passes \emph{all} paragraphs for the current coordinate, aggregated over the \texttt{s} and \texttt{g} dimensions, as a nested list.

\textbf{How it iterates.}
The \texttt{for} clause lists the dimensions iterated over.
Writing \texttt{for p, f} launches the task once per valid \texttt{(p,\,f)} pair as those pairs materialize.

As a first example, the single line
\begin{center}
    \begin{tabular}{c}
        \begin{lstlisting}[numbers=none,basicstyle=\small\ttfamily]
CaptionedFig<f> = extract_captioned_figures(ParsedPaper) for p;
\end{lstlisting}
    \end{tabular}
\end{center}
says: for each paper \texttt{p}, call \texttt{extract\_captioned\_figures} on the parsed PDF and assign the resulting figures to a new dimension \texttt{f}.
This dimension would then depend on \texttt{p}.
Figure~\ref{fig:operon-scicap} shows the full pipeline written in this style.

\begin{figure}[t]
    \centering
    \begin{lstlisting}[aboveskip=0pt,belowskip=0pt]
operon::define_operon! {
    sci_cap_enhanced = {
        PaperId<p>      = get_paper_id();
        ParsedPaper     = parse_paper(PaperId)                                  for p;
        CaptionedFig<f> = extract_captioned_figures(ParsedPaper)                for p;
        Section<s>      = extract_sections(ParsedPaper)                         for p;
        Paragraph<g>    = extract_paragraphs(Section)                           for p, s;
        Relevance       = vlm_evaluate(CaptionedFig, Paragraph)                 for p, f, s, g;
        RelevantPg<r>   = filter_aggregate(Paragraph<s, g>, Relevance<s, g>)    for p, f;
        OcrToken<t>     = ocr_extract(CaptionedFig)                             for p, f;
        Row             = collect_row(CaptionedFig, RelevantPg<r>, OcrToken<t>) for p, f;
    }
}
\end{lstlisting}
    \caption{Operon pipeline definition for the motivating example.
        Each line declares a data collection, the task that produces it, its inputs, and the dimensions
        it is iterated over (\texttt{for}).
        Angle brackets on the left-hand side denote the new dimension introduced by that collection;
        angle brackets on an argument denote aggregation over those dimensions before passing to the task.}
    \label{fig:operon-scicap}
\end{figure}

For each paper \texttt{p}, the pipeline extracts a variable number of figures (\texttt{f}), sections (\texttt{s}), and paragraphs (\texttt{g}).
The VLM then scores every \texttt{(f,\,s,\,g)} combination within a paper (\texttt{Relevance}).
Paragraphs meeting a relevance threshold are aggregated into \texttt{RelevantPg} for each figure.
This workflow is ragged at every level: figure, section, and paragraph counts all vary per paper.
As a result, no two papers drive the same set of VLM calls.
We use this pipeline as the running example throughout the paper.

\paragraph{Why a legible declaration can still be ill-formed}

Naming makes a declaration legible, but legibility is not correctness.
A pipeline can name every dimension and still fail to fit together.
To illustrate, suppose we try to score figure--section relevance directly, skipping the paragraph dimension:
\begin{center}
    \begin{tabular}{c}
        \begin{lstlisting}[numbers=none,xleftmargin=0pt,basicstyle=\small\ttfamily]
BadRelevance = vlm_evaluate(CaptionedFig, (*@\err{Section}@*)) for p, f;
\end{lstlisting}
    \end{tabular}
\end{center}
Every dimension here is named, and every entity declared, so nothing on the surface of the line is amiss.
The declaration is nevertheless ill-formed: \texttt{Section} is indexed by \texttt{\{p,\,s\}}, yet the iteration space \texttt{\{p,\,f\}} provides no binding for \texttt{s}.
Even a system that tracks these names can check the access only once the data exists.
A raw nested-container pipeline need not catch it at all, faulting out of bounds or, more insidiously, pairing the \(i\)-th figure with the \(i\)-th section---a semantically arbitrary match.
Operon's static checker rejects the declaration at build time.
The dimension \texttt{s}, contributed by \texttt{Section}'s characteristic dimensions, is not accounted for in the iteration space.
The user must either extend the iteration to \texttt{for p, f, s} (scoring each figure--section pair independently) or aggregate with \texttt{Section<s>} to collect all sections per paper per figure call.
Section~\ref{subsec:operon_overview} gives the formal condition behind this check.
The full static-checking criteria include structurally deeper errors that names alone cannot catch, such as collecting across a parent dimension while leaving a child dimension free.
In every case, the point is the same: no runtime data is needed to detect the error.

\section{Ragged and Named Dimensions}
\label{sec:ragged_and_named_dimensions}

In this section, we formalize the concepts to describe ragged data.
For this, we present a system of named dimensions equipped with an explicit dependency relation, and develop a generalization of multidimensional arrays on top of this system.

\subsection{Dimensions}
\label{subsec:dimensions}

\newcommand{\false}{\textcolor{gray}{\text{F}}}
\newcommand{\true}{\textbf{T}}

We define named dimensions, or simply dimensions, as identifiers for each axis of repeated data.
In rectangular arrays, the behavior of each dimension is invariant with respect to the others, and hence we may treat each dimension independently.
However, in ragged arrays, the size of one dimension may depend on the position along another dimension.
To make this dependency explicit, we introduce the following definition.

\begin{definition}[Dimension spaces]\label{def:dimension_spaces}
    A \emph{dimension space} \((\mathcal{D}, \prec)\) is a finite set \(\mathcal{D}\) of dimensions with a strict partial order \(\prec\).
    We denote the reflexive closure of \(\prec\) as \(\preceq\).
    The relation \(d \prec e\) means that \(e\) \emph{depends} on \(d\); in this relationship, \(d\) is the \emph{ancestor}, and \(e\) is the \emph{descendant}.
    If neither \(d \preceq e\) nor \(e \preceq d\), we write \(d \parallel e\) and say \(d\) and \(e\) are \emph{independent}.
\end{definition}

\begin{example}\label{ex:motex_dimension_space}
    Recall the motivating example shown in Section~\ref{subsec:motivating_example}.
    The dimension space induced by this example would be:
    \[
        \mathcal{D} = \set{\dimP, \dimF, \dimT, \dimR, \dimS, \dimG}. \quad
        \begin{array}{r|r cccccc l}
            \toprule
            \multicolumn{2}{c}{} & \multicolumn{6}{c}{e} & \\[-2pt]\cmidrule(lr){3-8}
            \multicolumn{2}{c}{d \prec e} & \dimP & \dimF & \dimT & \dimR & \dimS & \dimG & \text{Description of }d \\ \hline
            \multirow{6}{*}{\(d\)} & \dimP & \false & \true & \true & \true & \true & \true & \text{Papers} \\
            & \dimF & \false & \false & \true & \true & \false & \false & \text{Figures} \\
            & \dimT & \false & \false & \false & \false & \false & \false & \text{OCR Tokens} \\
            & \dimR & \false & \false & \false & \false & \false & \false & \text{Relevant paragraphs} \\
            & \dimS & \false & \false & \false & \false & \false & \true & \text{Sections} \\
            & \dimG & \false & \false & \false & \false & \false & \false & \text{Paragraphs} \\
            \bottomrule
        \end{array}
    \]
    We may confirm that the intuitive dependencies translate well into the relation \(\prec\): for example, \(\dimP \prec \dimF\) as figures depend on papers, \(\dimS \prec \dimG\) as paragraphs depend on sections, and \(\dimF \parallel \dimS\) as figures and sections are independent.
\end{example}

Since dimension spaces are finite posets, we adopt the following standard notions in our setting.

\begin{definition}[Structure of dimension spaces]\label{def:structure_of_dimension_spaces}
    Given a dimension space \((\mathcal{D}, \prec)\), we define the following terms:
    \begin{enumerate}
        \item A \emph{subspace} of \(\mathcal{D}\) is an induced subposet \((\mathcal{E}, \prec|_\mathcal{E})\) for any \(\mathcal{E} \subseteq \mathcal{D}\).
        All subsets \(\mathcal{E} \subseteq \mathcal{D}\) discussed here and below are assumed to be subspaces with the induced order \(\prec|_{\mathcal{E}}\).
        \item Downward closures and downward closed subposets are simply referred to as \emph{closures} and \emph{closed subspaces}.
        We use \emph{closedness} in place of \emph{downward closedness} since upward closedness is irrelevant to the discussion.
        A closure of a subspace \(\mathcal{E}\) is \(\mathcal{E}^\downarrow = \bigcup_{e \in \mathcal{E}} \set{d \in \mathcal{D} \mid d \preceq e}\), while a closed subspace \(\mathcal{F}\) satisfies \(\mathcal{F}^\downarrow = \mathcal{F}\); for a singleton subspace \(\set{d}\), we write \(d^\downarrow = \set{d}^\downarrow\) and call such closures \emph{principal}.
        \item The \emph{dependency space} of a subspace \(\mathcal{E}\) is defined as \(\mathcal{E}^\Downarrow = \mathcal{E}^\downarrow \setminus \mathcal{E}\).
        For a singleton subspace \(\set{d}\), we write \(d^\Downarrow = \set{d}^\Downarrow\) and also call such dependency spaces \emph{principal}.
    \end{enumerate}
\end{definition}

Intuitively, a dimension subspace is closed if it does not require reference to other dimensions.
In a closed subspace, all ancestors of each dimension can be found within itself, making the subspace self-contained.
Thus, discussions about closed subspaces typically do not require context beyond the subspace itself.

We conclude this section with a useful property of principal dependency spaces.

\begin{lemma}\label{lem:principal_deps_are_closed}
    Every principal dependency space is closed.
\end{lemma}

\subsection{Shapes and Coordinates}
\label{subsec:shapes_and_coordinates}

To associate the dimensions defined in Section~\ref{subsec:dimensions} with a structure that holds data, we must decide how to index data across dimensions.
The obvious answer is to use ``coordinates''---maps from dimensions to nonnegative integers---to specify which ``cell'' a piece of data belongs in.

In rectangular arrays, coordinates are confined by a simple tuple of lengths, the \emph{shape}, that specifies the acceptable indices along each dimension.
We generalize this notion to ragged arrays by introducing \emph{resolutions}, which specify lengths in a dependency-aware manner.

\begin{definition}[Resolutions]\label{def:resolutions}
    On a dimension space \((\mathcal{D}, \prec)\), a \emph{resolution} of a dimension \(d \in \mathcal{D}\) is a tuple \((d, c, \ell)\) where \(c \in [d^\Downarrow \to \nat]\) and \(\ell \in \nat\).\footnote{
        We use the notation \([\cdot \to \cdot]\) to denote the set of total functions from the domain to the codomain, and similarly, \([\cdot \rightharpoonup \cdot]\) for partial functions.
    }
    A \emph{resolution map} is a set of resolutions \(R\) that satisfies \((d, c, \ell_1), (d, c, \ell_2) \in R \rightarrow \ell_1 = \ell_2\).
\end{definition}

In this definition, a single resolution is a mapping that, given a dimension \(d\) and a total map \(c\) over the ancestor dimensions of \(d\), returns a nonnegative length \(\ell\).
A length \(\ell\) would accommodate values in \([0, \ell)\) along \(d\); we explicitly allow \(\ell = 0\) for ``empty'' dimensions with no valid values.
We henceforth interpret a resolution map \(R\) as a partial function \(\mathcal{D} \times [\mathcal{D} \rightharpoonup \nat] \rightharpoonup \nat\) (or, more precisely, \(\bigcup_{d \in \mathcal{D}} \left(\set{d} \times [d^\Downarrow \to \nat] \right) \rightharpoonup \nat\)).
In the motivating pipeline, for example, \((\dimF,\, \set{\dimP \mapsto 0},\, 3)\) is a resolution stating that paper~0 has 3 figures; \((\dimG,\, \set{\dimP \mapsto 0,\, \dimS \mapsto 2},\, 2)\) states that section~2 of paper~0 has 2 paragraphs.

\begin{figure}[t]
    \centering
    \begin{tikzpicture}[scale=0.5, every node/.style={font=\tiny}]
	\newcommand{\eparsedpaper}{
		\textcolor{gray}{\text{\ttfamily p}}
	}
	\newcommand{\esection}{
		\textcolor{red}{\text{\ttfamily s}}
	}
	\newcommand{\eparagraph}{
		\textcolor{violet}{\text{\ttfamily g}}
	}
	\newcommand{\ecaptionedfig}{
		\textcolor{Green4}{\text{\ttfamily f}}
	}
	\newcommand{\erelevance}{
	    \textcolor{orange}{\text{\ttfamily Relevance}}
    }
	\newcommand{\mparsedpaper}[1][]{
		\textcolor{gray}{\text{\ttfamily p} \mapsto #1}
	}
	\newcommand{\msection}[1][]{
		\textcolor{red}{\text{\ttfamily s} \mapsto #1}
	}
	\newcommand{\mparagraph}[1][]{
		\textcolor{violet}{\text{\ttfamily g} \mapsto #1}
	}
	\newcommand{\mcaptionedfig}[1][]{
		\textcolor{Green4}{\text{\ttfamily f} \mapsto #1}
	}
	\newcommand{\mrelevance}[1][]{
        \textcolor{orange}{\text{\ttfamily Relevance} \mapsto #1}
    }

	\newcommand{\cube}[5][]{%
		\begin{scope}[shift={(#2,#3,#4)}]
			\coordinate (O) at (0,0,0);
			\coordinate (A) at (1,0,0);
			\coordinate (B) at (1,1,0);
			\coordinate (C) at (0,1,0);
			\coordinate (D) at (0,0,1);
			\coordinate (E) at (1,0,1);
			\coordinate (F) at (1,1,1);
			\coordinate (G) at (0,1,1);

			\fill[fill=#1!50!white,opacity=0.9] (A)--(B)--(F)--(E)--cycle; 
			\fill[fill=#1!70!white,opacity=0.9] (B)--(C)--(G)--(F)--cycle; 
			\fill[fill=#1!90!white,opacity=0.9] (D)--(E)--(F)--(G)--cycle; 

			\draw[black,thin] (D)--(E)--(F)--(G)--cycle;
			\draw[black,thin] (E)--(A)--(B)--(C)--(G);
			\draw[black,thin] (B)--(F);

			\ifx&#5&
			\else
			\node[tdplot_screen_coords, font=\scriptsize\bfseries, below, align=right] at (0,0.5,0.75) {#5};
			\fi
		\end{scope}
	}

	\newcommand{\cubeedge}[5][]{%
		\begin{scope}[shift={(#2,#3,#4)}]
			\coordinate (O) at (0,0,0);
			\coordinate (A) at (1,0,0);
			\coordinate (B) at (1,1,0);
			\coordinate (C) at (0,1,0);
			\coordinate (D) at (0,0,1);
			\coordinate (E) at (1,0,1);
			\coordinate (F) at (1,1,1);
			\coordinate (G) at (0,1,1);

			\draw[#1, thick] (O)--(A)--(B)--(C)--cycle;
			\draw[#1, thick] (D)--(E)--(F)--(G)--cycle;
			\draw[#1, thick] (E)--(A)--(B)--(C)--(G);
			\draw[#1, thick] (B)--(F);
			\draw[#1, thick] (D)--(O);

			\ifx&#5&
			\else
			\node[tdplot_screen_coords, font=\scriptsize\bfseries, below, align=right] at (0,0.5,0.75) {#5};
			\fi
		\end{scope}
	}

	\newcommand{\dashedcubeedge}[5][]{
		\begin{scope}[shift={(#2,#3,#4)}]
			\coordinate (O) at (0,0,0);
			\coordinate (A) at (1,0,0);
			\coordinate (B) at (1,1,0);
			\coordinate (C) at (0,1,0);
			\coordinate (D) at (0,0,1);
			\coordinate (E) at (1,0,1);
			\coordinate (F) at (1,1,1);
			\coordinate (G) at (0,1,1);

			\draw[#1, thin, dashed] (D)--(E)--(F)--(G)--cycle;
			\draw[#1, thin, dashed] (E)--(A)--(B)--(C)--(G);
			\draw[#1, thin, dashed] (B)--(F);
			\draw[#1, thin, dashed] (D)--(O);
			\draw[#1, thick, dashed] (A)--(O)--(D);

			\ifx&#5&
			\else
			\node[tdplot_screen_coords, font=\scriptsize\bfseries, below, align=right] at (0,0.5,0.75) {#5};
			\fi
		\end{scope}
	}

	\cube[violet]{1.5}{0}{-4.5}{}
	\cube[violet]{1.5}{0}{-3.5}{}
	\cube[violet]{1.5}{0}{-2.5}{}
	\cube[violet]{1.5}{0}{-1.5}{}

	\cube[violet]{2.5}{0}{-3.5}{}
	\cube[violet]{2.5}{0}{-2.5}{}
	\cube[violet]{2.5}{0}{-1.5}{}

	\cube[violet]{3.5}{0}{-2.5}{}
	\cube[violet]{3.5}{0}{-1.5}{}

	\cube[violet]{5.5}{0}{-3.5}{}
	\cube[violet]{5.5}{0}{-2.5}{}
	\cube[violet]{5.5}{0}{-1.5}{}

	\cube[red]{1.5}{0}{0}{}
	\cube[red]{2.5}{0}{0}{}
	\cube[red]{3.5}{0}{0}{}
	\cube[red]{4.5}{0}{0}{}
	\cube[red]{5.5}{0}{0}{}

	\cube[gray]{0}{0}{0}{}

	\cube[Green4]{0}{1.5}{0}{}
	\cube[Green4]{0}{2.5}{0}{}
	\cube[Green4]{0}{3.5}{0}{}

	\cube[orange]{1.5}{1.5}{-4.5}{}
	\cube[orange]{1.5}{2.5}{-4.5}{}
	\cube[orange]{1.5}{3.5}{-4.5}{}

	\cube[orange]{1.5}{1.5}{-3.5}{}
	\cube[orange]{1.5}{2.5}{-3.5}{}
	\cube[orange]{1.5}{3.5}{-3.5}{}

	\cube[orange]{1.5}{1.5}{-2.5}{}
	\cube[orange]{1.5}{2.5}{-2.5}{}
	\cube[orange]{1.5}{3.5}{-2.5}{}

	\cube[orange]{1.5}{1.5}{-1.5}{}
	\cube[orange]{1.5}{2.5}{-1.5}{}
	\cube[orange]{1.5}{3.5}{-1.5}{}

	\cube[orange]{2.5}{1.5}{-3.5}{}
	\cube[orange]{2.5}{2.5}{-3.5}{}
	\cube[orange]{2.5}{3.5}{-3.5}{}

	\cube[orange]{2.5}{1.5}{-2.5}{}
	\cube[orange]{2.5}{2.5}{-2.5}{}
	\cube[orange]{2.5}{3.5}{-2.5}{}

	\cube[orange]{2.5}{1.5}{-1.5}{}
	\cube[orange]{2.5}{2.5}{-1.5}{}
	\cube[orange]{2.5}{3.5}{-1.5}{}

	\cube[orange]{3.5}{1.5}{-2.5}{}
	\cube[orange]{3.5}{2.5}{-2.5}{}
	\cube[orange]{3.5}{3.5}{-2.5}{}

	\cube[orange]{3.5}{1.5}{-1.5}{}
	\cube[orange]{3.5}{2.5}{-1.5}{}
	\cube[orange]{3.5}{3.5}{-1.5}{}

	\cube[orange]{5.5}{1.5}{-3.5}{}
	\cube[orange]{5.5}{2.5}{-3.5}{}
	\cube[orange]{5.5}{3.5}{-3.5}{}

	\cube[orange]{5.5}{1.5}{-2.5}{}
	\cube[orange]{5.5}{2.5}{-2.5}{}
	\cube[orange]{5.5}{3.5}{-2.5}{}

	\cube[orange]{5.5}{1.5}{-1.5}{}
	\cube[orange]{5.5}{2.5}{-1.5}{}
	\cube[orange]{5.5}{3.5}{-1.5}{}

	\draw[->, thick] (1.5,0,2) -- (6.5,0,2) node[anchor=north]{$R(\esection, \{\mparsedpaper[0]\}) = 5$};
	\draw[->, thick] (7,0,-0.5) -- (7,0,-3.5) node[anchor=west]{$R(\eparagraph, \{\mparsedpaper[0], \msection[4]\}) = 3$};
	\draw[->, thick] (-0.5,1.5,1) -- (-0.5,4.5,1) node[anchor=east]{$R(\ecaptionedfig, \{\mparsedpaper[0]\}) = 3$};

	\node[tdplot_screen_coords, anchor=east] at (0,0,0.5) {$R(\eparsedpaper, \emptyset) = 1$};

	\node[tdplot_screen_coords, anchor=north west] at (12, 7.5, 0) {
	    \begin{minipage}{5cm}
			\begin{align*}			
				R &= \{ \\
					&\qquad (\eparsedpaper, \emptyset, 1), \\
					&\qquad (\esection, \{\mparsedpaper[0]\}, 5), \\
					&\qquad (\eparagraph, \{\mparsedpaper[0], \msection[0]\}, 4), \\
					&\qquad (\eparagraph, \{\mparsedpaper[0], \msection[1]\}, 3), \\
					&\qquad (\eparagraph, \{\mparsedpaper[0], \msection[2]\}, 2), \\
					&\qquad (\eparagraph, \{\mparsedpaper[0], \msection[3]\}, 0), \\
					&\qquad (\eparagraph, \{\mparsedpaper[0], \msection[4]\}, 3), \\
					&\qquad (\ecaptionedfig, \{\mparsedpaper[0]\}, 3) \\
				\}
			\end{align*}
	    \end{minipage}
	};
\end{tikzpicture}
    \caption{
        A resolution map \(R\) on the dimension space \(\set{\dimP, \dimS, \dimG, \dimF}\) from Example~\ref{ex:motex_dimension_space} defining a ragged profile.
        For the single \textcolor{gray}{paper} shown, there are 3 \textcolor{Green4}{figure}s and 5 \textcolor{red}{section}s; each \textcolor{red}{section} contains 4, 3, 2, 0, and 3 \textcolor{violet}{paragraph}s, respectively.
        This configuration uniquely defines the 36 possible positions for \textcolor{orange}{relevance score}s, which are computed for each \textcolor{violet}{paragraph} and each \textcolor{Green4}{figure}.
    }
    \label{fig:good_resolution_map}
\end{figure}

As shown in Figure~\ref{fig:good_resolution_map}, a well-chosen resolution map may shape a ragged profile.
However, since each resolution carries information about the ancestor dimensions, we must verify that the resolutions do not contradict themselves.
Specifically, in each occurrence of a position \(c\), we must check each dimension \(d \in \dom(c)\) to see if the resolution of \(d\) allows the value of \(c(d)\) at that position.
For this, we define a condition that verifies whether a position \(c\) over some dimensions is valid under a resolution map \(R\).

\begin{definition}[In-bounds condition]\label{def:in-bounds_condition}
    For a resolution map \(R\) defined on a dimension space \((\mathcal{D}, \prec)\) and a partial function \(c: \mathcal{D} \rightharpoonup \nat\), we call the following the \emph{in-bounds condition}.
    \[
        c \ib R \iff \forall d \in \dom(c).\;\left(d, c|_{d^\Downarrow}\right) \in \dom(R) \wedge c(d) < R\left(d, c|_{d^\Downarrow}\right)
    \]
    Note that for \(c \ib R\) to hold true, \(\dom(c)\) must be closed, since it requires \(c|_{d^\Downarrow}\) be total over \(d^\Downarrow\) for all \(d\).
    When \(\dom(c)\) is a principal dependency space \((d')^\Downarrow\), we refer to each resolution \(\left(d, c|_{d^\Downarrow}, \ell\right)\) as an \emph{ancestor} of any resolution \((d', c, \ell')\) with \(\ell' \in \nat\), whereas \((d', c, \ell')\) is a \emph{descendant}.
\end{definition}

As the in-bounds condition provides the means to verify the validity of positions, we may now define which resolution maps are well-formed.

\begin{definition}[Shapes]\label{def:shapes}
    On a dimension space \((\mathcal{D}, \prec)\), a \emph{shape} is a resolution map \(R\) that satisfies the following condition.
    \[
        \forall d \in \mathcal{D}.\; \forall c : d^\Downarrow \to \nat.\; (d, c) \in \dom(R) \leftrightarrow c \ib R
    \]
\end{definition}

The \emph{only-if} direction \((d, c) \in \dom(R) \rightarrow c \ib R\) necessitates that ancestor resolutions constrain all \(c\).
The \emph{if} direction \(c \ib R \rightarrow (d, c) \in \dom(R)\) further enforces that there are no unresolved lengths; that is, if the ancestor resolutions rule that a position \(c: d^\Downarrow \to \nat\) is in-bounds, then there must be a resolution \((d, c, \ell)\).
Also note that for minimal dimensions \(d\) whose \(d^\Downarrow = \emptyset\), \(\emptyset \ib R\) is vacuously true for the only function \(\emptyset: \emptyset \to \nat\), so the definition maintains that there is precisely one resolution \((d, \emptyset, \ell)\) for such dimensions.

Coordinates are now naturally defined as maps within the bounds set by a shape.

\begin{definition}[Coordinates]\label{def:coordinates}
    Given a dimension space \((\mathcal{D}, \prec)\), a shape \(R\), and a closed subspace \(\mathcal{F} \subseteq \mathcal{D}\), we have the \emph{coordinate space} \(\coordsp{\mathcal{F}}{R}\):
    \[
        \coordsp{\mathcal{F}}{R} = \set{c: \mathcal{F} \to \nat \mid c \ib R},
    \]
    where each \(c: \coord{\mathcal{F}}{R}\) is a \emph{coordinate} over \(\mathcal{F}\).
\end{definition}

Collecting the coordinates with a shared ancestor coordinate \(c\) gives us a \emph{subcoordinate space}.

\begin{definition}[Siblings and Subcoordinates]\label{def:siblings_and_subcoordinates}
    Given a coordinate space \(\coordsp{\mathcal{F}}{R}\), a closed subspace \(\mathcal{F}' \subseteq \mathcal{F}\), and a coordinate \(c^\bot: \coord{\mathcal{F}'}{R}\), we have the \(c^\bot\)\emph{-sibling coordinates} \(\coordsp{\mathcal{F}}{R}_{c^\bot}\):
    \[
        \coordsp{\mathcal{F}}{R}_{c^\bot} = \set{c^+: \coord{\mathcal{F}}{R} \mid c^+|_{\mathcal{F}'} = c^\bot}.
    \]
    Writing \(\mathcal{E} = \mathcal{F} \setminus \mathcal{F}'\), removing the common ancestor coordinate \(c^\bot\) from each sibling coordinate gives us the \emph{subcoordinate space} \(\subcoordsp{\mathcal{E}}{c^\bot}{R}\):
    \[
        \subcoordsp{\mathcal{E}}{c^\bot}{R} = \set{c^+|_{\mathcal{E}} \mid c^+ \in \coordsp{\mathcal{F}}{R}_{c^\bot}}.
    \]
    Each \(c^\top: \subcoord{\mathcal{E}}{c^\bot}{R}\) is a \emph{subcoordinate} over \(\mathcal{E}\) and \(c^\bot\).
\end{definition}

Continuing the example, fixing \(c^\bot = \set{\dimP \mapsto 0, \dimF \mapsto 0}\) gives the \(c^\bot\)-sibling coordinates over \(\mathcal{F} = \set{\dimP, \dimF, \dimS, \dimG}\): every relevance-score coordinate for figure~0 of paper~0.
Visually, this would mean taking the lowermost slice of the 3D array of relevance scores in Figure~\ref{fig:good_resolution_map}.

We conclude by stating some properties of coordinates and subcoordinates to demonstrate their well-behavedness.

\begin{proposition}\label{prop:properties_of_coordinates}
    Given a dimension space \((\mathcal{D}, \prec)\) and a shape \(R\), we have the following:
    \begin{enumerate}
        \item For closed subspaces \(\mathcal{F}' \subseteq \mathcal{F} \subseteq \mathcal{D}\) and a coordinate \(c: \coord{\mathcal{F}}{R}\), \(c|_{\mathcal{F}'} \in \coordsp{\mathcal{F}'}{R}\).
        \item For a closed \(\mathcal{F} \subseteq \mathcal{D}\), \(\subcoordsp{\mathcal{F}}{\emptyset}{R} = \coordsp{\mathcal{F}}{R}\).
        \item For a closed \(\mathcal{F} \subseteq \mathcal{D}\), a subspace \(\mathcal{E} \subseteq \mathcal{F}\) with closed \(\mathcal{F} \setminus \mathcal{E}\), and a coordinate \(c^\bot: \coord{\mathcal{F} \setminus \mathcal{E}}{R}\), there exists a \emph{restricted shape} \(R|_{(\mathcal{E}, c^\bot)}\), a shape on \((\mathcal{E}, \prec|_\mathcal{E})\), such that \[
                  \subcoordsp{\mathcal{E}}{c^\bot}{R} = \coordsp{\mathcal{E}}{R|_{(\mathcal{E}, c^\bot)}}.
              \]
              That is, we can interpret each subcoordinate space as a coordinate space when the shape is appropriately restricted.
              We have \(R|_{(\mathcal{E}, \emptyset)} \subseteq R\) when \(\mathcal{E} = \mathcal{F}\).
    \end{enumerate}
\end{proposition}

\subsection{Arrays}
\label{subsec:arrays}

We arrive at the final step in formulating the dimension system, which is associating the system with real-life ragged arrays.
Since we have already established the shape of possible coordinates over dimensions, this process is straightforward.

\begin{definition}[Arrays]\label{def:arrays}
    For a dimension space \((\mathcal{D}, \prec)\), a shape \(R\), a closed subspace \(\mathcal{F} \subseteq \mathcal{D}\), and a space of values \(V\), an \emph{array} is a function \[
        \mathtt{arr}: \coordsp{\mathcal{F}}{R} \to V.
    \]
    When this function is not total, we call it a \emph{partial array} \(\overline{\mathtt{arr}}: \coordsp{\mathcal{F}}{R} \rightharpoonup V\).
\end{definition}

Also, when we fix a coordinate over the ancestor dimensions, we get a smaller array over the descendant dimensions.
For example, fixing \(c = \set{\dimP \mapsto 0}\) on the array of \texttt{CaptionedFig}s should give the subarray \(\texttt{CaptionedFig}[\set{\dimP \mapsto 0}]\).
This array represents the figures of paper~0, and should naturally be indexed by the subcoordinates over \(\set{\dimF}\).

\begin{definition}[Subarrays]\label{def:subarrays}
    For an array \(\mathtt{arr}: \coordsp{\mathcal{F}}{R} \to V\), a subspace \(\mathcal{E} \subseteq \mathcal{F}\) with closed \(\mathcal{F} \setminus \mathcal{E}\), and a coordinate \(c: \coord{\mathcal{F} \setminus \mathcal{E}}{R}\), the \emph{subarray} \(\mathtt{arr}[c]\) is a function
    \[
        \mathtt{arr}[c]: \subcoordsp{\mathcal{E}}{c}{R} \to V
    \]
    that satisfies\footnote{
        \(\sqcup\) denotes the disjoint union of sets when their disjointness is guaranteed by construction.
        We interpret functions as their graphs, i.e., sets of ordered pairs.
    }
    \[
        \forall c^\ast: \subcoord{\mathcal{E}}{c}{R}.\;\mathtt{arr}[c](c^\ast) = \mathtt{arr}(c^\ast \sqcup c).
    \]
    This is equivalent to restricting \(\mathtt{arr}\) to the \(c\)-sibling coordinates and then removing the common ancestor \(c\) from each coordinate in the domain.
\end{definition}

\section{Incremental Resolutions}
\label{sec:incremental_resolutions}

Operon handles data processing tasks and data entries as arrays on a global dimension space and a shared shape.
As such, we may understand Operon as a system that computes for the completion of all defined arrays \(\mathtt{arr}: \coordsp{\mathcal{F}}{R} \to V\), where \(\mathtt{arr}\) conceptually represents either instances of a processing task (e.g., \(\mathtt{vlm\_evaluate}\)) or a data collection (e.g., \(\mathtt{Relevance}\)).
Assuming that the variables in the signature \(\mathcal{D}, R, \mathcal{F}, V\) are known upfront, the system becomes a simple fill-in-the-blanks engine that populates the values for all coordinates in \(\coordsp{\mathcal{F}}{R}\).

However, while \(\mathcal{D}\), \(\mathcal{F}\), and \(V\) can indeed be determined statically (\textsection\ref{subsec:operon_overview}), the shape \(R\) does not follow suit.
The system does not have any knowledge of the desired shape until the execution of user-defined tasks.
Instead, specific tasks (e.g., \(\mathtt{extract\_sections}\) or \(\mathtt{filter\_aggregate}\)) produce new resolutions that would ideally accumulate to form a final shape.
For the runtime system to behave predictably, we must be able to express the intermediate states of \(R\) as new resolutions are added.
To this end, we establish which states are acceptable as intermediate resolution maps, provide a confluent and terminating algorithm that maintains this property, and extend our definitions of coordinates to handle unknowns.

\subsection{Partial Shapes}
\label{subsec:partial_shapes}

Recall the definition of shapes in Definition~\ref{def:shapes}.
For a resolution map to be a shape, coordinates mentioned in its domain must be in-bounds of itself (\((d, c) \in \dom(R) \rightarrow c \ib R\)), and all in-bounds coordinates must appear in its domain (\(c \ib R \rightarrow (d, c) \in \dom(R)\)).
While the former connotes noncontradiction, the latter condition requires the resolution map to be \emph{complete} in the sense that no obvious holes are left unfilled.
Relaxing the condition to allow incompleteness gives us a natural definition for \emph{partial shapes}.

\begin{definition}[Partial shapes]\label{def:partial_shapes}
    On a dimension space \((\mathcal{D}, \prec)\), a \emph{partial shape} is a resolution map \(\overline{R}\) that satisfies the following condition.
    \[
        \forall d \in \mathcal{D}.\; \forall c: d^\Downarrow \to \nat.\; (d, c) \in \dom(\overline{R}) \rightarrow c \ib \overline{R}
    \]
    In particular, if a partial shape \(\overline{R}\) is not a shape, we call \(\overline{R}\) an \emph{incomplete shape}.
    By contrast, we may use the terms \emph{shape} and \emph{complete shape} interchangeably.
\end{definition}

Partial shapes behave as valid intermediate states while building towards a complete shape, starting from the trivial empty map \(\emptyset\).
In the motivating pipeline, for instance, the map \(\overline{R} = \set{(\dimP, \emptyset) \mapsto 5}\) is a partial shape that occurs when the number of papers is known to be 5.
However, the number of figures, sections, or paragraphs for each paper is not yet known.
As tasks complete and emit their output lengths, the corresponding resolutions are added one by one, extending \(\overline{R}\) until it becomes a complete shape covering all dimensions.

From a top-down perspective, partial shapes are initial segments of complete shapes when topologically sorted with respect to the ancestor-descendant relation of resolutions (as defined in Def.~\ref{def:in-bounds_condition}).
Any such topological sorting would therefore list a sequence of resolutions whose cumulative addition produces a chain of partial shapes, eventually resulting in the desired complete shape.
However, since we do not have the final shape in advance, we take an approach where we start from a partial shape (often the empty map), repeatedly produce a resolution that does not contradict the current partial shape, and extend the partial shape with that resolution.
We call such resolutions \emph{compatible} with the given partial shape.

\begin{definition}[Compatible resolutions]\label{def:compatible_resolutions}
    For a partial shape \(\overline{R}\) on \((\mathcal{D}, \prec)\), if a dimension \(d \in \mathcal{D}\) and a function \(c^\Downarrow: d^\Downarrow \to \nat\) satisfy \[
        (d, c^\Downarrow) \bowtie \overline{R} \iff c^\Downarrow \ib \overline{R} \wedge (d, c^\Downarrow) \notin \dom(\overline{R}),
    \]
    then the pair \((d, c^\Downarrow)\) is \emph{compatible} with \(\overline{R}\).
    Any resolution \((d, c^\Downarrow, \ell)\) with \(\ell \in \nat\) is also said to be compatible with \(\overline{R}\).
\end{definition}

In the example above, \((\dimF, \set{\dimP \mapsto 0})\) is compatible with \(\overline{R} = \set{(\dimP, \emptyset) \mapsto \ell}\) for any \(\ell > 0\): the coordinate \(\set{\dimP \mapsto 0}\) is in-bounds (since paper~0 exists), and \((\dimF, \set{\dimP \mapsto 0})\) is not yet in \(\dom(\overline{R})\).
Running \texttt{extract\_captioned\_figures} on paper~0 and recording its output length (say, 10) adds the compatible resolution \((\dimF, \set{\dimP \mapsto 0}, 10)\) to \(\overline{R}\).

The following lemma states that a compatible resolution, as defined above, indeed preserves the partial shape property on extension.

\begin{lemma}\label{lem:compatibility}
    For a partial shape \(\overline{R}\) and a resolution \((d, c^\Downarrow, \ell)\), the extension \(\overline{R}\{(d, c^\Downarrow) \mapsto \ell\}\) stays a partial shape if \((d, c^\Downarrow) \bowtie \overline{R}\).
\end{lemma}

\subsection{Incremental Construction}
\label{subsec:incremental_construction}

As briefly mentioned in the previous subsection, we aim to incrementally construct a complete shape by starting from an initial partial shape \(\emptyset\) and repeatedly adding compatible resolutions.
A simple linear algorithm that performs this task is shown in Alg.~\ref{alg:incremental_construction}.
This algorithm repeatedly finds a compatible \((d, c)\) pair, queries an oracle function \(\pi\) for the desired length \(\ell\) at that coordinate, and extends the current partial shape \(\overline{R}\) with the new resolution \((d, c, \ell)\).
In practice, a query to \(\pi\) would represent the execution of a user-defined task that produces the desired length.
In the motivating pipeline, \(\pi(\dimF, \set{\dimP \mapsto i})\) corresponds to running \texttt{extract\_captioned\allowbreak\_figures} on paper~\(i\) and returning the number of figures produced; \(\pi(\dimG, \set{\dimP \mapsto i, \dimS \mapsto j})\) corresponds to running \texttt{extract\_paragraphs} on section~\(j\) of paper~\(i\).

We assume the oracle function \(\pi: \bigcup_{d \in \mathcal{D}} \left(\set{d} \times [d^\Downarrow \to \nat] \right) \to \nat\) is a total and deterministic function for this discussion; we loosen this assumption to stochastic functions later in this section.

\begin{algorithm}[t]
    \caption{Incremental construction of a shape}
    \label{alg:incremental_construction}
    \small
    \Input{A dimension space \((\mathcal{D}, \prec)\) and a function \(\pi: (d, c) \mapsto \ell\)}
    \Output{A complete shape \(R\)}

    \Proc{main}{
        \(\overline{R} \longleftarrow \emptyset\)\;
        \While{\(\overline{R}\) is incomplete}{
            \(C \longleftarrow \set{(d, c) \mid (d, c) \bowtie \overline{R}}\)\tcp*{never empty due to Thm.~\ref{thm:progress}}
            \((d, c) \longleftarrow\) element in \(C\)\;
            \(\ell \longleftarrow \pi(d, c)\)\;
            \(\overline{R} \longleftarrow \overline{R}\{(d, c) \mapsto \ell\}\)\;
        }
        \Return{\(\overline{R}\)}\;
    }
\end{algorithm}

The correctness of Alg.~\ref{alg:incremental_construction} relies on two assumptions: first, that there is always at least one compatible resolution to add to an incomplete shape, and second, that the process of adding compatible resolutions eventually leads to a complete shape.
We formalize these assumptions in Thms.~\ref{thm:progress} and \ref{thm:termination}, respectively.

\begin{theorem}[Progress]\label{thm:progress}
    A partial shape has a compatible resolution if and only if it is incomplete.
\end{theorem}

\begin{theorem}[Termination]\label{thm:termination}
    There is no infinite sequence of partial shapes where each step adds a resolution.
\end{theorem}

We may further extend the above algorithm for parallel execution, as shown in Alg.~\ref{alg:parallel_incremental_construction}.
In this version, multiple worker threads each own a compatible \((d, c)\) pair to process, which allows concurrent queries to the oracle function \(\pi\).
To avoid duplicate work, the main thread keeps track of the \((d, c)\) pairs that workers are currently processing in a set \(S\) local to the main thread.
If all compatible pairs are being processed, the main thread waits for any worker to finish and update the shared shape \(\overline{R}\) before proceeding.

For this parallelization to be correct, each worker's \((d, c)\) pair must remain compatible with the shared shape \(\overline{R}\) regardless of other workers' actions.
Thm.~\ref{thm:local_commutativity} ensures this by stating that adding a compatible resolution does not invalidate other compatible resolutions.

\begin{algorithm}[t]
    \caption{Parallel incremental construction of a shape}
    \label{alg:parallel_incremental_construction}
    \small
    \Input{A dimension space \((\mathcal{D}, \prec)\) and a function \(\pi: (d, c) \mapsto \ell\)}
    \Output{A complete shape \(R\)}

    \Proc{main}{
        \(\overline{R} \longleftarrow \emptyset\)\tcp*{as a shared reference with concurrent appends}
        \(S \longleftarrow \emptyset\)\tcp*{"seen" \((d, c)\) pairs, thread-local}
        \Loop{
            \(\overline{R}_s \longleftarrow\) snapshot of \(\overline{R}\)\;
            \lIf{\(\overline{R}_s\) is complete}{\Return{\(\overline{R}_s\)}}
            \(C \longleftarrow \set{(d, c) \mid (d, c) \bowtie \overline{R}_s}\)\tcp*{never empty due to Thm.~\ref{thm:progress}}
            \uIf(\tcp*[f]{all compatible pairs are being processed}){\(C \subseteq S\)}{
                wait until \(\overline{R} \neq \overline{R}_s\)\;
                \KwSty{continue}\;
            }
            \Else{
                \((d, c) \longleftarrow\) element in \(C \setminus S\)\;
                \(S \longleftarrow S \cup \set{(d, c)}\)\;
                \KwSty{spawn} \ProgSty{worker}\((\overline{R}, d, c)\)\;
            }
        }
    }
    \Proc{worker \((\overline{R}, d, c)\)}{
        \(\ell \longleftarrow \pi(d, c)\)\;
        \(\overline{R} \longleftarrow \overline{R}\{(d, c) \mapsto \ell\}\)\tcp*{atomic append to shared reference}
        \Return{}\;
    }
\end{algorithm}

\begin{theorem}[Local commutativity]\label{thm:local_commutativity}
    If \((d, c) \bowtie \overline{R}\) and \((d', c') \bowtie \overline{R}\) with \((d, c) \neq (d', c')\), then \((d', c') \bowtie \overline{R}\{(d, c) \mapsto \ell\}\) for any \(\ell\).
\end{theorem}

Local commutativity here plays the same structural role as commutativity in concurrent program verification~\cite{farzan2022sound,farzan2023stratified} and in mechanized reasoning about structured data~\cite{gladshtein2024mechanised}: it ensures that independent resolution steps can be reordered without affecting the outcome.

The following corollary follows immediately from the above theorems and states that the incremental construction process yields a deterministic complete shape for non-stochastic user functions.

\begin{corollary}[Determinism]\label{cor:determinism}
    Under a fixed function \(\pi: \bigcup_{d \in \mathcal{D}} \left(\set{d} \times [d^\Downarrow \to \nat] \right) \to \nat\), any non-starving execution of Alg.~\ref{alg:incremental_construction} or Alg.~\ref{alg:parallel_incremental_construction} terminates and returns the same complete shape.
\end{corollary}

Finally, loosening the constraint on \(\pi\) to independently stochastic functions, we derive a central result that supports the confluent property of our incremental algorithm.
Here, \(\dist(X)\) denotes the set of finitely supported probability distributions over a set \(X\)~\cite{DIAZCARO2018115}.

\begin{theorem}[Distribution confluence of shapes]\label{thm:distribution_confluence_of_shapes}
    Under a stochastic function \(\pi\) that samples independently from a fixed function \(\rho: \bigcup_{d \in \mathcal{D}} \left(\set{d} \times [d^\Downarrow \to \nat] \right) \to \dist(\nat)\), any non-starving execution of Alg.~\ref{alg:incremental_construction} or Alg.~\ref{alg:parallel_incremental_construction} terminates and returns a complete shape whose distribution is invariant to the order of scheduling.
\end{theorem}

\subsection{Coordinates with Unknowns}
\label{subsec:coordinates_with_unknowns}

Incomplete shapes naturally lead to the question of how to define coordinates over them.
Operon might find some data entries before the shape for that data is fully known, and those entries must still be addressable.
For example, the \textcolor{orange}{relevance score} at \(\set{\dimP \mapsto 0, \dimS \mapsto 0, \dimG \mapsto 0, \dimF \mapsto 0}\) in Figure~\ref{fig:good_resolution_map} could be computed before the resolution \((\dimG, \set{\dimP \mapsto 0, \dimS \mapsto 4}, 3)\) becomes known.
Since the array of \textcolor{orange}{relevance score}s is defined over \(\mathcal{F} = \set{\dimP, \dimS, \dimG, \dimF}\), the coordinate space \(\coordsp{\mathcal{F}}{R}\) cannot be defined as in Def.~\ref{def:coordinates} without the above resolution.
We wish to extend the definition of coordinates to allow such \emph{unknown} values while keeping fully resolved coordinates accessible.

The following definition achieves this by permitting coordinates to be partial functions over \(\mathcal{F}\).
The coordinate must be in-bounds for its domain, but when a resolution is missing for some dimension \(d\), the coordinate must also omit \(d\) from its domain.

\begin{definition}[Coordinates with unknowns]\label{def:coordinates_with_unknowns}
    For a partial shape \(\overline{R}\) on \((\mathcal{D}, \prec)\) and a closed \(\mathcal{F} \subseteq \mathcal{D}\), a \emph{coordinate} \(c: \ucoord{\mathcal{F}}{\overline{R}}\) is a partial function \(c: \mathcal{F} \rightharpoonup \nat\) that satisfies the following.
    \begin{enumerate}
        \item \(c \ib \overline{R}\).
        \item \(\forall d \in \mathcal{F} \setminus \dom(c).\; \left(d, c|_{d^\Downarrow}\right) \notin \dom(\overline{R})\).
    \end{enumerate}
    The \emph{coordinate space} \(\ucoordsp{\mathcal{F}}{\overline{R}}\) denotes the set of such coordinates.

    Note that when \(c\) is total over \(\mathcal{F}\), the second condition is vacuously satisfied.
    We call such coordinates \emph{fully resolved} and, by slight abuse of notation, write \(c: \coord{\mathcal{F}}{\overline{R}}\), since in this case the above definition coincides with Def.~\ref{def:coordinates}.
    Naturally, \(\coordsp{\mathcal{F}}{\overline{R}} \subseteq \ucoordsp{\mathcal{F}}{\overline{R}}\).
\end{definition}

In the motivating pipeline, consider the state after paper~0's figures are known (\((\dimF, \set{\dimP \mapsto 0}) \in \dom(\overline{R})\)) but before sections are resolved.
A coordinate such as \(\set{\dimP \mapsto 0, \dimF \mapsto 0}\) over \(\mathcal{F} = \set{\dimP, \dimF, \dimS, \dimG}\) is then a coordinate with unknowns: the \(\dimS\) and \(\dimG\) values are absent because \((\dimS, \set{\dimP \mapsto 0}) \notin \dom(\overline{R})\).
Recall that \texttt{vlm\_evaluate} jobs must iterate over all \((\dimP, \dimF, \dimS, \dimG)\)-tuples, which is not yet enumerable because the section and paragraph counts are unknown.
The coordinate with unknowns \(\set{\dimP \mapsto 0, \dimF \mapsto 0}\) becomes a sensible placeholder to represent all the \texttt{vlm\_evaluate} jobs at figure~0 of paper~0 in this state of partial knowledge.
This idea to represent unresolved jobs with coordinates with unknowns is central to the design of Operon, as it allows tracking the state of jobs as concrete objects (called \emph{tickets}; see Def.~\ref{def:tickets}) without enumerating the entire coordinate space.

This extension aligns well with the original definition, as the definition without unknowns becomes a special case where \(\overline{R}\) is complete.
We therefore characterize sibling coordinates, subcoordinates, arrays, and subarrays without change in their definitions, except that the shapes in those definitions may now be partial.

\begin{proposition}\label{prop:unknown_extension}
    Def.~\ref{def:coordinates_with_unknowns} is a strict extension of the original definition of coordinates in Def.~\ref{def:coordinates}.
    That is, \(\ucoordsp{\mathcal{F}}{R} = \coordsp{\mathcal{F}}{R}\) when \(R\) is a complete shape.
\end{proposition}

The following theorem portrays how the coordinate space changes as we add new compatible resolutions to a partial shape.
Intuitively, adding a new resolution for a dimension \(d\) at a coordinate \(c^\Downarrow\) effectively \emph{explodes} the \(c^\Downarrow\)-sibling coordinates along that dimension, where each \(c^\Downarrow\)-sibling \(c\) is replaced by \(\ell\) new coordinates \(c\{d \mapsto 0\}, c\{d \mapsto 1\}, \ldots, c\{d \mapsto \ell - 1\}\).

\begin{theorem}[Coordinate explosion]\label{thm:coordinate_explosion}
    For a coordinate space \(\ucoordsp{\mathcal{F}}{\overline{R}}\) and a resolution \((d, c^\Downarrow, \ell)\) with \(d \in \mathcal{F}\), \(c^\Downarrow: \coord{d^\Downarrow}{\overline{R}}\), and \((d, c^\Downarrow) \bowtie \overline{R}\), only the \(c^\Downarrow\)-sibling coordinates \(U = \ucoordsp{\mathcal{F}}{\overline{R}}_{c^\Downarrow}\) are affected by the addition of \((d, c^\Downarrow) \mapsto \ell\) into \(\overline{R}\) as follows.
    \begin{enumerate}
        \item \(\forall c' \in U.\; d \notin \dom(c')\).
              I.e., none of the \(c^\Downarrow\)-siblings already have a value for \(d\) filled in.
        \item Writing \(\overline{R}_\ell = \overline{R}\{(d, c^\Downarrow) \mapsto \ell\}\), \[
                  \ucoordsp{\mathcal{F}}{\overline{R}_\ell} = \left(\ucoordsp{\mathcal{F}}{\overline{R}} \setminus U\right) \sqcup \bigsqcup_{c' \in U} {\bigsqcup_{i=0}^{\ell-1} \set{c'\{d \mapsto i\}}}.
              \]
    \end{enumerate}
\end{theorem}

Returning to the example, when \texttt{extract\_sections} runs on paper~0 and emits 5 sections, we add \((\dimS, \set{\dimP \mapsto 0}, 5)\) to \(\overline{R}\).
This explodes all coordinates whose \(\dimS\) slot was previously absent at \(\dimP \mapsto 0\).
The coordinate \(\set{\dimP \mapsto 0, \dimF \mapsto 0}\) is therefore replaced by 5 ``exploded'' coordinates \(\set{\dimP \mapsto 0, \dimF \mapsto 0, \dimS \mapsto i}\) for \(i \in \set{0, 1, 2, 3, 4}\), each still awaiting paragraph-count resolution for their respective section.

The following result follows immediately from the above theorem.
It acts as a central tool for calculating the number of dependencies of a job in Operon, as will be described in Section~\ref{subsec:implementation}.
This provides a way to lazily count the number of siblings over a specific coordinate without explicitly enumerating the coordinate space.

\begin{corollary}[Sibling counts]\label{cor:sibling_counts}
    Consider the same setting as in Thm.~\ref{thm:coordinate_explosion}.
    For a closed subspace \(\mathcal{F}' \subseteq \mathcal{F}\) and a coordinate \(c^\bot: \ucoord{\mathcal{F}'}{\overline{R}}\), the number of \(c^\bot\)-siblings changes as follows.
    \begin{enumerate}
        \item If \(c^\bot\) is irrelevant to the explosion's position \(c^\Downarrow\), the count of \(c^\bot\)-siblings is unchanged: \[
                  c^\bot|_{d^\Downarrow} \not\subseteq c^\Downarrow \implies \left|\ucoordsp{\mathcal{F}}{\overline{R}_\ell}_{c^\bot}\right| - \left|\ucoordsp{\mathcal{F}}{\overline{R}}_{c^\bot}\right| = 0.
              \]
        \item If the explosion affects \(c^\bot\) directly, then \(c^\bot\) is no longer a coordinate under \(\overline{R}_\ell\), so we count the siblings of the exploded versions of \(c^\bot\) instead. Each new coordinate \(c^\bot\set{d \mapsto i}\) inherits the sibling count without increment: \[
                  c^\bot|_{d^\Downarrow} \subseteq c^\Downarrow \wedge d \in \mathcal{F}' \implies \forall i \in \nat \cap [0, \ell).\; \left|\ucoordsp{\mathcal{F}}{\overline{R}_\ell}_{c^\bot \set{d \mapsto i}} \right| - \left|\ucoordsp{\mathcal{F}}{\overline{R}}_{c^\bot}\right| = 0.
              \]
        \item If the explosion affects the descendants but does not explode \(c^\bot\) directly, the count is incremented by \(\ell - 1\) for each \(c^\Downarrow\)-sibling: \[
                  c^\bot|_{d^\Downarrow} \subseteq c^\Downarrow \wedge d \notin \mathcal{F}' \implies \left|\ucoordsp{\mathcal{F}}{\overline{R}_\ell}_{c^\bot}\right| - \left|\ucoordsp{\mathcal{F}}{\overline{R}}_{c^\bot}\right| = (\ell - 1) \left|\ucoordsp{\mathcal{F}}{\overline{R}}_{c^\Downarrow}\right|.
              \]
    \end{enumerate}
\end{corollary}

\section{Operon}
\label{sec:operon}

In this section, we present the system design for Operon.

\subsection{Overview}
\label{subsec:operon_overview}

Operon takes advantage of Rust's procedural macro feature to accept pipeline definitions in a concise DSL.
During macro expansion, it inspects the declared pipeline and generates the code necessary for execution.
Figure~\ref{fig:operon_syntax} describes the formal syntax of our DSL.
A concrete task definition, as presented in Figure~\ref{fig:operon-scicap}, corresponds one-to-one with a task \(\left< f, s_\text{out}, \overrightarrow{s_{\text{in, }i}}, \mathcal{F}, n \right>\) with \(s_\text{out} = \left< \tau_\text{out}, \mathcal{E}_\text{out} \right>\) and \(s_{\text{in, }i} = \left< \tau_{\text{in, }i}, \mathcal{E}_{\text{in, }i} \right>\).
For example, the task definition
\begin{center}
\begin{tabular}{c}
\begin{lstlisting}[numbers=none,basicstyle=\small\ttfamily]
RelevantPg<r> = filter_aggregate(Paragraph<s, g>, Relevance<s, g>) for(8) p, f;
\end{lstlisting}
\end{tabular}
\end{center}
corresponds to the task \[
    \left<
        \texttt{filter\_aggregate},
        \left< \texttt{RelevantPg}, \set{\texttt{r}} \right>,
        \left[ \left< \texttt{Paragraph}, \set{\texttt{s}, \texttt{g}} \right>, \left< \texttt{Relevance}, \set{\texttt{s}, \texttt{g}} \right> \right],
        \set{\texttt{p}, \texttt{f}},
        8
    \right>.
\]

\begin{figure}[t]
\centering\small
\begin{align*}
    \textbf{Pipeline} &\quad p ::= \vec{t} \\
    \textbf{Task} &\quad t ::= \left< f, s_{\text{out}}, \overrightarrow{s_{\text{in, }i}}, \mathcal{F}, n \right> &&\text{unique }\tau_{\text{in, }i} \\
    \textbf{Entity signature} &\quad s ::= \left< \tau, \mathcal{E} \right> \\
    \textbf{Entity signature map} &\quad \Sigma ::= \vec{s} &&\text{unique }\tau \\
    \textbf{Dim. space} &\quad \mathcal{D}, \mathcal{E}, \mathcal{F} ::= \vec{d} \\
\end{align*}
\[
    \textbf{Concurrency} \enspace n \in \mathbb{Z}^+ \qquad
    \textbf{Entity type} \enspace \tau \in \textbf{Type vars} \qquad
    \textbf{Dimension} \enspace d, e \in \textbf{Idents}
\]
\[
    \textbf{Function} \enspace f : \prod_{i} \Bigl(\underbrace{\texttt{list ... list}}_{|\mathcal{E}_{\text{in, }i}|} \tau_{\text{in, }i}\Bigr) \rightarrow \underbrace{\texttt{list ... list}}_{|\mathcal{E}_{\text{out}}|} \tau_{\text{out}} \qquad
    \textbf{Dep. rel.} \enspace \prec \;\subseteq \mathcal{D} \times \mathcal{D}
\]
\caption{Syntax of the Operon domain-specific language.}
\label{fig:operon_syntax}
\end{figure}

\begin{figure}[t]
\centering\small
\begin{equation*}\boxed{
    \frac{\vphantom{d}}{(\emptyset, \emptyset, \emptyset) \mid () \vdash (\emptyset, \emptyset, \emptyset)} \enspace (\textsc{Unit})
}\end{equation*}
\begin{equation*}\boxed{
    \frac{(\emptyset, \emptyset, \emptyset) \mid \vec{t} \vdash (\mathcal{D}_1, \prec_1, \Sigma_1) \qquad (\mathcal{D}_1, \prec_1, \Sigma_1) \mid t' \vdash (\mathcal{D}_2, \prec_2, \Sigma_2)}{(\emptyset, \emptyset, \emptyset) \mid \vec{t} :: t' \vdash (\mathcal{D}_2, \prec_2, \Sigma_2)} \enspace (\textsc{Chain})
}\end{equation*}
\begin{equation*}\boxed{
    \frac{\begin{array}{c}
        \tau_{\text{out}} \notin \dom(\Sigma) \qquad \mathcal{E}_{\text{out}} \cap \mathcal{D} = \emptyset \qquad |\mathcal{E}_{\text{out}}| \leq 1 \\
        \forall i.\; \tau_{\text{in, }i} \in \dom(\Sigma) \\
        \forall i.\; \mathcal{E}_{\text{in, }i} \subseteq \Sigma(\tau_{\text{in, }i}) \\
        \forall i.\; \Sigma(\tau_{\text{in, }i}) \setminus \mathcal{E}_{\text{in, }i} \text{ closed under } \prec \\
        \mathcal{F} = \bigcup_{i} \left(\Sigma(\tau_{\text{in, }i}) \setminus \mathcal{E}_{\text{in, }i}\right) \\
    \end{array}}
    {\begin{array}{l}
        (\mathcal{D}, \prec, \Sigma) \mid \left<
        f, 
        \left< \tau_{\text{out}}, \mathcal{E}_{\text{out}} \right>,
        \overrightarrow{\left< \tau_{\text{in, }i}, \mathcal{E}_{\text{in, }i} \right>},
            \mathcal{F},
            n
        \right> \vdash \\
        \Bigl(\mathcal{D} \sqcup \mathcal{E}_{\text{out}}, \; \prec \sqcup\; \mathcal{F} \times \mathcal{E}_{\text{out}}, \; \Sigma \set{\tau_{\text{out}} \mapsto \mathcal{F} \sqcup \mathcal{E}_{\text{out}}}\Bigr)
    \end{array}} \enspace (\textsc{TaskDef})
}\end{equation*}
\caption{Static checking rules for the DSL.}
\label{fig:operon_checking_rules}
\end{figure}

\begin{table}[t]
    \centering\small
    \caption{
        Concrete examples of errors that Operon's static checking rule \textsc{TaskDef} rejects.
        The specific error locations are \colorbox{red!30!white}{highlighted} in red.
    }
    \begin{tabular}{l l l}
    \toprule[0.15em]
    \multicolumn{1}{c}{\textbf{\textsc{TaskDef} condition}} & \multicolumn{1}{c}{\textbf{Violating example}} & \multicolumn{1}{c}{\textbf{Rationale}} \\ \midrule[0.15em]
    \(
    \tau_{\text{out}} \notin \dom(\Sigma)
    \)
                                                            &
    \(
    \begin{aligned}
         & \texttt{A} = \texttt{alpha}(); \\[-0.5em]
         & \err{A} = \texttt{beta}();
    \end{aligned}
    \)
                                                            &
    Cannot redefine output entity types.

    \\ \midrule

    \(
    \mathcal{E}_{\text{out}} \cap \mathcal{D} = \emptyset
    \)
                                                            &
    \(
    \begin{aligned}
         & \texttt{A<i>} = \texttt{alpha}();               \\[-0.5em]
         & \texttt{B<}\err{i}\texttt{>} = \texttt{beta}();
    \end{aligned}
    \)
                                                            &
    Cannot redefine dimensions.

    \\ \midrule

    \(
    |\mathcal{E}_{\text{out}}| \leq 1
    \)
                                                            &
    \(
    \texttt{A<i,}\err{j}\texttt{>} = \texttt{alpha}();
    \)
                                                            &
    Must output at most one dimension.

    \\ \midrule

    \(
    \tau_{\text{in, }i} \in \dom(\Sigma)
    \)
                                                            &
    \(
    \begin{aligned}
         & \texttt{A} = \texttt{alpha}(\err{B}); \\[-0.2em]
         & \texttt{B} = \texttt{beta}();
    \end{aligned}
    \)
                                                            &
    Must define input entities before use.

    \\ \midrule

    \(
    \mathcal{E}_{\text{in, }i} \subseteq \Sigma(\tau_{\text{in, }i})
    \)
                                                            &
    \(
    \begin{aligned}
         & \texttt{A<i>} = \texttt{alpha}();                           \\[-0.5em]
         & \texttt{B} = \texttt{beta}(\texttt{A<i,}\err{j}\texttt{>});
    \end{aligned}
    \)
                                                            &
    \begin{tabular}{@{}l@{}}
        Can only collect input entities across \\
        the characterizing dimensions.         \\
        Here, \(\mathcal{E}_{\text{in, }0} = \set{\texttt{i}, \texttt{j}} \not\subseteq \Sigma(\texttt{A}) = \set{\texttt{i}}\).
    \end{tabular}

    \\ \midrule

    \(
    \begin{aligned}
         & \Sigma(\tau_{\text{in, }i}) \setminus \mathcal{E}_{\text{in, }i} \\
         & \quad\text{closed under } \prec
    \end{aligned}
    \)
                                                            &
    \(
    \begin{aligned}
         & \texttt{A<i>} = \texttt{alpha}();                                      \\[-0.2em]
         & \texttt{B<j>} = \texttt{beta}(\texttt{A}) \;\texttt{for}\; \texttt{i}; \\[-0.5em]
         & \texttt{C} = \texttt{gamma}(\err{B<i>}) \;\texttt{for}\; \texttt{j};
    \end{aligned}
    \)
                                                            &
    \begin{tabular}{@{}l@{}}
        Cannot collect across parent dimensions                                                             \\
        while leaving some child dimensions free.                                                           \\
        We have \(\Sigma(\texttt{B}) = \set{\texttt{i}, \texttt{j}}\) with \(\texttt{i} \prec \texttt{j}\), \\
        and collecting \(\texttt{B}\) over \(\mathcal{E}_{\text{in, }0} = \set{\texttt{i}}\) would          \\
        result in a nonsense list with null values.
    \end{tabular}

    \\ \midrule

    \multirow{3.2}{*}{\(
        \mathcal{F} = \bigcup_{i} \left(\Sigma(\tau_{\text{in, }i}) \setminus \mathcal{E}_{\text{in, }i}\right)
        \)}
                                                            &
    \(
    \begin{aligned}
         & \texttt{A<i>} = \texttt{alpha}();                                  \\[-0.5em]
         & \texttt{B} = \texttt{beta}(\texttt{A})\err{\vphantom{\texttt{i}}};
    \end{aligned}
    \)
                                                            &
    \begin{tabular}{@{}l@{}}
        Must at least iterate over all free dimensions \\
        of input entities.
    \end{tabular}

    \\ \cmidrule{2-3}
                                                            &
    \(
    \begin{aligned}
         & \texttt{A<i>} = \texttt{alpha}();                                       \\[-0.5em]
         & \texttt{B} = \texttt{beta}(\texttt{A<i>})\err{\texttt{for} \texttt{i}};
    \end{aligned}
    \)
                                                            &
    \begin{tabular}{@{}l@{}}
        Cannot have redundant iterations.
    \end{tabular}
    \\ \bottomrule[0.15em]
\end{tabular}
\label{tab:operon_rejected_examples}
\end{table}

The pipeline consists of one or more tasks that collectively define the overarching data flow.
In a static analysis as shown in Figure~\ref{fig:operon_checking_rules}, we check whether the pipeline \(p\) is well-formed according to these rules, i.e., whether \((\emptyset, \emptyset, \emptyset) \mid p \vdash (\mathcal{D}, \prec, \Sigma)\) holds for some \((\mathcal{D}, \prec, \Sigma)\).
We provide some concrete examples of malformed pipelines that violate the \textsc{TaskDef} rule in Table~\ref{tab:operon_rejected_examples}.

For a well-formed pipeline, the triple \((\mathcal{D}, \prec, \Sigma)\) from this analysis gains meaning as the global dimension space, the dependency relation, and the map from entity type to its \emph{characteristic} dimension subspace, respectively.
\(\Sigma\), in particular, acts as the dimension-aware typing context: for each entity type \(\tau\) mentioned in the pipeline, \(\Sigma(\tau)\) specifies which dimensions \(\tau\) can be collected or indexed over.

The checking rules ensure that the inferred \((\mathcal{D}, \prec, \Sigma)\) satisfies several well-formedness properties, as stated in the following lemma.
\begin{lemma}\label{lem:operon_well-formedness}
    Given \((\emptyset, \emptyset, \emptyset) \mid p \vdash (\mathcal{D}, \prec, \Sigma)\),
    \begin{enumerate}
        \item the relation \(\prec\) is a strict partial order over \(\mathcal{D}\);
        \item for all entity types \(\tau \in \dom(\Sigma)\), the characteristic dimension space \(\Sigma(\tau)\) is closed under \(\prec\);
        \item for all tasks \(t = \left< f, s_\text{out}, \overrightarrow{\left< \tau_{\text{in, }i}, \mathcal{E}_{\text{in, }i} \right>}, \mathcal{F}, n \right>\) in \(p\), the dimension spaces \(\mathcal{F}\) and \(\Sigma(\tau_{\text{in, }i}) \setminus \mathcal{E}_{\text{in, }i}\) are closed under \(\prec\).
    \end{enumerate}
\end{lemma}

Once the checks are complete, \emph{entities} can be defined based on the inferred information.
Entities are the data units that Operon aims to produce and process, and they are organized into (partial) arrays according to their types and their characteristic dimensions.

\begin{definition}[Entities]
    Given \((\emptyset, \emptyset, \emptyset) \mid p \vdash (\mathcal{D}, \prec, \Sigma)\), for all entity types \(\tau \in \dom(\Sigma)\), an \emph{entity array} \(\overline{E}(\tau)\) is defined as a partial array over the fully resolved coordinates \[
        \overline{E}(\tau): \coordsp{\Sigma(\tau)}{\overline{R}} \rightharpoonup \tau
    \] for some partial shape \(\overline{R}\).
    Elements of this array are called \emph{entities}.
\end{definition}

Now, the goal of Operon can be restated as: given a pipeline definition \(p\), run the pipeline to incrementally construct a partial shape \(\overline{R}\) and fully populate the entity arrays \(\overline{E}(\tau)\) for all \(\tau \in \dom(\Sigma)\).
Therefore, at its core, Operon is a state machine that continuously transforms the state \((\overline{R}, \overline{E})\) by executing user-defined functions specified in the pipeline.
Starting from the trivial state \((\emptyset, \lambda\tau.\;\emptyset)\), Operon undergoes the following state transitions, known as \emph{jobs}, until it reaches a terminal state where \(\overline{R}\) is complete and all entity arrays are total.

\begin{definition}[Jobs]\label{def:operon_jobs}
    Consider a task \(\left< f, \left< \tau_{\text{out}}, \mathcal{E}_{\text{out}} \right>, \overrightarrow{\left< \tau_{\text{in, }i}, \mathcal{E}_{\text{in, }i} \right>}, \mathcal{F}, n \right>\) in a pipeline \(p\), the inferred \((\mathcal{D}, \prec, \Sigma)\), and a current state \((\overline{R}, \overline{E})\).
    For a fully resolved coordinate \(c : \coord{\mathcal{F}}{\overline{R}}\), if the subarrays \(\overline{E}(\tau_{\text{in, }i})[c|_{\Sigma(\tau_{\text{in, }i}) \setminus \mathcal{E}_{\text{in, }i}}]\) are all total over their respective domains, the function \(f\) can be invoked with these subarrays as inputs.
    We refer to this call as a \emph{job} at coordinate \(c\), denoted as \(j_t(c)\).
    A job \(j_t(c)\) transforms the current state \((\overline{R}, \overline{E})\) into a new state \((\overline{R}', \overline{E}')\) as follows.
    This transition is valid exactly once for each \(t\) and \(c\).
    \begin{itemize}
        \item If \(|\mathcal{E}_{\text{out}}| = 0\), then \(\overline{R}' = \overline{R}\).
        Otherwise, write \(\mathcal{E}_{\text{out}} = \set{e}\) and let \(\ell\) be the length of the output array returned by \(f\).
        Then, \(\overline{R}' = \overline{R}\set{(e, c) \mapsto \ell}\).
        \item \(\overline{E}'\) is identical to \(\overline{E}\) except for the subarray \(\overline{E}'(\tau_{\text{out}})[c]\), which is assigned the output array.
    \end{itemize}
\end{definition}

The static checking rules and Lemma~\ref{lem:operon_well-formedness} are what allow this transition to be well-defined.
The notation \(\ucoordsp{\mathcal{F}}{\overline{R}}\) requires \(\mathcal{F}\) to be closed under \(\prec\), and each subarray \(\overline{E}(\tau_{\text{in, }i})[c|_{\Sigma(\tau_{\text{in, }i}) \setminus \mathcal{E}_{\text{in, }i}}]\) is well-defined only when \(\Sigma(\tau_{\text{in, }i}) \setminus \mathcal{E}_{\text{in, }i} \subseteq \mathcal{F}\) is closed under \(\prec\); Lemma~\ref{lem:operon_well-formedness}(3) covers both.
Because \(\mathcal{E}_{\text{out}}\) is disjoint from previously defined dimensions, the coordinate \((e, c)\) is written exactly once across all \((t, c)\) pairs, so the update \(\overline{R}' = \overline{R}\set{(e, c) \mapsto \ell}\) is valid.
Finally, the output subarray \(\overline{E}(\tau_{\text{out}})[c]: \usubcoordsp{\mathcal{E}_{\text{out}}}{c|_{\mathcal{E}_{\text{out}}^\Downarrow}}{\overline{R}} \rightharpoonup \tau_{\text{out}}\) is the empty function before the job and, after the partial-shape update, its subcoordinate space explodes to fit \(f(\ldots)\).
Assigning the values from \(f\) while leaving all other entities unchanged completes the transition.

The normalizing constraint \(|\mathcal{E}_{\text{out}}| \leq 1\) in \(\textsc{TaskDef}\) is a usability choice.
Allowing multiple output dimensions is possible with minimal changes to Definition~\ref{def:operon_jobs}, but would require the user to supply and enforce the dimensional dependencies within \(\mathcal{E}_{\text{out}}\), since no standard array structure encodes our ragged formulation.
Restricting \(\mathcal{E}_{\text{out}}\) to at most one dimension avoids this: a 0- or 1-dimensional output translates trivially to a ragged subarray, and a multi-dimensional output can still be produced by splitting it across single-dimensional tasks, at some cost in usability or performance.

\subsection{Implementation}
\label{subsec:implementation}

Recall that one of the primary goals of Operon, \emph{work-conservation}, is to launch each job as soon as it becomes executable.
Jobs become ready to run when (1) their coordinates have fully resolved over \(\mathcal{F}\) and (2) all their input entities have been computed.
Since both resolutions and input entities are produced by some other jobs in the pipeline, the readiness of a job relies on the status of others.

Operon manages this by using \emph{tickets}, which are lightweight objects that represent the state of each job in the system.

\begin{definition}[Tickets]\label{def:tickets}
    Consider a task \(t = \left< f, s_{\text{out}}, \overrightarrow{s_{\text{in, }i}}, \mathcal{F}, n \right>\) in a pipeline \((\emptyset, \emptyset, \emptyset) \mid p \vdash (\mathcal{D}, \prec, \Sigma)\).
    The \emph{ticket array} for task \(t\) is a \emph{total} array
    \[
        \overline{j}_t: \ucoordsp{\mathcal{F}}{\overline{R}} \rightarrow \nat \times \nat \times \{\texttt{Waiting}, \texttt{Queued}, \texttt{Done}\},
    \]
    where each entry \(\overline{j}_t(c) = (\texttt{count}, \texttt{quota}, \texttt{status})\) is called a \emph{ticket}.
\end{definition}

A ticket \(\overline{j}_t(c)\) conceptually corresponds to a job \(j_t(c)\), even though the latter is not defined unless the coordinate \(c\) is total over \(\mathcal{F}\).
For a partially defined coordinate \(c\), the ticket represents all potential jobs that could arise from \(c\) as more resolutions are added to \(\overline{R}\).
In the initial state \((\overline{R}, \overline{E}) = (\emptyset, \lambda\tau.\;\emptyset)\), the coordinate space with unknowns \(\ucoordsp{\mathcal{F}}{\overline{R}}\) starts as \(\set{\emptyset}\), and hence there exists a single ticket \(\overline{j}_t(\emptyset)\) representing all jobs of task \(t\).
As resolutions for dimensions in \(\mathcal{F}\) are added to \(\overline{R}\), the coordinate space expands according to Theorem~\ref{thm:coordinate_explosion}, and tickets are duplicated along the newly resolved dimensions.
When a ticket's coordinate becomes total over \(\mathcal{F}\), it corresponds to exactly one job, and we say it is \emph{fully resolved}.
If the task has no input entities, as in the first task in all pipelines, its ticket is fully resolved from the start and can have the corresponding job launched immediately; otherwise, the ticket must wait for upstream jobs to resolve its coordinates.

Throughout the ticket's lifetime, its \texttt{count} and \texttt{quota} fields are updated to track the number of completed dependencies and the total number of dependencies, respectively.
While the dependencies are most intuitively explained as the completion of all input entities, i.e., \(\overline{E}(\tau_{\text{in, }i})[c|_{\Sigma(\tau_{\text{in, }i}) \setminus \mathcal{E}_{\text{in, }i}}]\) being total for all \(i\), we can track back to the tasks that produce these entities and count their tickets as dependencies instead.
Expressing \(t_i = \left< \_, \left< \tau_{\text{in, }i}, \mathcal{E}_i \right>, \_, \mathcal{F}_i, \_ \right>\) as the task that produces the input entity \(\tau_{\text{in, }i}\), the dependencies of ticket \(\overline{j}_t(c)\) can be defined as the set of tickets \[
    \bigsqcup_{i} \overline{j}_{t_i}\left[c|_{\mathcal{F}_i \setminus \mathcal{E}_{\text{in, }i}}\right].
\]
The size of this set, \[
    \sum_{i} \left| \ucoordsp{\mathcal{F}_i}{\overline{R}}_{c|_{\mathcal{F}_i \setminus \mathcal{E}_{\text{in, }i}}}\right|,
\] determines the \texttt{quota} of the ticket, while the \texttt{count} is incremented each time one of these dependent tickets reaches the \texttt{Done} status.
Once the two counts are equal, it can be assumed that the ticket is fully resolved (Lemma~\ref{lem:ticket_ready}), and the ticket becomes ready for execution (\texttt{Queued}).
The ticket is further updated to \texttt{Done} when the corresponding job completes.

\begin{lemma}\label{lem:ticket_ready}
    When a ticket's \emph{\texttt{count}} equals its \emph{\texttt{quota}}, the ticket is fully resolved.
\end{lemma}

While the readiness of a ticket can be reduced to a two-integer system as described above, computing the \texttt{quota} of a ticket can be expensive upfront, as it requires counting the number of sibling coordinates for each input entity.
We therefore adopt a lazy approach to tracking the \texttt{quota} of tickets, where each exploded ticket sends down \emph{ticket explosion events} that indicate to add \(\ell - 1\) to the \texttt{quota} field of each ticket that satisfies the condition in Corollary~\ref{cor:sibling_counts}(3).
Since the number of exploded tickets matches the right-hand side multiplier of the equation in Corollary~\ref{cor:sibling_counts}(3), the \texttt{quota} field will eventually reach the correct value after all events have been processed.

\begin{figure}[t]
    \begin{tikzpicture}[
  >=Latex,
  node distance=4mm and 0mm,
  font=\tiny,
  block/.style = {rectangle, draw, align=center, minimum width=20mm, minimum height=7mm},
  decision/.style = {diamond, draw, aspect=2.5, align=center, inner sep=0.5pt, minimum width=20mm, minimum height=7mm},
  terminator/.style = {rectangle, rounded corners, draw, align=center},
  line/.style = {draw, -{Stealth[]}},
  line dashed/.style = {draw, dashed, -{Stealth[]}},
  subroutine/.style = {
    draw,
    dashed,
    rounded corners,
    inner sep=6pt,
    label={[anchor=south west, font=\tiny\bfseries]north west:#1},
  },
  ]


  \node[terminator] (init)
    {Initialize scheduler};

  \node[block, below=of init] (loop)
    {Main event loop};

  \node[decision, below=of loop] (evkind)
    {event kind?};


  \node[decision, below=5mm of evkind] (peerkind)
    {peer event type?};

  \node[block, below=5mm of peerkind] (res_resolve)
    {resolve blank\\coordinates};

  \node[block, below=of res_resolve] (res_prop)
    {propagate \\ explosion events};

  \node[block, below=of res_prop] (enqueue)
    {enqueue newly\\ready tickets};

  \node[block, left=5mm of res_resolve] (jobf_update)
    {update dependency\\counters};

  \node[block, right=5mm of res_resolve] (pred_handle)
    {handle coordinate\\explosion};


    \node[block, right=30mm of peerkind] (int_emit)
      {propagate\\job/resolution events};

    \node[decision, below=of int_emit] (quiescent)
      {no pending tickets\\and events?};

    \node[terminator, below=of quiescent] (retok)
      {signal completion and terminate};


  \node[block, left=30mm of peerkind, align=center] (spawn_worker)
    {spawn worker task};

\node[block, below=13mm of enqueue] (worker_storage_ops)
    {perform storage ops};
    \node[block, left=5mm of worker_storage_ops] (worker_execute_job)
      {execute job};
\node[block, right=5mm of worker_storage_ops] (worker_emit_event)
    {emit internal event};
    \node[subroutine={Worker Task}, fit=(worker_execute_job)(worker_storage_ops)(worker_emit_event)] (taskbox) {};


  \path[line] (init) -- (loop);
  \path[line] (loop) -- (evkind);

  \path[line] (evkind) -| node[above]{internal event} (int_emit);
  \path[line] (evkind) -- node[left]{peer event} (peerkind);
  \path[line] (evkind) -| node[above, align=center]
    {polled worker permit \\ \textit{with nonempty queue}}
    (spawn_worker);

  \path[line] (int_emit) -- (quiescent);
  \path[line] (quiescent) -- node[left]{Yes} (retok);
  \path[line] (quiescent.east) -- node[above]{No} ($(quiescent.east)+(5mm,0)$) |- (loop.east);

  \path[line] (peerkind) -| node[above]{job finished} (jobf_update);
  \path[line] (peerkind) -- node[right]{resolution found} (res_resolve);
  \path[line] (peerkind) -| node[above]{predecessor exploded} (pred_handle);

  \path[line] (jobf_update) |- (enqueue);

  \path[line] (pred_handle) |- (enqueue);

  \path[line] (res_resolve) -- (res_prop);
  \path[line] (res_prop) -- (enqueue);

    \path[line] (enqueue.south) -- ($(enqueue.south)+(0,-4mm)$) -| ($(quiescent.east)+(5mm,0)$) |- (loop.east);

  \path[line] (worker_execute_job) -- (worker_storage_ops);
    \path[line] (worker_storage_ops) -- (worker_emit_event);
  \path[line dashed] (spawn_worker.west) -- ($(spawn_worker.west)+(-4mm,0)$) |- (taskbox);
  \path[line] (spawn_worker.south) |- ($(enqueue.south)+(0,-4mm)$) -| ($(quiescent.east)+(5mm,0)$) |- (loop.east);

\end{tikzpicture}
    \caption{Idealized flowchart of an individual scheduler.}
    \label{fig:operon_flowchart}
\end{figure}

A dedicated scheduler for each task manages these tickets and events, as illustrated in Figure~\ref{fig:operon_flowchart}.
These schedulers operate independently and communicate solely through the following kinds of \emph{peer events}:
\begin{itemize}
    \item \emph{Job completion}: Downstream tickets should increment their \texttt{count}.
    \item \emph{Resolution}: Downstream ticket arrays whose \(\mathcal{F}\) contains this dimension should explode.
    \item \emph{Ticket explosion}: Downstream tickets depending on the exploded tickets should update their \texttt{quota}s accordingly.
\end{itemize}
Each scheduler listens to peer events, processes them to discover newly runnable jobs, and dispatches internal workers to run the computations.
Internal workers then execute the following:
\begin{enumerate}
    \item Pull the input entities from the database and pass them into the user-defined function \(f\).
    \item Execute the function.
    \item Put any output entities and resolutions into the database.
    \item Mark its ticket as \texttt{Done} and commit the transaction.
    \item Signal the job completion to the scheduler.
\end{enumerate}

This manner of execution ensures the atomic invariant that a ticket is marked \texttt{Done} if and only if all its corresponding output entities and resolutions have been successfully persisted in the underlying storage.
Entities are never passed between workers or schedulers, and instead are always persisted in the database and pulled by the workers when needed.
This incurs a performance cost tradeoff, but it separates data management from scheduling and allows the concurrency limit to prevent too many entities from residing in memory at any given time.

The loop continues until there are no pending tickets or events, at which point the scheduler signals completion of its task and terminates.
The theorems and corollary in Section~\ref{subsec:incremental_construction} guarantee the well-behavedness of the progress of the partial shape \(\overline{R}\).

The entities \(\overline{E}\), resolutions \(\overline{R}\), and tickets \(\overline{j}_t\) are all stored in an underlying storage system, currently implemented in PostgreSQL.
The persistence allows straightforward pause-resume and crash-recovery functionality, as the entire state of the schedulers can be reconstructed from the database.
We describe the mechanism in more detail in Section~\ref{subsec:fault-tolerance}.

\subsection{Fault-tolerance}
\label{subsec:fault-tolerance}

Operon provides two recovery primitives, \texttt{check} and \texttt{rebuild}, that together guarantee correct resumption after unexpected termination.

\paragraph{Check}

The \texttt{check} command verifies the atomic invariant that every ticket marked \texttt{Done} has its corresponding output entities and resolution successfully persisted in the underlying storage.
The job completion process naturally enforces this invariant by updating the ticket status only after inserting the entities and resolutions.
However, external failures (e.g., storage crashes) may violate it, so the user can verify it at varying levels of strictness:
skip entirely (\texttt{-{}-mode trust-all}),
verify only the existence of resolutions (\texttt{-{}-mode metadata-only}),
check for the presence of entities at the boundary (\texttt{-{}-mode quick}),
or perform a full consistency check of all data (\texttt{-{}-mode exhaustive}).
If the check detects an inconsistency, it rejects the recovery attempt.

\paragraph{Rebuild}

Once the invariant is verified, \texttt{rebuild} constructs the maximal well-formed state \((\overline{R}, \overline{E}, \overline{j}_t)\) that can be derived solely from the persisted results.
Concretely, it treats \texttt{Done} tickets as a reliable base: the rebuilt state is exactly the one in which all jobs corresponding to \texttt{Done} tickets have been completed, and no others.
Any data not corresponding to a \texttt{Done} ticket is ignored.
Since the \texttt{count} and \texttt{quota} values of non-\texttt{Done} tickets may be inaccurate after a crash, they are recomputed by emulating the completion of the \texttt{Done} tickets' corresponding jobs.

\paragraph{Recovery guarantees}

Table~\ref{tab:recovery_scenarios} summarizes the recovery behavior across representative failure and corruption scenarios.
Under the assumption that the database is not corrupted, every termination scenario recovers successfully with results identical to a clean run.
Since \texttt{rebuild} precisely reconstructs the state of all completed jobs and discards any partial results, no job marked as \texttt{Done} is re-executed upon resumption.

\begin{table}[t]
    \centering\footnotesize
    \caption{
        Recovery behavior under representative scenarios.
        Each column denotes the outcome of a recovery step.
        \cmark: success, \xmark: rejected / error, ${\boldsymbol{-}}$: not applicable.
    }
    \begin{tabular}{@{}l|cccc@{\hspace{1.5em}}l|cccc@{}}
        \toprule
        Scenario                                               & \texttt{check}                                   & \texttt{rebuild}   & \texttt{run}       & Result             &
        Scenario                                               & \texttt{check}                                   & \texttt{rebuild}   & \texttt{run}       & Result               \\

        \hline
        \multicolumn{5}{c}{\emph{Termination (no corruption)}} &
        \multicolumn{5}{c}{\emph{Metadata corruption}}                                                                                                                             \\

        \hline
        Graceful quit                                          & ${\boldsymbol{-}}$                               & ${\boldsymbol{-}}$ & \cmark             & \cmark             &
        Missing tickets (\texttt{Done})                        & \cmark                                           & \cmark             & \cmark             & \cmark               \\

        Force quit                                             & \cmark                                           & \cmark             & \cmark             & \cmark             &
        Fake tickets (\texttt{Done})                           & \xmark                                           & ${\boldsymbol{-}}$ & ${\boldsymbol{-}}$ & ${\boldsymbol{-}}$   \\

        User function error                                    & \cmark                                           & \cmark             & \cmark             & \cmark             &
        Missing resolutions                                    & \xmark                                           & ${\boldsymbol{-}}$ & ${\boldsymbol{-}}$ & ${\boldsymbol{-}}$   \\

        Storage plug-out                                       & \cmark                                           & \cmark             & \cmark             & \cmark             &
        Corrupt resolutions                                    & \xmark                                           & ${\boldsymbol{-}}$ & ${\boldsymbol{-}}$ & ${\boldsymbol{-}}$   \\

        Crash during rebuild                                   & \cmark                                           & \cmark             & \cmark             & \cmark             &
        Fake resolutions                                       & \cmark                                           & \cmark             & \cmark             & \cmark               \\

        \hline
        \multicolumn{5}{c}{\emph{Data corruption}}             & \multicolumn{5}{c}{\emph{Configuration changes}}                                                                  \\

        \hline
        Missing entities                                       & \xmark                                           & ${\boldsymbol{-}}$ & ${\boldsymbol{-}}$ & ${\boldsymbol{-}}$ &
        New/removed task                                       & \cmark                                           & \cmark             & \cmark             & \cmark               \\

        Corrupt entities                                       & \cmark                                           & \cmark             & \cmark             & \xmark             &
        Altered task signature                                 & \xmark                                           & ${\boldsymbol{-}}$ & ${\boldsymbol{-}}$ & ${\boldsymbol{-}}$   \\

        Fake entities                                          & \cmark                                           & \cmark             & \cmark             & \cmark             &
                                                               &                                                  &                    &                    &                      \\
        \bottomrule
    \end{tabular}
    \label{tab:recovery_scenarios}
\end{table}

\subsection{Evaluation}
\label{sec:evaluation}
\label{subsec:performance}

We evaluate Operon against Prefect~\cite{prefecthq2025prefect}, the closest prior art in scope.
The workload is a mock version of the motivating pipeline (Section~\ref{subsec:motivating_example}).
Every task preserves the full dimensional structure but performs negligible computation, except for the heavy tasks (\texttt{parse\_paper}, \texttt{vlm\_evaluate}, \texttt{ocr\_extract}), which are assigned a configurable sleep interval \(t_{\text{sleep}}\) to simulate model-inference latency.\footnote{All runs use a local PostgreSQL backing store on a single Mac mini (M4, 16\,GB RAM), with heavy-task concurrency capped at 64 and three trials per configuration. Full setup and raw data are in the appendix.}
Because Operon is implemented in Rust and Prefect in Python, raw wall-clock differences in execution time are not a reliable basis for claims of algorithmic superiority.
Our goal instead is to establish \emph{practical feasibility} against an industry-grade engine, and \emph{work-conservation} as a property of the scheduling architecture.

\begin{figure}[t]
    \centering
    \begin{subfigure}[t]{0.42\textwidth}
        \centering
        \includegraphics[width=\textwidth]{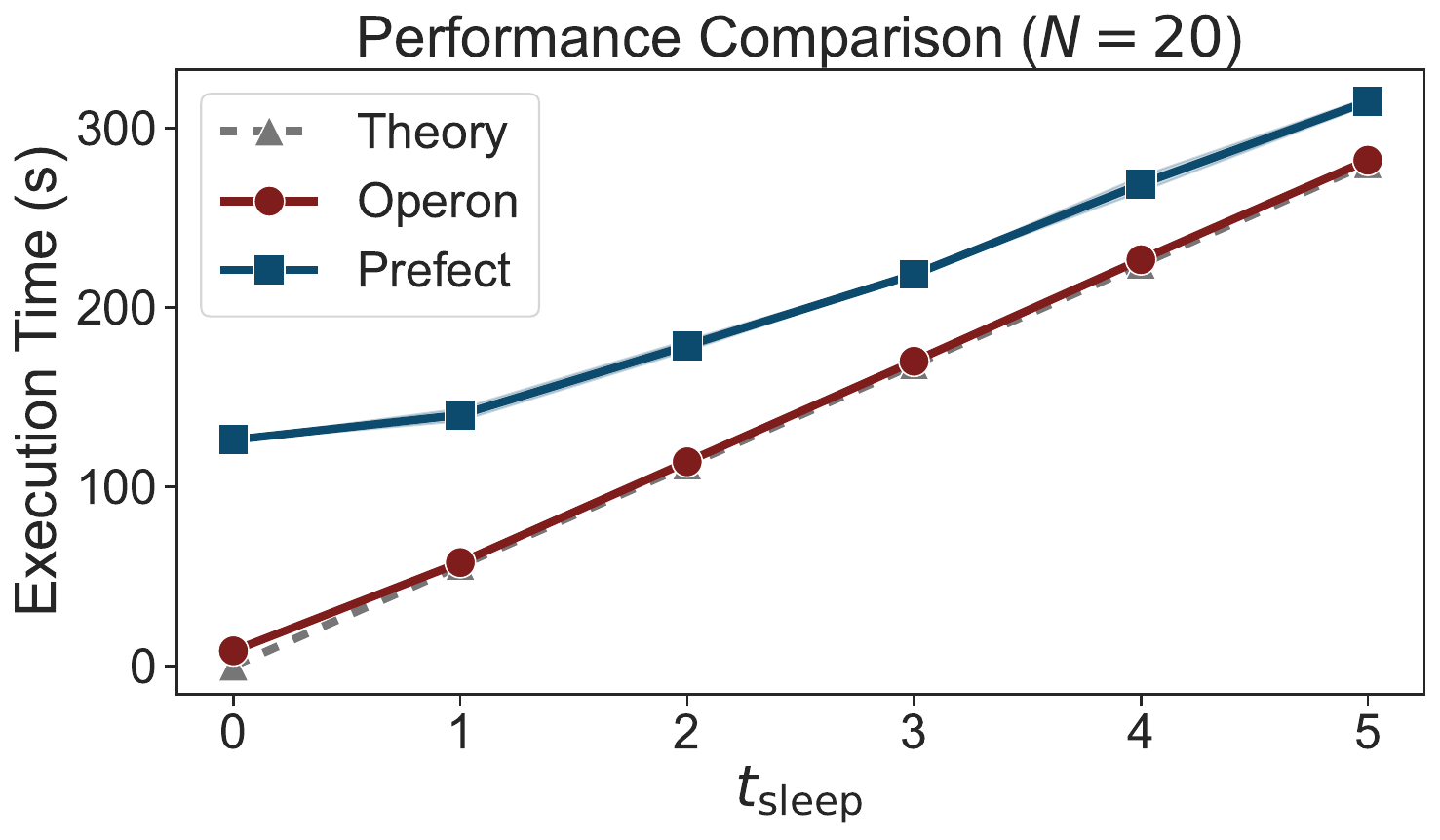}
        \caption{}
        \label{fig:experiment_sleep}
    \end{subfigure}
    \hfill
    \begin{subfigure}[t]{0.52\textwidth}
        \centering
        \includegraphics[width=\textwidth]{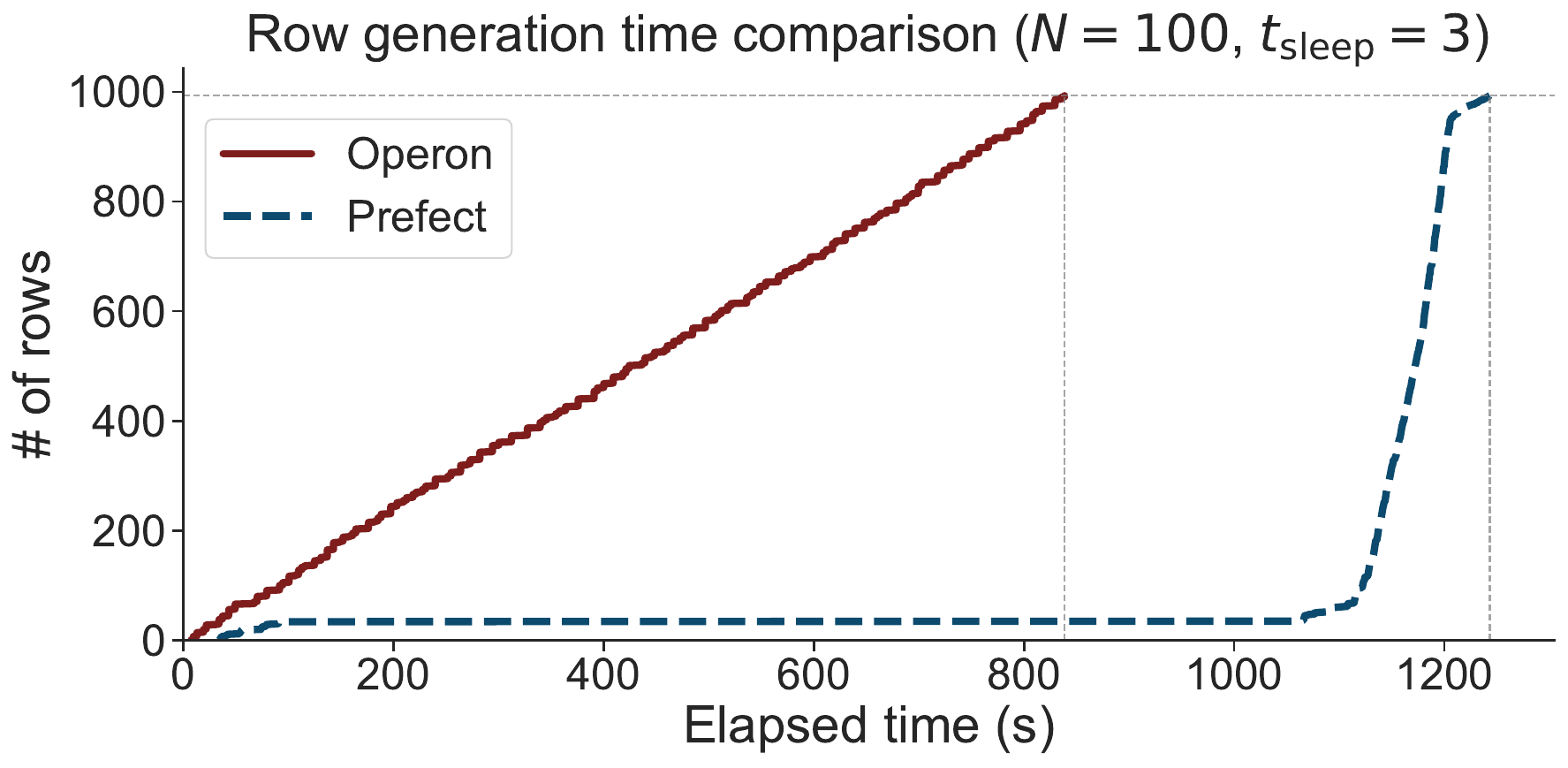}
        \caption{}
        \label{fig:experiment2}
    \end{subfigure}
    \caption{Empirical evaluation of Operon (red) against Prefect (blue).
        (\subref{fig:experiment_sleep}) Total execution time versus the sleep interval \(t_{\text{sleep}}\) at \(N = 20\); the reference line ``Theory'' (gray) is the theoretical minimum \(\left(\left\lceil\frac{\text{\# of vlm\_evaluate}}{64}\right\rceil + 1\right)\times t_\text{sleep}\).
        (\subref{fig:experiment2}) Generated \texttt{Row} entries over time at \(N=100\), \(t_{\text{sleep}}=3\) s.}
    \label{fig:experiments}
\end{figure}

\paragraph{Practical feasibility}
Figure~\ref{fig:experiment_sleep} shows that Operon's total execution time tracks the theoretical minimum closely across sleep intervals, confirming that its scheduling decisions add no meaningful overhead relative to task execution time.
As \(t_{\text{sleep}}\) grows, the gap to Prefect narrows, indicating that the overhead is a fixed cost rather than one that scales with the workload.
Even at \(t_{\text{sleep}} = 0\), this overhead does not preclude deployment on realistic workloads, and the appendix shows that the same tracking holds as the workload size \(N\) scales.

\paragraph{Work-conservation}
A work-conserving scheduler dispatches each instance as soon as its inputs are available, which, for the motivating pipeline, predicts a roughly linear accumulation of \texttt{Row} entries over time.
Figure~\ref{fig:experiment2} confirms this for Operon, whereas Prefect's output stagnates early and surges near the end.
The cause is structural: its \texttt{ThreadPoolTaskRunner} enqueues the tens of thousands of instances a ragged pipeline generates behind a fixed-size thread pool, so lighter downstream tasks such as \texttt{collect\_row} sit starved behind heavy ones such as \texttt{vlm\_evaluate}.
This is exactly the work-conservation failure anticipated in Section~\ref{subsec:problem_statement}.

Operon's fault-tolerance primitives also carry a recovery overhead, which we measure across workflow scales in the appendix.
In brief, \texttt{check} stays under one second up to \({\sim}10^5\) tickets, except in \texttt{exhaustive} mode, which is roughly an order of magnitude costlier; \texttt{rebuild} grows roughly linearly with the total ticket count.

\subsection{Limitations}
\label{subsec:limitations}

\paragraph{Database overhead}
Operon's pointwise-persistence model carries a concrete cost.
It keeps the runtime state, e.g., the shape, entities, and tickets of all tasks, in a running PostgreSQL instance.
Every scheduler event therefore must open a database transaction and incurs a round trip.
This is the price of the persistence-versus-performance tradeoff set out in Section~\ref{subsec:problem_statement}: the same durable storage backs crash recovery (Section~\ref{subsec:fault-tolerance}), bounds memory in large-scale workflows, and suits CPU-based data-parallel processing~\cite{cugola2012low}.
The round trip is affordable only in Operon's target regime, where tasks run on the order of milliseconds to seconds (Section~\ref{subsec:performance}).

\paragraph{Structural constraints}
Operon supports only DAG-structured workflows, making it impossible to express workflows requiring cyclic structures directly.
Cases where the number of cycles is statically determined can be rewritten into a DAG through loop unrolling; however, when the number of cycles is dynamically determined at runtime, it cannot be expressed in Operon's declarative model.
Imperative frameworks like Prefect would be more suitable for such scenarios.

\section{Related Work}
\label{sec:related_work}

\paragraph{Named dimensions and ragged tensors}
The demand for named dimensions is demonstrated by packages such as xarray~\cite{hoyer2017xarray}, einops~\cite{rogozhnikov2022einops}, and Awkward~\cite{pivarski2018awkward}.
Dex~\cite{paszke2021getting} goes further by making index membership a type-level property, so that an array's dimension structure is statically enforced rather than a runtime annotation.
In particular, modern deep-learning workloads rely on operations between multidimensional tensors~\cite{chiang2021named}, where unnamed axes are a common source of errors.
Growing array programming research addresses this with static shape safety: Futhark~\cite{henriksen2021towards} uses size parameters to enforce shape invariants in regular arrays, and Star~\cite{bachurski2025star} generalizes this to algebraic shape expressions checked at compile time.
Ragged tensors are typically implemented via padding into rectangular tensors~\cite{fegade2022cora} or pointer-based layouts such as Iliffe vectors~\cite{iliffe1961use}.
More recent systems---TensorFlow ragged tensors~\cite{abadi2015tensorflow} and Awkward~\cite{pivarski2018awkward}---support ragged tensors natively.
Awkward provides a design for ragged data over a totally ordered set of named axes, focusing primarily on memory layout and low-level representation.
Finch~\cite{ahrens2025finch} supports flexible control flow over sparse and structured tensors, though its focus remains on compiling individual array operations rather than multi-task pipelines.

The above systems treat named dimensions as a representation concern and preserve arbitrary array indexing.
Adopting the totality constraint motivated in Section~\ref{subsec:problem_statement}, Operon instead forgoes dynamic \emph{access} in exchange for a further degree of dynamism in \emph{generation}, specializing the named-dimension idea to the workflow-scheduling domain.
This lets Operon statically verify that every data edge connects compatible coordinate spaces before any coordinate is known, which is a guarantee the representation-focused systems above cannot offer, since they can validate an index only once the surrounding data exists.

\paragraph{Workflow engines}
Among existing workflow engines, Prefect~\cite{prefecthq2025prefect} is the closest in scope: it supports dynamic task generation and a data-oriented execution model.
Its thread-pool scheduler, however, exhibits the work-conservation failure characterized in Sections~\ref{subsec:problem_statement} and~\ref{sec:evaluation}.
Apache Airflow~\cite{apache2025airflow} primarily targets static DAG scheduling; its mapped task mechanism offers limited dynamic expansion but does not generalize to multidimensional, data-driven patterns.
Luigi~\cite{spotify2025luigi} defines task graphs statically and relies on file-based target checks for fault tolerance, offering no dynamic workflow generation or structured state persistence.
FlumeJava~\cite{chambers2010flumejava} introduced the deferred-evaluation style these systems share, representing a data-parallel pipeline as a lazily fused dataflow graph.
Dask~\cite{rocklin2015dask} and Ray~\cite{moritz2018ray} carry the same lazy-graph construction to dynamic task graphs at scale, materializing them as inputs become available.
None of the three provides named dimensions, ragged structure within a collection, or static well-formedness checks.
Because their scheduling is driven by a global task queue rather than per-dimension readiness, all three exhibit the same work-conservation failure under heterogeneous ragged fanout that Section~\ref{sec:evaluation} demonstrates in Prefect.
None of these systems derives the confluence or completeness guarantees that Operon obtains from its pipeline structure.

\paragraph{HPC runtime systems and parallel language design}
Operon's tickets play a role close to Realm's events~\cite{treichler2014realm}: both are non-blocking handles used to build a dependence graph at runtime rather than compile time, placing both in what Realm's own taxonomy calls \emph{dynamic explicit representation} runtimes.
They differ in where the graph's structure comes from.
A Realm event is opaque, so a client (typically a higher-level runtime such as Legion) must wire events together by hand, and Realm's contribution is making that wiring cheap.
Operon instead derives a ticket's creation, merging, and discharge rules automatically from the named-dimension coordinate space.
This lets Operon check ticket generation against the declared dimensions statically, unlike Realm, whose runtime never inspects the meaning of the graph it defers.
The two also target different regimes, discussed further in Section~\ref{subsec:limitations}.

Chapel~\cite{chamberlain2007parallel} names several principles in parallel programming.
First, it argues that productive parallel languages should offer a \emph{global view of computation}.
An algorithm is written once, over a data structure as a whole, rather than \emph{fragmented} by hand into per-task pieces coordinating through explicit messages.
Operon's DSL takes the same stance for workflows: a task is declared once over its named dimensions, and every instance is derived from the coordinate space rather than hand-mapped.
\emph{General, composable parallelism} is the property that calls for the unification of task and data parallelism.
We have expanded on this property as \emph{work-conservation} and investigated its impact throughout this paper.
Its \emph{separation of algorithm and implementation} has a direct analogue in Operon: a task definition's well-formedness depends only on the declared pipeline, not on how or when its instances are scheduled and executed.
The two diverge past these principles.
A Chapel domain is a single array's index set, resizable at runtime but carrying no declared dependency to other domains and no static check that two operations' domains are compatible.
Operon's named dimensions form a declared partial order across an entire pipeline, checked for well-formedness before any data exists.
Operon also offers durable, crash-recoverable execution through tickets and effectively-once semantics (Section~\ref{subsec:fault-tolerance}), whereas Chapel assumes a single in-memory run.

Swift/T~\cite{armstrong2014compiler} occupies a comparable position in the scientific-workflow space.
It uses compiler analysis to derive implicit task parallelism from dataflow programs targeting massively parallel HPC machines.
However, unlike Operon, the data model is regular and non-ragged, and there is no inter-task dimension compatibility check.
Further along the in-memory end of the spectrum, task-based HPC runtimes---Charm++~\cite{kale1993charm}, PaRSEC~\cite{bosilca2013parsec}, and StarPU~\cite{augonnet2011starpu}---hold minimum effective task granularities (METG(50\%)) from a few hundred nanoseconds to roughly 100\,\textmu{}s at scale as measured by Task Bench~\cite{slaughter2020taskbench}.
These numbers are achieved by keeping their dependence graphs entirely in memory and transient.
Operon trades the raw performance observed in these systems for durability.
Its database-backed tickets survive crashes at the cost of a round trip per operation, which is affordable because Operon's tasks run several orders of magnitude coarser, on the order of milliseconds to seconds (Section~\ref{subsec:performance}).
Adapting this construction to the in-memory, high-frequency regime and evaluating it with METG(50\%) is future work.

\section{Conclusion}
\label{sec:conclusion}

We have presented Operon, a statically verifiable workflow engine that implements the view that ragged data is a flat function over a coordinate space.
This perspective directly resolves the three core challenges of ragged data pipelines: explicit dimensional dependency relations eliminate shape-semantic ambiguity at static verification time; the order-theoretic coordinate model provides a natural representation for partially-known data states; and the incremental scheduler dispatches each task instance as soon as its coordinate becomes available.
Together, these properties are guaranteed by construction for every well-formed Operon program, independent of the actual lengths produced at runtime.

DAG-based agentic frameworks that coordinate multiple LLM or VLM calls increasingly exhibit the same challenges that motivate Operon.
Each agent invocation may produce a variable number of downstream tasks, failure rates are high, and productive execution requires dispatching downstream calls as soon as any upstream result is ready rather than waiting for the full batch~\cite{park2025practical,niu2025flow,yang2025agentnet,masters2025orchestrating}.
These frameworks are natural application targets for Operon: the pipeline structure they define is already acyclic and data-parallel, the unpredictable output cardinalities are the ragged dimensions Operon was designed to model, and the high failure rates are the scenario for which Operon's fault-tolerance machinery was built.
Systems such as AFlow~\cite{zhang2025aflow} and MacNet~\cite{qian2025scaling} illustrate the engineering overhead of managing this structure by hand; Operon's named-dimension model offers a principled alternative that eliminates the bookkeeping while providing static guarantees the current frameworks do not.

A natural direction for future work is developing a type system for ragged arrays within the named dimension framework, together with a corresponding algebra of array operations beyond the elementary aggregation, slicing, and Cartesian products currently supported.
Such an algebra would allow Operon programs to express a richer class of ragged computations while preserving the static guarantees that make the system's runtime behavior predictable.


%

\bibliographystyle{ACM-Reference-Format}
\bibliography{references}

\newpage
\appendix
\section{Compatibility with Nested Containers}

We show that our design of arrays is compatible with traditional multidimensional arrays.
Nested linear containers serve as natural baselines for this comparison, as they are most commonly used in many languages to express ragged data.
In our formulation, such arrays correspond to systems with a \emph{totally ordered} dimension space.
Each resolution in a shape then represents an individual container, with the \(d\), \(c\), and \(\ell\) values denoting the depth, the indices to this container, and the length, respectively.
As such, we discuss how shapes and arrays transform when we linearly extend the dimension space to a total order.

Hereafter, we understand a linear extension \(L\) as a permutation of \(\mathcal{D}\) that preserves the original order \(\prec\); the extended order \(d \prec_L e\) means that \(d\) appears before \(e\) in \(L\).

\begin{definition}[Canonical expansions]\label{def:canonical_expansions}
    For a shape \(R\) on \((\mathcal{D}, \prec)\) with a linear extension \(L\), if a shape \(R_L\) on \((\mathcal{D}, \prec_L)\) satisfies
    \[
        (d, c_L, \ell) \in R_L \implies (d, c_L|_{d^\Downarrow}, \ell) \in R,
    \]
    we call \(R_L\) a \emph{canonical expansion} of \(R\) relative to \(L\).
\end{definition}

The following theorem asserts that all arrays can be rewritten into a nested container while preserving their data.

\begin{theorem}\label{thm:canonical_expansions}
    For a shape \(R\) on \((\mathcal{D}, \prec)\) with a linear extension \(L\),
    \begin{enumerate}
        \item the canonical expansion \(R_L\) uniquely exists;
        \item \(R_L\) preserves coordinate spaces, that is, if \(\mathcal{F} \subseteq \mathcal{D}\) is closed in both \((\mathcal{D}, \prec)\) and \((\mathcal{D}, \prec_L)\), then \(\coordsp{\mathcal{F}}{R} = \coordsp{\mathcal{F}}{R_L}\).
    \end{enumerate}
\end{theorem}

Finally, by comparing \(|R|\) to \(|R_L|\), we obtain an upper bound on the additional storage required by our design.
Here, we exclude zero-length resolutions to avoid degenerate expansions that make \(|R_L|\) artificially small: an independent dimension with zero length could nullify everything else if \(L\) has it as the first element.
The following theorem states that the number of resolutions never exceeds the number of nested containers, provided the above assumption holds.

\begin{theorem}\label{thm:shape_size_upper_bound}
    Consider a shape \(R\) on \((\mathcal{D}, \prec)\) and a linear extension \(L\).
    Assuming that \(\ell > 0\) for all \((d, c, \ell) \in R\), the number of resolution entries \(|R|\) satisfies \(|R| \leq |R_L|\).
\end{theorem}

\section{Proofs}
\label{sec:appx_proofs}

\begin{lemma*}[\ref{lem:principal_deps_are_closed}]
    Every principal dependency space is closed.
\end{lemma*}

\begin{proof}
    Suppose \(d^\Downarrow\) is not closed.
    Then, there must exist some \(e \in d^\Downarrow\) such that \(e^\downarrow \not\subseteq d^\Downarrow\).
    Since \(e \in d^\downarrow\) and \(d^\downarrow\) is closed, we have \(e^\downarrow \subseteq d^\downarrow\).
    Also, since \(e \prec d\) and \(\prec\) is asymmetric, we have \(d \not\prec e\) and in turn \(d \notin e^\downarrow\).
    Thus, \(e^\downarrow \subseteq d^\downarrow \setminus \set{d} = d^\Downarrow\), which is a contradiction.
\end{proof}

\begin{proposition*}[\ref{prop:properties_of_coordinates}(1)]
    For closed subspaces \(\mathcal{F}' \subseteq \mathcal{F} \subseteq \mathcal{D}\) and a coordinate \(c: \coord{\mathcal{F}}{R}\), \(c|_{\mathcal{F}'} \in \coordsp{\mathcal{F}'}{R}\).
\end{proposition*}

\begin{proof}
    Consider a dimension \(d \in \mathcal{F}'\).
    From the in-bounds condition \(c \ib R\), we have \((d, c|_{d^\Downarrow}) \in \dom(R)\) and \(R(d, c|_{d^\Downarrow}) > c(d)\).
    Since \(\mathcal{F}'\) is closed, \(d^\Downarrow \subseteq \mathcal{F}'\), from which we have \((c|_{\mathcal{F}'})|_{d^\Downarrow} = c|_{d^\Downarrow}\).
    Then, we can rewrite the statements as \((d, (c|_{\mathcal{F}'})|_{d^\Downarrow})\in \dom(R)\) and \(c|_{\mathcal{F}'}(d) < R(d, (c|_{\mathcal{F}'})|_{d^\Downarrow})\).
    This means that \(c|_{\mathcal{F}'} \ib R\), and therefore \(c|_{\mathcal{F}'} \in \coordsp{\mathcal{F}'}{R}\).
\end{proof}

\begin{proposition*}[\ref{prop:properties_of_coordinates}(2)]
    For a closed \(\mathcal{F} \subseteq \mathcal{D}\), \(\subcoordsp{\mathcal{F}}{\emptyset}{R} = \coordsp{\mathcal{F}}{R}\).
\end{proposition*}

\begin{proof}
    \[
        \begin{aligned}
        \subcoordsp{\mathcal{F}}{\emptyset}{R}
        &= \set{c|_{\mathcal{F}} \mid c \in \coordsp{\mathcal{F}}{R} \wedge c|_{\emptyset} = \emptyset} \\
        &= \set{c \mid c \in \coordsp{\mathcal{F}}{R}} \\
        &= \coordsp{\mathcal{F}}{R}.
        \end{aligned}
    \]
\end{proof}

\begin{proposition*}[\ref{prop:properties_of_coordinates}(3)]
    For a closed \(\mathcal{F} \subseteq \mathcal{D}\), a subspace \(\mathcal{E} \subseteq \mathcal{F}\) with closed \(\mathcal{F} \setminus \mathcal{E}\), and a coordinate \(c^\bot: \coord{\mathcal{F} \setminus \mathcal{E}}{R}\), there exists a \emph{restricted shape} \(R|_{(\mathcal{E}, c^\bot)}\), a shape on \((\mathcal{E}, \prec|_\mathcal{E})\), such that \[
        \subcoordsp{\mathcal{E}}{c^\bot}{R} = \coordsp{\mathcal{E}}{R|_{(\mathcal{E}, c^\bot)}}.
    \]
    That is, we can interpret each subcoordinate space as a coordinate space when the shape is appropriately restricted.
    We have \(R|_{(\mathcal{E}, \emptyset)} \subseteq R\) when \(\mathcal{E} = \mathcal{F}\).
\end{proposition*}

\begin{proof}
    We show that the proposed equality holds for
    \[
        R|_{(\mathcal{E}, c^\bot)} = \set{
            \left(e, c|_{e^\Downarrow \setminus (\mathcal{F} \setminus \mathcal{E})}, \ell\right)
            \;\middle\vert\;
            \begin{array}{l}
                (e, c, \ell) \in R \;\wedge \\
                e \in \mathcal{E} \;\wedge \\
                \forall d \in e^\Downarrow \cap (\mathcal{F} \setminus \mathcal{E}).\;c(d) = c^\bot(d)
            \end{array}
        }.
    \]

    First, we show that \(R|_{(\mathcal{E}, c^\bot)}\) is a shape on \((\mathcal{E}, \prec|_\mathcal{E})\):
    \begin{itemize}
        \item \(e^\Downarrow\) on \((\mathcal{E}, \prec|_\mathcal{E})\) is equal to \(e^\Downarrow \setminus (\mathcal{F} \setminus \mathcal{E})\) on \((\mathcal{D}, \prec)\).
        \item For all \((e, c, \ell) \in R|_{(\mathcal{E}, c^\bot)}\), we have \(R(e, (c \sqcup c^\bot)|_{e^\Downarrow}) = \ell\).
        Since \(R\) is a shape, \((c \sqcup c^\bot)|_{e^\Downarrow} \ib R\), and this implies that \(\forall d \in e^\Downarrow.\; (c \sqcup c^\bot)(d) < R(d, (c \sqcup c^\bot)|_{d^\Downarrow})\).
        Whenever \(d \in \mathcal{E}\), the left-hand side is \(c(d)\) and the right-hand side is \(R|_{(\mathcal{E}, c^\bot)}(d, c|_{d^\Downarrow \setminus (\mathcal{F} \setminus \mathcal{E})})\), and thus \(c \ib R|_{(\mathcal{E}, c^\bot)}\).
        \item Suppose \(c \ib R|_{(\mathcal{E}, c^\bot)}\) with some \(e \in \mathcal{E}\) and \(c: e^\Downarrow \setminus (\mathcal{F} \setminus \mathcal{E}) \to \nat\).
        Then,
        \begin{itemize}
            \item for any \(d \in e^\Downarrow \setminus (\mathcal{F} \setminus \mathcal{E})\), we have \((c \sqcup c^\bot)|_{e^\Downarrow}(d) = c(d) < R|_{(\mathcal{E}, c^\bot)}(d, (c \sqcup c^\bot)|_{d^\Downarrow \setminus (\mathcal{F} \setminus \mathcal{E})}) = R(d, (c \sqcup c^\bot)|_{d^\Downarrow})\);
            \item for any \(d \in e^\Downarrow \cap (\mathcal{F} \setminus \mathcal{E})\), we have \((c \sqcup c^\bot)|_{e^\Downarrow}(d) = c^\bot(d) < R(d, c^\bot|_{d^\Downarrow}) = R(d, (c \sqcup c^\bot)|_{d^\Downarrow})\).
        \end{itemize}
        Therefore, \((c \sqcup c^\bot)|_{e^\Downarrow} \ib R\), and since \(R\) is a shape, \((e, (c \sqcup c^\bot)|_{e^\Downarrow}) \in \dom(R)\).
        Accordingly, \((e, (c \sqcup c^\bot)|_{e^\Downarrow \setminus (\mathcal{F} \setminus \mathcal{E})}) = (e, c)\) is in \(\dom(R|_{(\mathcal{E}, c^\bot)})\).
    \end{itemize}

    Now that we have established that \(R|_{(\mathcal{E}, c^\bot)}\) is a shape, we show the proposed equality by proving two inclusions.

    \begin{itemize}
        \item \(\subcoordsp{\mathcal{E}}{c^\bot}{R} \subseteq \coordsp{\mathcal{E}}{R|_{(\mathcal{E}, c^\bot)}}\).

        For \(c^\top: \subcoord{\mathcal{E}}{c^\bot}{R}\), write \(c^+ = c^\top \sqcup c^\bot\).
        We can write \(c^+: \coord{\mathcal{F}}{R}\) from the definition of subcoordinates.
        Now, for any \(d \in \mathcal{E}\), we have \(c^+(d) < R(d, c^+|_{d^\Downarrow}) = R|_{(\mathcal{E}, c^\bot)}(d, c^+|_{d^\Downarrow \setminus (\mathcal{F} \setminus \mathcal{E})})\) from \(c^+ \ib R\).
        Since \(c^+(d) = c^\top(d)\) and \(c^+|_{d^\Downarrow \setminus (\mathcal{F} \setminus \mathcal{E})} = c^\top|_{d^\Downarrow \setminus (\mathcal{F} \setminus \mathcal{E})}\), we can rewrite the above as \(c^\top(d) < R|_{(\mathcal{E}, c^\bot)}(d, c^\top|_{d^\Downarrow \setminus (\mathcal{F} \setminus \mathcal{E})})\), which implies \(c^\top \ib R|_{(\mathcal{E}, c^\bot)}\), and thus \(c^\top \in \coordsp{\mathcal{E}}{R|_{(\mathcal{E}, c^\bot)}}\).

        \item \(\subcoordsp{\mathcal{E}}{c^\bot}{R} \supseteq \coordsp{\mathcal{E}}{R|_{(\mathcal{E}, c^\bot)}}\).

        For \(c^\top: \coord{\mathcal{E}}{R|_{(\mathcal{E}, c^\bot)}}\), write \(c^+ = c^\top \sqcup c^\bot\).
        We wish to show that \(c^+ \ib R\).

        \begin{itemize}
            \item for any \(d \in \mathcal{E}\), from \(c^\top \ib R|_{(\mathcal{E}, c^\bot)}\), we have \[
                (d, c^\top|_{d^\Downarrow \setminus (\mathcal{F} \setminus \mathcal{E})}) \in \dom(R|_{(\mathcal{E}, c^\bot)})
            \]
            and therefore from the definition of \(R|_{(\mathcal{E}, c^\bot)}\), \[
                (d, (c^\top|_{d^\Downarrow \setminus (\mathcal{F} \setminus \mathcal{E})} \sqcup c^\bot|_{d^\Downarrow \cap (\mathcal{F} \setminus \mathcal{E})})) = (d, c^+|_{d^\Downarrow}) \in \dom(R).
            \]
            \item for any \(d \in (\mathcal{F} \setminus \mathcal{E})\), from \(c^\bot \ib R\), we have \[(d, c^\bot|_{d^\Downarrow}) \in \dom(R).\]
            Also, since \(\mathcal{F} \setminus \mathcal{E}\) is closed, we have \(d^\Downarrow \subseteq (\mathcal{F} \setminus \mathcal{E})\), which means that \(c^+|_{d^\Downarrow} = c^\bot|_{d^\Downarrow}\).
            Therefore, we can conclude that \((d, c^+|_{d^\Downarrow}) \in \dom(R)\).
        \end{itemize}

        Either way, \((d, c^+|_{d^\Downarrow}) \in \dom(R)\) for all \(d \in \mathcal{F}\), so \(c^+ \ib R\), and thus \(c^+ \in \coordsp{\mathcal{F}}{R}\).
        Then, since \(c^+|_{(\mathcal{F} \setminus \mathcal{E})} = c^\bot\), \(c^+|_{\mathcal{E}} = c^\top \in \subcoordsp{\mathcal{E}}{c^\bot}{R}\).
    \end{itemize}
    From the two inclusions, the stated equality holds.
\end{proof}

\begin{lemma*}[\ref{lem:compatibility}]
    For a partial shape \(\overline{R}\) and a resolution \((d, c, \ell)\), the extension \(\overline{R}\{(d, c) \mapsto \ell\}\) stays a partial shape if \((d, c) \bowtie \overline{R}\).
\end{lemma*}

\begin{proof}
    Write \(\overline{R}' = \overline{R}\{(d, c) \mapsto \ell\}\).
    For any pair \((d^\ast, c^\ast) \in \dom(\overline{R}')\), \(c^\ast \ib \overline{R}\) holds:
    \begin{itemize}
        \item If \((d^\ast, c^\ast) \in \dom(\overline{R})\), we have \(c^\ast \ib \overline{R}\) since \(\overline{R}\) is a partial shape.
        \item Otherwise, if it is the newly added \((d, c)\), \(c \ib \overline{R}\) holds by the definition of \((d, c) \bowtie \overline{R}\).
    \end{itemize}

    Now, consider a dimension \(e \in \dom(c^\ast)\).
    From \(c^\ast \ib \overline{R}\), we have \((e, c^\ast|_{e^\Downarrow}) \in \dom(\overline{R})\) and \(c^\ast(e) < \overline{R}(e, c^\ast|_{e^\Downarrow}) = \overline{R}'(e, c^\ast|_{e^\Downarrow})\) (since \((e, c^\ast|_{e^\Downarrow})\) was already in \(\dom(\overline{R})\) and was unchanged in \(\overline{R}'\)).
    This means that \(c^\ast \ib \overline{R}'\).

    Since \(c^\ast \ib \overline{R}'\) holds for any pair \((d^\ast, c^\ast) \in \dom(\overline{R}')\), the extended resolution map \(\overline{R}' = \overline{R}\{(d, c) \mapsto \ell\}\) is also a partial shape.
\end{proof}

\begin{theorem*}[\ref{thm:progress}]
    A partial shape has a compatible resolution if and only if it is incomplete.
\end{theorem*}

\begin{proof}
    Let \(\overline{R}\) denote the partial shape on \((\mathcal{D}, \prec)\).

    \begin{itemize}
        \item (\(\Rightarrow\)) If \(\exists(d, c) \bowtie \overline{R}\), then \(\overline{R}\) is incomplete.

        Since \((d, c) \bowtie \overline{R}\), \(c \ib \overline{R} \land (d, c) \notin \dom(\overline{R})\).
        This is a counterexample to \(c \ib \overline{R} \rightarrow (d, c) \in \dom(\overline{R})\), which is a required condition for \(\overline{R}\) to be complete.
        Therefore, \(\overline{R}\) is incomplete.

        \item (\(\Leftarrow\)) If \(\overline{R}\) is incomplete, it has a compatible resolution.

        Assume for contradiction that for all \(d \in \mathcal{D}\) and all \(c: d^\Downarrow \rightarrow \nat\), if \(c \ib \overline{R}\), then \((d, c) \in \dom(\overline{R})\).
        From the definition of partial shapes, we already know that for all \(d \in \mathcal{D}\) and all \(c: d^\Downarrow \rightarrow \nat\), if \((d, c) \in \dom(\overline{R})\), then \(c \ib \overline{R}\).
        Combining these two yields: \[
            \forall d \in \mathcal{D}. \; \forall c: d^\Downarrow \rightarrow \nat. \; (d, c) \in \dom(\overline{R}) \Leftrightarrow c \ib \overline{R}.
        \]
        This is the defining condition for \(\overline{R}\) to be complete, a contradiction.
        Therefore, there exists \((d, c)\) such that \(c \ib \overline{R}\) and \((d, c) \notin \dom(\overline{R})\), and \((d, c) \bowtie \overline{R}\) for such a pair.
    \end{itemize}
\end{proof}

\begin{theorem*}[\ref{thm:termination}]
    There is no infinite sequence of partial shapes where each step adds a resolution.
\end{theorem*}

\begin{proof}
    We show this by well-founded induction on the downset lattice of \((\mathcal{D}, \prec)\), where our hypothesis \(\Phi(\mathcal{F})\) is that there is no infinite sequence of resolutions \(S = \seq{r_n}\) where each resolution \(r_n = (d_n, c_n, \ell_n)\) has \(d_n \in \mathcal{F}\) and \((d_n, c_n) \bowtie \overline{R} = \bigsqcup_{k<n} \set{r_k}\).
    The theorem then follows from \(\Phi(\mathcal{D})\).
    We preface with that the lattice is well-founded because \(\mathcal{D}\) is finite.
    \begin{itemize}
        \item For the base case, let \(\mathcal{F} = \emptyset\).
        Since the only valid partial shape on \(\emptyset\) is \(\emptyset\) which is already complete, the hypothesis \(\Phi(\emptyset)\) holds.

        \item For the inductive step, let \(\mathcal{F} \subseteq \mathcal{D}\) be nonempty and closed, and assume that for all closed \(\mathcal{F}' \subsetneq \mathcal{F}\), the hypothesis \(\Phi(\mathcal{F}')\) holds.
        Suppose for contradiction that there exists an infinite sequence \(S = \seq{r_n}\) such that \(d_n \in \mathcal{F}\) and \((d_n, c_n) \bowtie \overline{R} = \bigsqcup_{k<n} \set{r_k}\).

        Since \(\mathcal{F}\) is nonempty and finite, let us choose a maximal element \(d^\top \in \mathcal{F}\).
        Because \(d^\top\) is maximal, the inclusion or exclusion of any resolutions of \(d^\top\) does not affect the compatibility of any other dimensions in \(\mathcal{F}\), and therefore the sequence must still be valid if we remove all resolutions of \(d^\top\) from the sequence.
        From \(\Phi(\mathcal{F} \setminus \set{d^\top})\), we can conclude that this modified sequence \(S'\) is finite, and thus the number of resolutions in \(S\) is finite for all \(d \in \mathcal{F} \setminus \set{d^\top}\).

        Since \(S\) is infinite, there must have been infinitely many resolutions of the form \((d^\top, c^\top, \ell)\) in \(S\), and thus infinitely many such \emph{distinct} coordinates \(c^\top\).
        It follows that \(\dom(c^\top) = (d^\top)^\Downarrow\) is nonempty.
        Writing \(L = \max \set{\ell \mid e \in (d^\top)^\Downarrow \land (e, c, \ell) \in S}\), we see that since \(L^{|(d^\top)^\Downarrow|}\) is finite, there must be at least one coordinate \(c^\top\) such that \((d^\top, c^\top, \ell) \in S\) and \(c^\top(e) \geq L\) for some \(e \in (d^\top)^\Downarrow\).
        However, from \((d^\top, c^\top) \bowtie \overline{R}\), we obtain \(\overline{R}(e, c^\top|_{e^\Downarrow}) > c^\top(e) \geq L\), which contradicts the maximality of \(L\).
        We conclude that there is no such infinite sequence \(S\), and thus \(\Phi(\mathcal{F})\) holds.
    \end{itemize}
    Therefore, \(\Phi(\mathcal{D})\) holds by induction, which proves the theorem.
\end{proof}

\begin{theorem*}[\ref{thm:local_commutativity}]
    If \((d, c) \bowtie \overline{R}\) and \((d', c') \bowtie \overline{R}\) with \((d, c) \neq (d', c')\), then \((d', c') \bowtie \overline{R}\{(d, c) \mapsto \ell\}\).
\end{theorem*}

\begin{proof}
    From \((d', c') \bowtie \overline{R}\), we have \(c' \ib \overline{R}\) and \((d', c') \notin \dom (\overline{R})\).

    Now, consider a dimension \(e \in \dom(c')\).
    From the in-bounds condition \(c' \ib \overline{R}\), we have \((e, c'|_{e^\Downarrow}) \in \dom(\overline{R})\) and \(c'(e) < \overline{R}(e, c'|_{e^\Downarrow})\).
    Since \((d, c) \bowtie \overline{R}\) requires \((d, c) \notin \dom(\overline{R})\), \(\overline{R}' = \overline{R}\{(d, c) \mapsto \ell\}\) has the resolution for \((e, c'|_{e^\Downarrow})\) unchanged.
    Therefore, \((e, c'|_{e^\Downarrow}) \in \dom(\overline{R}')\) and \(c'(e) < \overline{R}'(e, c'|_{e^\Downarrow})\), that is, \(c' \ib \overline{R}'\).

    Also, since \((d', c') \notin \dom (\overline{R})\), it is trivial that \((d', c') \notin \dom (\overline{R}\{(d, c) \mapsto \ell\})\).

    Therefore, we can conclude that \((d', c') \bowtie \overline{R}\{(d, c) \mapsto \ell\}\).
\end{proof}

\begin{corollary*}[\ref{cor:determinism}]
    Under a fixed function \(\pi: \bigcup_{d \in \mathcal{D}} \left(\set{d} \times [d^\Downarrow \to \nat] \right) \to \nat\), any non-starving execution of Alg.~\ref{alg:incremental_construction} or Alg.~\ref{alg:parallel_incremental_construction} terminates and returns the same complete shape.
\end{corollary*}

\begin{proof}
    Define the reduction relation \(\overline{R}_1 \to \overline{R}_2\) when \(\exists (d, c)\) such that \(\overline{R}_2 = \overline{R}_1 \set{(d, c) \mapsto \pi(d, c)}\) and \((d, c) \bowtie \overline{R}_1\).
    We may thus view both Alg.~\ref{alg:incremental_construction} and Alg.~\ref{alg:parallel_incremental_construction} as implementations of a reduction system starting from the empty partial shape \(\emptyset\).
    From Thm.~\ref{thm:termination} and Thm.~\ref{thm:local_commutativity}, this system is terminating and locally confluent, which implies that it is globally confluent.
    Therefore, any non-starving execution of either algorithm must terminate at the same normal form, i.e., the complete shape.
\end{proof}

\begin{theorem*}[\ref{thm:distribution_confluence_of_shapes}]
    Under a stochastic function \(\pi\) that samples independently from a fixed function \(\rho: \bigcup_{d \in \mathcal{D}} \left(\set{d} \times [d^\Downarrow \to \nat] \right) \to \dist(\nat)\), any non-starving execution of Alg.~\ref{alg:incremental_construction} or Alg.~\ref{alg:parallel_incremental_construction} terminates and returns a complete shape whose distribution is invariant to the order of scheduling.
\end{theorem*}

\begin{proof}
    We use the framework of \cite{DIAZCARO2018115} for the explicit representation of \(\dist\) and the notion of confluence in probabilistic abstract rewriting systems (PARS), adapting their notation for consistency with our domain.
    We reproduce some of their definitions and results that we use here:
    \begin{itemize}
        \item A distribution over a set \(A\), \(D \in \dist^\ast(A)\), is a finite list of pairs \((p, a) \in \mathbb{R}^+ \times A\).
        Furthermore, \(D\) is a normalized distribution \(D \in \dist(A)\) if it additionally meets \(\sum_{(p, a) \in D} p = 1\).
        \item Two distributions \(D, E \in \dist^\ast(A)\) are equivalent, denoted \(D \approx E\), if \(\forall a \in A. \sum_{(p, a) \in D} p = \sum_{(q, a) \in E} q\).
        \(\approx\) is a congruence relation on \(\dist^\ast(A)\).
        \item A probabilistic abstract rewriting system (PARS) is a pair \((A, \hookrightarrow)\) with a set \(A\) and a relation that maps elements to their potential results \(\hookrightarrow \;\subseteq A \times \dist(A)\) (called \emph{pointwise evolution}).
        \item Given a PARS \((A, \hookrightarrow)\), there exists the associated ARS \((\dist^\ast(A), \twoheadrightarrow)\), where \(\twoheadrightarrow \;=\; \twoheadrightarrow_P \cup \approx\) with \(\twoheadrightarrow_P\) (called \emph{parallel evolution}) being defined by the rules: \[
            \frac{a \hookrightarrow A \quad ds \twoheadrightarrow_P ds'}{[(p, a)]::ds \twoheadrightarrow_P pA :: ds'} \qquad
            \frac{ds \twoheadrightarrow_P ds'}{[(p, a)]::ds \twoheadrightarrow_P [(p, a)]::ds'} \qquad
            \frac{}{[] \twoheadrightarrow_P []}
        \]
        where \(::\) denotes list concatenation and \(pA\) denotes the distribution obtained by multiplying each probability in \(A\) by \(p\).
        \item A PARS is \emph{distribution confluent} or \emph{confluent} if the associated ARS is confluent.
        A confluent PARS has a unique terminal distribution from any initial element.
        \item If every element \(a\) in \(A\) has the \emph{diamond property} for the relation \(\twoheadrightarrow_P/\approx\), i.e., if \[D \hookleftarrow a \hookrightarrow E \implies \exists F.\; D (\twoheadrightarrow_P/\approx) F (\twoheadleftarrow_P/\approx) E,\] then the PARS \((A, \hookrightarrow)\) is confluent.
    \end{itemize}
    Define the PARS \((\mathcal{R}, \hookrightarrow)\) where \(\mathcal{R} = \left[\bigcup_{d \in \mathcal{D}} \left(\set{d} \times [d^\Downarrow \to \nat]\right) \rightharpoonup \nat\right]\) is the set of resolution maps and the relation \(\hookrightarrow\) is defined by the distributional function \(\rho\) as follows: \[
        \frac{(d, c) \bowtie \overline{R} \quad \rho(d, c) = [(p_i, \ell_i)]_i}{\overline{R} \hookrightarrow [(p_i, \overline{R}\set{(d, c) \mapsto \ell_i})]_i}
    \]
    Note that \((d, c) \bowtie \overline{R}\) guarantees that \(\overline{R}\) is an incomplete partial shape.
    We may now identify both Alg.~\ref{alg:incremental_construction} and Alg.~\ref{alg:parallel_incremental_construction} as implementations of this PARS, starting from the trivial partial shape \(\emptyset\) and terminating at a complete shape.

    We show that the PARS \((\mathcal{R}, \hookrightarrow)\) is confluent by showing the diamond property for any partial shape \(\overline{R}\).
    Suppose \(\overline{R} \hookrightarrow D\) and \(\overline{R} \hookrightarrow E\) for some distributions \(D, E \in \dist(\mathcal{R})\).
    Due to the definition of \(\hookrightarrow\), there exist \((d_1, c_1) \bowtie \overline{R}\) and \((d_2, c_2) \bowtie \overline{R}\) such that \begin{align*}
        \rho(d_1, c_1) &= [(p_i, \ell_i)]_i, \qquad D = [(p_i, \overline{R}\set{(d_1, c_1) \mapsto \ell_i})]_i; \\
        \rho(d_2, c_2) &= [(q_j, m_j)]_j, \qquad E = [(q_j, \overline{R}\set{(d_2, c_2) \mapsto m_j})]_j.
    \end{align*}
    In the trivial case where \((d_1, c_1) = (d_2, c_2)\), we have \(D = E\) and the diamond property trivially holds with \(F = D = E\).
    Otherwise, let \[
        F = \left[\left(p_i q_j, \overline{R} \set{(d_1, c_1) \mapsto \ell_i} \set{(d_2, c_2) \mapsto m_j}\right)\right]_{ij}.
    \]
    We have \[
        \forall i.\; \dfrac{
            \dfrac{
                (d_1, c_1) \bowtie \overline{R} \quad (d_1, c_1) \neq (d_2, c_2)
            }{
                (d_2, c_2) \bowtie \overline{R}\set{(d_1, c_1) \mapsto \ell_i}
            } \;(\text{Thm.~\ref{thm:local_commutativity}}) \quad
            \rho(d_2, c_2) = [(q_j, m_j)]_j
        }{
            \overline{R}\set{(d_1, c_1) \mapsto \ell_i} \hookrightarrow \left[\left(q_j, \overline{R}\set{(d_1, c_1) \mapsto \ell_i} \set{(d_2, c_2) \mapsto m_j}\right)\right]_j
        }
    \]
    and therefore by applying the first rule of \(\twoheadrightarrow_P\) to every element in \(D\), we obtain \[
        D \twoheadrightarrow_P \text{concat}_i \left(p_i \left[\left(q_j, \overline{R}\set{(d_1, c_1) \mapsto \ell_i} \set{(d_2, c_2) \mapsto m_j}\right)\right]_j\right) \approx F,
    \]
    and vice versa for \(E (\twoheadrightarrow_P/\approx) F\).
    The confluence of the PARS \((\mathcal{R}, \hookrightarrow)\) follows, and we conclude that any non-starving algorithm that implements this PARS terminates in finite time (Thm.~\ref{thm:termination}) with a unique terminal distribution over complete shapes.
\end{proof}

\begin{proposition*}[\ref{prop:unknown_extension}]
    Def.~\ref{def:coordinates_with_unknowns} is a strict extension of the original definition of coordinates in Def.~\ref{def:coordinates}.
    That is, \(\coord{\mathcal{F}}{R}\) is unchanged under either definition when \(R\) is a complete shape.
\end{proposition*}

\begin{proof}
    Note that throughout the proof, we refer to the two conditions in Def.~\ref{def:coordinates_with_unknowns} as conditions (1) and (2).

    Assume \(R\) is a complete shape on \((\mathcal{D}, \prec)\), i.e. \[
        \forall d \in \mathcal{D}.\; \forall c \in d^\Downarrow \rightarrow \nat.\; (d, c) \in \dom(R) \leftrightarrow c \ib R.
    \]

    Let \(c: \mathcal{F} \rightharpoonup \nat\) satisfy the conditions in Def.~\ref{def:coordinates_with_unknowns}.
    We must show that \(c\) must be total on \(\mathcal{F}\) and satisfy \(c \ib R\).

    Suppose for contradiction that \(c\) is partial, i.e. \(\dom(c) \subsetneq \mathcal{F}\).
    Then, there exists a \(\prec\)-minimal element \(d^\ast \in \mathcal{F} \setminus \dom(c)\).
    From the minimality of \(d^\ast\), we can infer that \((d^\ast)^\Downarrow \subseteq \dom(c)\), and thus \(c|_{(d^\ast)^\Downarrow}\) is total on \((d^\ast)^\Downarrow\).
    Now, consider a dimension \(e \in (d^\ast)^\Downarrow\).
    Applying the condition (1), we have \((e, c|_{e^\Downarrow}) \in \dom(R)\) and \(c(e) < R(e, c|_{e^\Downarrow})\), and in turn \(c|_{(d^\ast)^\Downarrow} \ib R\).
    Since \(R\) is a complete shape, this means that \((d^\ast, c|_{(d^\ast)^\Downarrow}) \in \dom(R)\), contradicting condition (2).
    Hence, \(c\) is total on \(\mathcal{F}\).

    With totality, Def.~\ref{def:coordinates_with_unknowns} simplifies to \(c \ib R\), which is precisely the definition of coordinates in Def.~\ref{def:coordinates}.
    Therefore, the coordinate space remains unchanged under either definition.
\end{proof}

\begin{theorem*}[\ref{thm:coordinate_explosion}(1)]
    For a coordinate space \(\ucoordsp{\mathcal{F}}{\overline{R}}\) and a resolution \((d, c^\Downarrow, \ell)\) with \(d \in \mathcal{F}\), \(c^\Downarrow: \coord{d^\Downarrow}{\overline{R}}\), and \((d, c^\Downarrow) \bowtie \overline{R}\), only the \(c^\Downarrow\)-sibling coordinates \(U = \ucoordsp{\mathcal{F}}{\overline{R}}_{c^\Downarrow}\) are affected by the addition of \((d, c^\Downarrow) \mapsto \ell\) into \(\overline{R}\) as follows.

    \(\forall c' \in U.\; d \notin \dom(c')\).
    I.e., none of the \(c^\Downarrow\)-siblings already have a value for \(d\) filled in.
\end{theorem*}

\begin{proof}
    Suppose for contradiction that \(d \in \dom(c')\).
    Then, by definition of \(\ucoordsp{\mathcal{F}}{\overline{R}}_{c^\Downarrow}\), we have \(c' \ib \overline{R}\) and thus \((d, c'|_{d^\Downarrow}) = (d, c^\Downarrow) \in \dom(\overline{R})\).
    This contradicts \((d, c^\Downarrow) \bowtie \overline{R}\), which states that \((d, c^\Downarrow) \notin \dom(\overline{R})\).
    Therefore, \(d \notin \dom(c')\).
\end{proof}

\begin{theorem*}[\ref{thm:coordinate_explosion}(2)]
    Writing \(\overline{R}_\ell = \overline{R}\{(d, c^\Downarrow) \mapsto \ell\}\), \[
        \ucoordsp{\mathcal{F}}{\overline{R}_\ell} = \left(\ucoordsp{\mathcal{F}}{\overline{R}} \setminus U\right) \sqcup \bigsqcup_{c' \in U} {\bigsqcup_{i=0}^{\ell-1} \set{c'\{d \mapsto i\}}}.
    \]
\end{theorem*}

\begin{proof}
    Let us partition both sides of the wanted equality by whether the coordinate is a \(c^\Downarrow\)-sibling or not.
    We then wish to show:
    \begin{align*}
        \ucoordsp{\mathcal{F}}{\overline{R}_\ell}_{c^\Downarrow} &= \bigsqcup_{c' \in U} {\bigsqcup_{i=0}^{\ell-1} \set{c'\{d \mapsto i\}}}; \\
        \ucoordsp{\mathcal{F}}{\overline{R}_\ell} \setminus \ucoordsp{\mathcal{F}}{\overline{R}_\ell}_{c^\Downarrow} &= \ucoordsp{\mathcal{F}}{\overline{R}} \setminus \ucoordsp{\mathcal{F}}{\overline{R}}_{c^\Downarrow}.
    \end{align*}
    Rewriting both set equalities as conditionals for an arbitrary \(c: \mathcal{F} \rightharpoonup \nat\), we get:
    \begin{align*}
        \text{(A) } c|_{d^\Downarrow} = c^\Downarrow &\implies \left(c: \ucoord{\mathcal{F}}{\overline{R}_\ell} \iff c(d) < \ell \land c|_{\mathcal{F} \setminus \set{d}}: \ucoord{\mathcal{F}}{\overline{R}}\right); \\
        \text{(B) } c|_{d^\Downarrow} \neq c^\Downarrow &\implies \left(c: \ucoord{\mathcal{F}}{\overline{R}_\ell} \iff c: \ucoord{\mathcal{F}}{\overline{R}}\right).
    \end{align*}
    (B) is trivial since the inclusion or exclusion of \((d, c^\Downarrow, \ell)\) does not affect the truthfulness of neither the conditions in Def.~\ref{def:coordinates_with_unknowns} for any coordinate \(c\) with \(c|_{d^\Downarrow} \neq c^\Downarrow\).
    We now prove (A) in two directions.
    \begin{itemize}
        \item (\(\Rightarrow\)).
        Let \(c: \ucoord{\mathcal{F}}{\overline{R}_\ell}\) with \(c|_{d^\Downarrow} = c^\Downarrow\).
        From \(c \ib \overline{R}_\ell\), \(\dom(c)\) is closed.
        \begin{enumerate}
            \item Since \((d, c^\Downarrow) \in \dom(\overline{R}_\ell)\), Def.~\ref{def:coordinates_with_unknowns}(2) requires \(d \in \dom(c)\) and therefore \(c(d) < \overline{R}_\ell(d, c^\Downarrow) = \ell\).
            \item Write \(c' = c|_{\mathcal{F} \setminus \set{d}}\).
            \begin{enumerate}
                \item We first show that \(\dom(c')\) is closed.
                It is enough to verify that \(d\) was a maximal element in \(\dom(c)\), i.e., for any \(e \in \mathcal{F}\) with \(d \prec e\), \(e \notin \dom(c)\).
                Indeed, if there were such an \(e\) that \(d \prec e\) and \(e \in \dom(c)\), then from \(c \ib \overline{R}_\ell\), we would have \(c(e) < \overline{R}_\ell(e, c|_{e^\Downarrow}) = \overline{R}(e, c|_{e^\Downarrow})\).
                Since \(\overline{R}\) is a partial shape, \(c|_{e^\Downarrow} \ib \overline{R}\) and thus \((d, (c|_{e^\Downarrow})|_{d^\Downarrow}) = (d, c^\Downarrow) \in \overline{R}\), which contradicts \((d, c^\Downarrow) \bowtie \overline{R}\).
                \item For any \(d^\ast \in \dom(c) \setminus \set{d} = \dom(c')\), we have \(c'(d^\ast) = c(d^\ast) < \overline{R}_\ell(d^\ast, c|_{(d^\ast)^\Downarrow})\) from \(c \ib \overline{R}_\ell\).
                Since \(\overline{R} = \overline{R}_\ell\) except at \((d, c^\Downarrow)\) and \(d \notin (d^\ast)^\Downarrow\) from the closedness of \(\dom(c')\), we can write \(c'(d^\ast) < \overline{R}(d^\ast, c'|_{(d^\ast)^\Downarrow})\) and thus \(c' \ib \overline{R}\).
                \item For \(d^\ast = d\), we have \(d \in \mathcal{F} \setminus \dom(c')\), and since \((d, c^\Downarrow) \bowtie \overline{R}\), we have \((d, c'|_{d^\Downarrow}) = (d, c^\Downarrow) \notin \dom(\overline{R})\).
                \item For any \(d^\ast \in \mathcal{F} \setminus \dom(c) = (\mathcal{F} \setminus \dom(c')) \setminus \set{d}\), we have \((d^\ast, c|_{(d^\ast)^\Downarrow}) \notin \dom(\overline{R}_\ell)\) from Def.~\ref{def:coordinates_with_unknowns}(2).
                Since \(\overline{R} = \overline{R}_\ell\) except at \((d, c^\Downarrow)\), we can write \((d^\ast, c|_{(d^\ast)^\Downarrow}) \notin \dom(\overline{R})\).
                \begin{itemize}
                    \item If \(d \in (d^\ast)^\Downarrow\), \(c'|_{(d^\ast)^\Downarrow}\) is not total on \((d^\ast)^\Downarrow\), and thus \((d^\ast, c'|_{(d^\ast)^\Downarrow}) \notin \dom(\overline{R})\).
                    \item Otherwise, \(c|_{(d^\ast)^\Downarrow} = c'|_{(d^\ast)^\Downarrow}\), and thus \((d^\ast, c'|_{(d^\ast)^\Downarrow}) \notin \dom(\overline{R})\).
                \end{itemize}
                Either way, \((d^\ast, c'|_{(d^\ast)^\Downarrow}) \notin \dom(\overline{R})\).
            \end{enumerate}
            Since Def.~\ref{def:coordinates_with_unknowns}(1) is satisfied from (b) and Def.~\ref{def:coordinates_with_unknowns}(2) is satisfied from (c) and (d), \(c': \ucoord{\mathcal{F}}{\overline{R}}\).
        \end{enumerate}
        Combining the above, we have (1) \(c(d) < \ell\) and (2) \(c|_{\mathcal{F} \setminus \set{d}}: \ucoord{\mathcal{F}}{\overline{R}}\).
        \(\square\)
        \item (\(\Leftarrow\)).
        Write \(c' = c|_{\mathcal{F} \setminus \set{d}}: \ucoord{\mathcal{F}}{\overline{R}}\).
        From \(c \ib \overline{R}\), \(\dom(c)\) is closed.
        \begin{enumerate}[label=(\alph*)]
            \item For any \(d^\ast \in \dom(c') = \dom(c) \setminus \set{d}\), we have \(c(d^\ast) = c'(d^\ast) < \overline{R}(d^\ast, c'|_{(d^\ast)^\Downarrow})\) from \(c' \ib \overline{R}\).
            Since \(\overline{R} = \overline{R}_\ell\) except at \((d, c^\Downarrow)\) and \(d \notin (d^\ast)^\Downarrow\) from the closedness of \(\dom(c')\), we can write \(c(d^\ast) < \overline{R}_\ell(d^\ast, c|_{(d^\ast)^\Downarrow})\).
            \item For \(d^\ast = d\), we have \(c(d) < \ell = \overline{R}_\ell(d, c|_{d^\Downarrow})\) from our assumption.
            \item For \(d^\ast \in (\mathcal{F} \setminus \dom(c')) \setminus \set{d} = \mathcal{F} \setminus \dom(c)\), we have \((d^\ast, c'|_{(d^\ast)^\Downarrow}) \notin \dom(\overline{R})\) from Def.~\ref{def:coordinates_with_unknowns}(2).
            Since \(\overline{R} = \overline{R}_\ell\) except at \((d, c^\Downarrow)\), we can write \((d^\ast, c'|_{(d^\ast)^\Downarrow}) \notin \dom(\overline{R}_\ell)\).
            \begin{itemize}
                \item If \(d \in (d^\ast)^\Downarrow\), assume that \((d^\ast, c|_{(d^\ast)^\Downarrow}) \in \dom(\overline{R}_\ell)\) for contradiction.
                Since \(d^\ast \neq d\), \((d^\ast, c|_{(d^\ast)^\Downarrow})\) is also present in \(\overline{R}\).
                By the definition of partial shapes, this implies that \(c|_{(d^\ast)^\Downarrow} \ib \overline{R}\).
                Since \(d \in (d^\ast)^\Downarrow\), this requires \((d, c|_{d^\Downarrow}) = (d, c^\Downarrow) \in \overline{R}\), which contradicts \((d, c^\Downarrow) \bowtie \overline{R}\).
                Therefore, \((d^\ast, c|_{(d^\ast)^\Downarrow}) \notin \dom(\overline{R}_\ell)\).
                \item If \(d \notin (d^\ast)^\Downarrow\), \(c|_{(d^\ast)^\Downarrow} = c'|_{(d^\ast)^\Downarrow}\), and thus \((d^\ast, c|_{(d^\ast)^\Downarrow}) \notin \dom(\overline{R}_\ell)\).
            \end{itemize}
            Either way, \((d^\ast, c|_{(d^\ast)^\Downarrow}) \notin \dom(\overline{R}_\ell)\).
        \end{enumerate}
        Since Def.~\ref{def:coordinates_with_unknowns}(1) is satisfied from (a) and (b) and Def.~\ref{def:coordinates_with_unknowns}(2) is satisfied from (c), \(c: \ucoord{\mathcal{F}}{\overline{R}_\ell}\).
    \end{itemize}
\end{proof}

\begin{lemma*}[\ref{lem:operon_well-formedness}]
    Given \((\emptyset, \emptyset, \emptyset) \mid p \vdash (\mathcal{D}, \prec, \Sigma)\),
    \begin{enumerate}
        \item the relation \(\prec\) is a strict partial order over \(\mathcal{D}\);
        \item for all entity types \(\tau \in \dom(\Sigma)\), the characteristic dimension space \(\Sigma(\tau)\) is closed under \(\prec\);
        \item for all tasks \(t = \left< f, s_\text{out}, \overrightarrow{\left< \tau_{\text{in, }i}, \mathcal{E}_{\text{in, }i} \right>}, \mathcal{F}, n \right>\) in \(p\), the dimension spaces \(\mathcal{F}\) and \(\Sigma(\tau_{\text{in, }i}) \setminus \mathcal{E}_{\text{in, }i}\) are closed under \(\prec\).
    \end{enumerate}
\end{lemma*}

\begin{proof}
    We prove the three items simultaneously by induction on the derivation of \((\emptyset, \emptyset, \emptyset) \mid p \vdash (\mathcal{D}, \prec, \Sigma)\) generated by the rules in Fig.~\ref{fig:operon_checking_rules}.
    Let the invariant \(\text{Inv}(\mathcal{D}, \prec, \Sigma)\) be the conjunction of the items (1)-(3) in Lemma~\ref{lem:operon_well-formedness}.

    \paragraph{Base step (\textsc{Unit})}
    For \(p=()\), the rule \textsc{Unit} yields \((\emptyset, \emptyset, \emptyset)\).
    The empty relation is a strict partial order on \(\emptyset\).
    \(\dom(\Sigma) = \emptyset\), so (2) is vacuous.
    There are no tasks in \(p\), so (3) is also vacuous.
    Therefore, \(\text{Inv}\) holds.

    \paragraph{Auxiliary step (\textsc{TaskDef})}
    Assume \(\text{Inv}(\mathcal{D}, \prec, \Sigma)\) and the premises of \(\textsc{TaskDef}\).
    Since \(|\mathcal{E}_\text{out}| \leq 1\), we can say that \(\mathcal{E}_\text{out} = \emptyset\) or \(\set{d_\text{out}}\).

    Define \[
        \mathcal{D}' = \mathcal{D} \sqcup \mathcal{E}_\text{out}; \qquad
        \prec' = \; \prec \sqcup \; \mathcal{F} \times \mathcal{E}_{\text{out}}; \qquad
        \Sigma' = \Sigma \set{\tau_{\text{out}} \mapsto \mathcal{F} \sqcup \mathcal{E}_{\text{out}}}.
    \]
    We show that \(\text{Inv}(\mathcal{D}', \prec', \Sigma')\) holds.

    We start by making couple claims about the properties of \(\prec'\):
    \begin{enumerate}[label=(\roman*)]
        \item For any \(d, e \in \mathcal{D}'\), if \((d, e) \in \; \prec'\), then \(d \in \mathcal{D}\).
        \item For any \(d, e \in \mathcal{D}\), if \((d, e) \in \; \prec'\), then \((d, e) \in \; \prec\).
        \item For any \(\mathcal{F}^\ast \subseteq \mathcal{D}\), if \(\mathcal{F}^\ast\) is closed under \(\prec\), then it is also closed under \(\prec'\).
    \end{enumerate}

    The proof for (i) is trivial.
    Consider \((d, e) \in \; \prec'\).
    If \((d, e) \in \; \prec\), then naturally \(d \in \mathcal{D}\).
    Otherwise, \((d, e) \in \mathcal{F} \times \mathcal{E}_{\text{out}}\), in which case \(d \in \mathcal{F} \subseteq \mathcal{D}\).
    In both cases, \(d \in \mathcal{D}\).
    The proof for (ii) is even simpler. Since \((d, e) \notin \mathcal{F} \times \mathcal{E}_{\text{out}}\), it must be that \((d, e) \in \; \prec\).
    Finally, for (iii), assume for contradiction that there exists \(\mathcal{F}^\ast \subseteq \mathcal{D}\) that is closed under \(\prec\) but not under \(\prec'\).
    Then, there exists \(d \in \mathcal{D}' \setminus \mathcal{F}^\ast, e \in \mathcal{F}^\ast\) such that \((d, e) \in \; \prec'\).
    By (i), \(d \in \mathcal{D}\).
    Then, by (ii) \((d, e) \in \; \prec\), which contradicts the closedness of \(\mathcal{F}^\ast\) under \(\prec\).
    Therefore, \(\mathcal{F}^\ast\) is closed under \(\prec'\).

    Now, we prove the three items that constitute \(\text{Inv}(\mathcal{D}', \prec', \Sigma')\).

    \emph{(1) \(\prec'\) is a strict partial order on \(\mathcal{D}'\).}

    We prove this by showing that \(\prec'\) is irreflexive and transitive.

    Irreflexivity:
    Assume for contradiction that there exists \(d \in \mathcal{D}'\) such that \((d, d) \in\; \prec'\).
    By (i), \(d \in \mathcal{D}\), and then by (ii), \((d, d) \in \; \prec\).
    This contradicts the irreflexivity of \(\prec\) by the inductive hypothesis.
    Therefore, there is no such \(d\), and thus \(\prec'\) is irreflexive.

    Transitivity:
    Suppose \(a \prec' b\) and \(b \prec' c\).
    By (i), \(a, b \in \mathcal{D}\), and then by (ii), \((a, b) \in \; \prec\).
    If \(c \in \mathcal{D}\), then by (ii), \((b, c) \in \; \prec\).
    Since \(\prec\) is transitive by the inductive hypothesis, we have \((a, c) \in \; \prec \; \subseteq \; \prec'\).
    Otherwise, \(c \in \mathcal{E}_\text{out}\).
    In which case, \((a, c) \in \mathcal{D} \times \mathcal{E}_{\text{out}}\), so we have \((a, c) \in \; \prec'\).
    Therefore, \(\prec'\) is transitive.

    \emph{(2) For all \(\tau \in \dom(\Sigma')\), \(\Sigma'(\tau)\) is closed under \(\prec'\).}

    For \(\tau = \tau_\text{out}\), \(\Sigma'(\tau_\text{out}) = \mathcal{F} \sqcup \mathcal{E}_{\text{out}}\).
    Assume for contradiction that this is not closed under \(\prec'\).
    Then, there exists \(d \in \mathcal{D}' \setminus \Sigma'(\tau_\text{out}), e \in \Sigma'(\tau_\text{out})\) such that \(d \prec' e\).
    Note that by (i), we have \(d \in \mathcal{D}\).
    If \(e \in \mathcal{F} \subseteq \mathcal{D}\), then by (ii) we have \((d, e) \in \; \prec\), which contradicts the closedness of \(\mathcal{F}\) under \(\prec\).
    Otherwise, \(e \in \mathcal{E}_{\text{out}}\).
    However, since \((d, e) \notin \; \prec\) and \((d, e) \notin \mathcal{F} \times \mathcal{E}_{\text{out}}\), \((d, e) \notin \; \prec'\), a contradiction.
    Therefore, \(\Sigma'(\tau_\text{out})\) is closed under \(\prec'\).

    For other entity types \(\tau \neq \tau_\text{out}\), \(\Sigma'(\tau) = \Sigma(\tau)\).
    By the inductive hypothesis, \(\Sigma(\tau)\) is closed under \(\prec\), and by (iii), it should also be closed under \(\prec'\).

    \emph{(3) For all tasks \(t^\ast = \left< f, s_\text{out}, \overrightarrow{\left< \tau_{\text{in, }i}, \mathcal{E}_{\text{in, }i} \right>}, \mathcal{F}, n \right>\) in \(p\), the dimension spaces \(\mathcal{F}\) and \(\Sigma(\tau_{\text{in, }i}) \setminus \mathcal{E}_{\text{in, }i}\) are closed under \(\prec\).}

    For \(t^\ast = t\), \(\Sigma(\tau_{\text{in, }i}) \setminus \mathcal{E}_\text{in, }i\) are closed under \(\prec\) by the premises of \textsc{TaskDef}, and \(\mathcal{F}\), the union of these sets, is also closed under \(\prec\).
    For other tasks \(t^\ast \neq t\), the dimension spaces are closed under \(\prec\) by the inductive hypothesis.
    By (iii), these dimension spaces are also closed under \(\prec'\).

    \paragraph{Inductive step (\textsc{Chain})}
    Assume \((\emptyset, \emptyset, \emptyset) \mid \vec{t} \vdash (\mathcal{D}_1, \prec_1, \Sigma_1)\), \((\mathcal{D}_1, \prec_1, \Sigma_1) \mid t' \vdash (\mathcal{D}_2, \prec_2, \Sigma_2)\), and \(\text{Inv}(\mathcal{D}_1, \prec_1, \Sigma_1)\).
    Since the only step that yields \((\mathcal{D}_1, \prec_1, \Sigma_1) \mid t' \vdash (\mathcal{D}_2, \prec_2, \Sigma_2)\) is \(\textsc{TaskDef}\), \(\text{Inv}(\mathcal{D}_2, \prec_2, \Sigma_2)\) also holds from applying the auxiliary step above.

    Therefore, \(\text{Inv}(\mathcal{D}, \prec, \Sigma)\) holds for the final triple \((\mathcal{D}, \prec, \Sigma)\).
\end{proof}

\begin{lemma*}[\ref{lem:ticket_ready}]
    When a ticket's \emph{\texttt{count}} equals its \emph{\texttt{quota}}, the ticket is fully resolved.
\end{lemma*}

\begin{proof}
    Consider a task \(t = \left< f, s_\text{out}, \overrightarrow{\left< \tau_{\text{in, }i}, \mathcal{E}_{\text{in, }i} \right>}, \mathcal{F}, n \right>\) and its ticket \(\overline{j}_t(c)\).
    Furthermore, let \((\overline{R}, \overline{E})\) be the current state of resolutions and entities.

    Base step: If \(|\overrightarrow{\left< \tau_{\text{in, }i}, \mathcal{E}_{\text{in, }i} \right>}| = 0\), the ticket is always fully resolved. Therefore, the proposition holds.

    Inductive step: Expressing \(t_i = \left< \_, \left< \tau_{\text{in, }i}, \mathcal{E}_i \right>, \_, \mathcal{F}_i, \_ \right>\) as the task that produces input entity \(\tau_{\text{in, }i}\), assume that the proposition holds for each \(t_i\).

    In order for the ticket's \texttt{count} to equal its \texttt{quota}, all tickets in \(\overline{j}_t(c)\)'s dependencies \[
        \bigsqcup_{i} \overline{j}_{t_i}\left[c|_{\mathcal{F}_i \setminus \mathcal{E}_{\text{in, }i}}\right].
    \] has to be in the \texttt{Done} state.
    This means that each of these tickets must have their \texttt{count} equal to their \texttt{quota}.
    By the inductive hypothesis, each ticket in the dependencies is fully resolved.

    Consider a dimension \(d \in \mathcal{F}\). From the parsing rule, we have \(d \in \bigcup_i \Sigma(\tau_{t_i})\).
    \begin{itemize}
        \item If \(\exists i. \; d \in \mathcal{E}_i\), then \(j_{t_i}(c|_{d^\Downarrow})\) is the job responsible for creating the resolution \((d, c|_{d^\Downarrow})\).
        The ticket corresponding to this job is present in the dependencies set of \(j_t(c)\), and thus is marked as done.
        Therefore, \((d, c|_{d^\Downarrow})\) must be present in \(\overline{R}\).

        \item Otherwise, \(\exists i. \; d \in \mathcal{F}_i\).
        Since the tickets \(\overline{j}_{t_i}\) are fully resolved, the resolution \((d, c|_{d^\Downarrow})\) must also be present in \(\overline{R}\).
    \end{itemize}
    Since for all \(d \in \mathcal{F}\), the resolution \((d, c|_{d^\Downarrow})\) is present in \(\overline{R}\), \(\overline{j}_t(c)\) is also fully resolved.

    By induction on the task in the order they are introduced, we can conclude that for all task \(t\), the proposition holds for \(t\).
    This induction is valid as the each the parsing rule \textsc{TaskDef} from Fig.~\ref{fig:operon_checking_rules} necessitates that all input entities for a task must have been introduced already.
\end{proof}

\begin{theorem*}[\ref{thm:canonical_expansions}(1)]
    For a shape \(R\) on \((\mathcal{D}, \prec)\) with a linear extension \(L\), the canonical expansion \(R_L\) uniquely exists.
\end{theorem*}

\begin{proof}

Let \(d^{\Downarrow_L}\) denote the dependency space of dimension \(d \in \mathcal{D}\) under \(\prec_L\).
Since \(\prec \subseteq \prec_L\), we have
\[
    d^\Downarrow \subseteq d^{\Downarrow_L} \quad \text{for all } d \in \mathcal{D}.
\]

\noindent (Existence):
    Define a resolution map \(R_L\) on \((\mathcal{D}, \prec_L)\)  as follows:
    \[
        \dom(R_L) := \set { (d, c_L) \mid d \in \mathcal{D}, c_L: d^{\Downarrow_L} \rightarrow \nat, c_L \ib R}
    \]
    and for every \((d, c_L) \in \dom(R_L)\), set
    \[
        R_L(d, c_L) := R(d, c_L|_{d^\Downarrow}).
    \]

    For all \(d \in \mathcal{D}\) and a partial function \(c: \mathcal{D} \rightharpoonup \nat\), we have:
    \begin{itemize}
        \item If \((d, c|_{d^{\Downarrow_L}}) \in \dom(R_L)\), then \(R_L(d, c|_{d^{\Downarrow_L}}) = R(d, (c|_{d^{\Downarrow_L}})|_{d^\Downarrow}) = R(d, c|_{d^\Downarrow})\).
        \item \(c \ib R \Rightarrow c|_{d^{\Downarrow_L}} \ib R \Rightarrow (d, c|_{d^{\Downarrow_L}}) \in \dom(R_L)\).
        \item \[
            \begin{aligned}
            (d, c|_{d^{\Downarrow_L}}) \in \dom(R_L)
            & \Leftrightarrow c|_{d^{\Downarrow_L}} \ib R \\
            & \Rightarrow c|_{d^\Downarrow} \ib R &(\because d^\Downarrow \subseteq d^{\Downarrow_L}) \\
            & \Leftrightarrow (d, c|_{d^\Downarrow}) \in \dom(R) &(\because R \text{ is a shape})
            \end{aligned}
        \]
    \end{itemize}

    From these, we get: \[
    \begin{aligned}
        (d, c) \in \dom(R_L)
        & \Leftrightarrow c \ib R \\
        & \Leftrightarrow \forall e \in \dom(c).\; (e, c|_{e^\Downarrow}) \in \dom(R) \wedge c(e) < R(e, c|_{e^\Downarrow})\\
        & \Leftrightarrow \forall e \in \dom(c).\; (e, c|_{e^{\Downarrow_L}}) \in \dom(R_L) \wedge c(e) < R_L(e, c|_{e^{\Downarrow_L}}) \\
        & \Leftrightarrow c \ib R_L.
        \end{aligned}
    \]

    This proves that \(R_L\) is a valid shape.
    Since \(R_L\) obviously satisfies the condition \((d, c_L, \ell) \in R_L \Rightarrow (d, c_L, \ell) \in R\), \(R_L\) is a canonical expression.

\noindent(Uniqueness):
    Assume for contradiction that two canonical expansions \(R_1 \neq R_2\) exist.

    By the definition of canonical expansions, whenever \((d, c) \in \dom(R_1) \cap \dom(R_2)\), \[
        R_1(d, c) = R(d, c|_{d^\Downarrow}) = R_2(d, c),
    \]
    so any disagreement must come from the domains.

    Hence \(\dom(R_1) \neq \dom(R_2)\).
    Without loss of generality, pick \[
        (d^\ast, c^\ast) \in \dom(R_1) \setminus \dom(R_2)
    \]
    with \(\prec_L\)-minimal \(d^\ast\) among all such witnesses.

    Since \(R_1\) and \(R_2\) are both shapes, \((d^\ast, c^\ast) \in \dom(R_1)\) means that \(c^\ast \ib R_1\) holds, and \((d^\ast, c^\ast) \notin \dom(R_2)\) means that \(c^\ast \ib R_2\) does not.
    This means that there exists \(e \in \dom (c^\ast)\) such that either
    \begin{itemize}
        \item \((e, c^\ast|_{e^\Downarrow}) \in \dom(R_1) \land (e, c^\ast|_{e^\Downarrow}) \notin \dom(R_2)\), or
        \item \(R_1(e, c^\ast|_{e^\Downarrow}) > c^*(e) \geq R_2(e, c^\ast|_{e^\Downarrow})\).
    \end{itemize}

    The latter is impossible since \(R_1\) and \(R_2\) cannot disagree on the shared domain.
    However, the former is also impossible since \(e \prec_L d^\ast\) and \((e,c^\ast|_{e^\Downarrow}) \in \dom(R_1) \setminus \dom(R_2)\), conflicting the \(\prec_L\)-minimality of \(d^\ast\).

    Therefore, the canonical expansion is unique.
\end{proof}

\begin{theorem*}[\ref{thm:canonical_expansions}(2)]
    For a shape \(R\) on \((\mathcal{D}, \prec)\) with a linear extension \(L\), \(R_L\) preserves coordinate spaces, that is, if \(\mathcal{F} \subseteq \mathcal{D}\) is closed in both \((\mathcal{D}, \prec)\) and \((\mathcal{D}, \prec_L)\), then \(\coordsp{\mathcal{F}}{R_L} = \coordsp{\mathcal{F}}{R}\).
\end{theorem*}

\begin{proof}
    In the proof for Theorem~\ref{thm:canonical_expansions}(1), we already established that the proposed canonical expansion \(R_L\) satisfies \(\forall c.\; c \ib R \Leftrightarrow c \ib R_L\).
    Since a canonical expansion uniquely exists, it follows that this is always true.
    Therefore, \(\coordsp{\mathcal{F}}{R_L} = \coordsp{\mathcal{F}}{R}\).
\end{proof}

\begin{theorem*}[\ref{thm:shape_size_upper_bound}]
    Consider a shape \(R\) on \((\mathcal{D}, \prec)\) and a linear extension \(L\).
    Assuming that \(\ell > 0\) for all \((d, c, \ell) \in R\), the number of resolution entries \(|R|\) satisfies \(|R| \leq |R_L|\).
\end{theorem*}

\begin{proof}
    From the unique canonical extension established in proof for Theorem~\ref{thm:canonical_expansions}(1), we have: \[
        \dom(R_L) := \set{(d, c_L) \mid d \in \mathcal{D}, c_L: d^{\Downarrow_L} \rightarrow \nat, c_L \ib R}
    \]
    and from the definition of shape: \[
        \dom(R) := \set{(d, c) \mid d \in \mathcal{D}, c \in d^\Downarrow \rightarrow \nat, c \ib R}.
    \]

    For each element \((d, c) \in \dom(R)\), let us say that \(c^+ = c \sqcup ((d^{\Downarrow_L} \setminus d^\Downarrow) \times \set{0})\).
    Since we can incrementally construct \(d^{\Downarrow_L} \setminus d^\Downarrow\) by appending the dimensions in the order defined by \(L\), we can apply Lemma~\ref{lem:shape_size_upper_bound_helper} to get \(c_L \ib R\).
    Then, it follows that \((d, c_L) \in \dom(R_L)\).

    Since there exists at least one element of \(\dom(R_L)\) for each element of \(\dom(R)\), we can conclude that \(|R| \leq |R_L|\).
\end{proof}

\begin{lemma}[Helper for Theorem~\ref{thm:shape_size_upper_bound}]\label{lem:shape_size_upper_bound_helper}
    Let \(R\) be a shape on \((\mathcal{D}, \prec)\). Fix a coordinate \(c\) that satisfies \(c \ib R\).
    Suppose moreover that every resolvable next dimension at \(c\) has positive length: \[
        \forall(d, c|_{d^\Downarrow}) \in \dom(R).\; R(d, c|_{d^\Downarrow}) > 0
    \]
    Then for any \(\mathcal{E} \subseteq \set{d \in \mathcal{D} \mid (d, c|_{d^\Downarrow}) \in \dom(R)} \setminus \dom(c)\), the in-bounds condition \(c \sqcup (\mathcal{E} \times \set{0}) \ib R\) holds.
\end{lemma}

\begin{proof}[Proof of Lemma~\ref{lem:shape_size_upper_bound_helper}]
    Consider a dimension \(e \in \dom (c) \sqcup \mathcal{E}\) and let \(c^+ = c \sqcup (\mathcal{E} \times \set{0})\).
    \begin{itemize}
        \item If \(e \in \dom(c)\), since the in-bounds condition \(c \ib R\) holds, we have \((e, c^+|_{e^\Downarrow}) \in \dom(R)\) and \(R(e, c^+|_{e^\Downarrow}) > c^+(e)\).
        \item Otherwise, \(e \in \mathcal{E}\). Then, by choice of \(\mathcal{E}\), we have \((e, c^+|_{e^\Downarrow}) \in \dom(R)\) and \(R(e, c^+|_{e^\Downarrow})\) \(> c^+(e) = 0\).
    \end{itemize}
    Thus, \(c^+ \ib R\).
\end{proof}

\section{Evaluation Data}
\label{sec:appx_evaluation}

This appendix collects the recovery-overhead analysis deferred from Section~\ref{subsec:performance} and the raw per-trial measurements behind the figures of Section~\ref{sec:evaluation}.

\subsection{Recovery Overhead}
\begin{figure}[htbp]
    \centering
    \begin{subfigure}[h]{\textwidth}
        \begin{center}
            \begin{tabular}{c}
                \begin{lstlisting}[basicstyle=\normalsize\ttfamily,numbers=none,xleftmargin=0pt]
A<i> = alpha();
B<j> = beta(A) for i;
C<k> = gamma(A) for i;
D  = delta(B, C) for i, j, k;\end{lstlisting}
            \end{tabular}
        \end{center}
        \caption{Workflow specification. The size of dimension \texttt{i} is \(N\), and the sizes of dimensions \texttt{j} and \texttt{k} are \(M\), resulting in a total of \(NM^2 + 2N + 1\) jobs.}
        \label{fig:recovery_workflow}
    \end{subfigure}
    \begin{subfigure}[h]{0.48\textwidth}
        \centering
        \includegraphics[width=\textwidth]{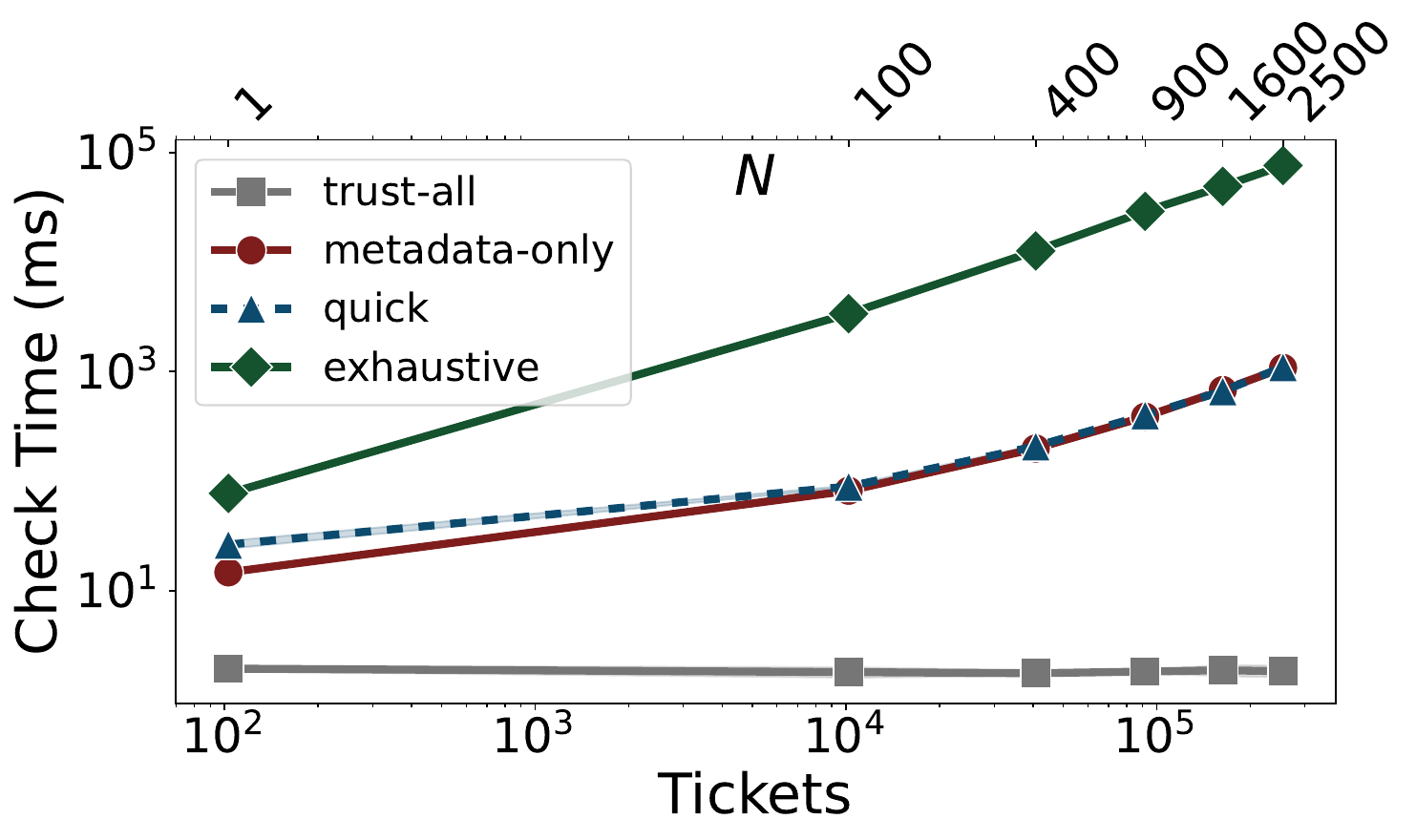}
        \caption{Check time vs.\ $N$ ($M=10$).}
        \label{fig:check_N}
    \end{subfigure}
    \hfill
    \begin{subfigure}[h]{0.48\textwidth}
        \centering
        \includegraphics[width=\textwidth]{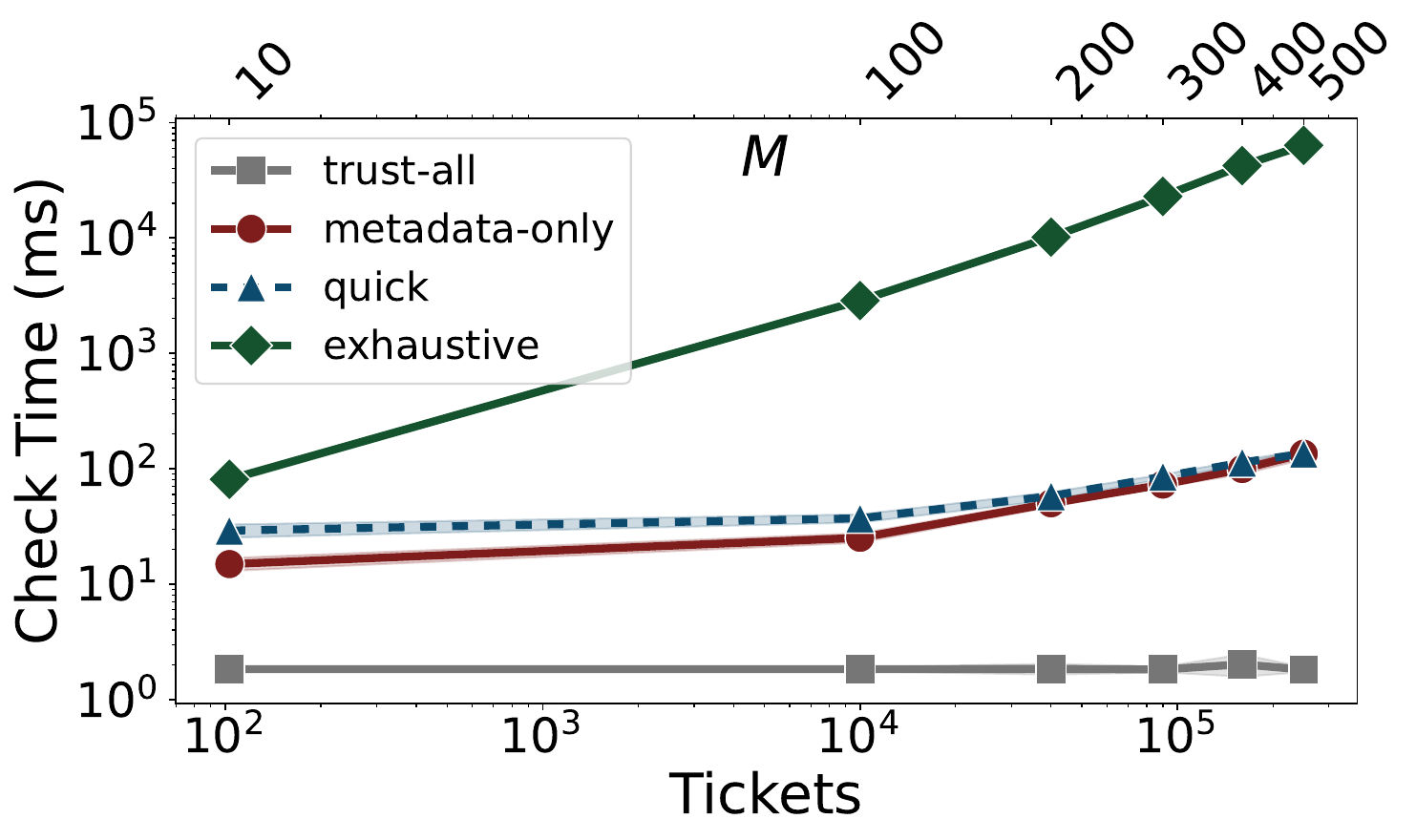}
        \caption{Check time vs.\ $M$ ($N=1$).}
        \label{fig:check_M}
    \end{subfigure}
    \hfill
    \begin{subfigure}[h]{0.48\textwidth}
        \centering
        \includegraphics[width=\textwidth]{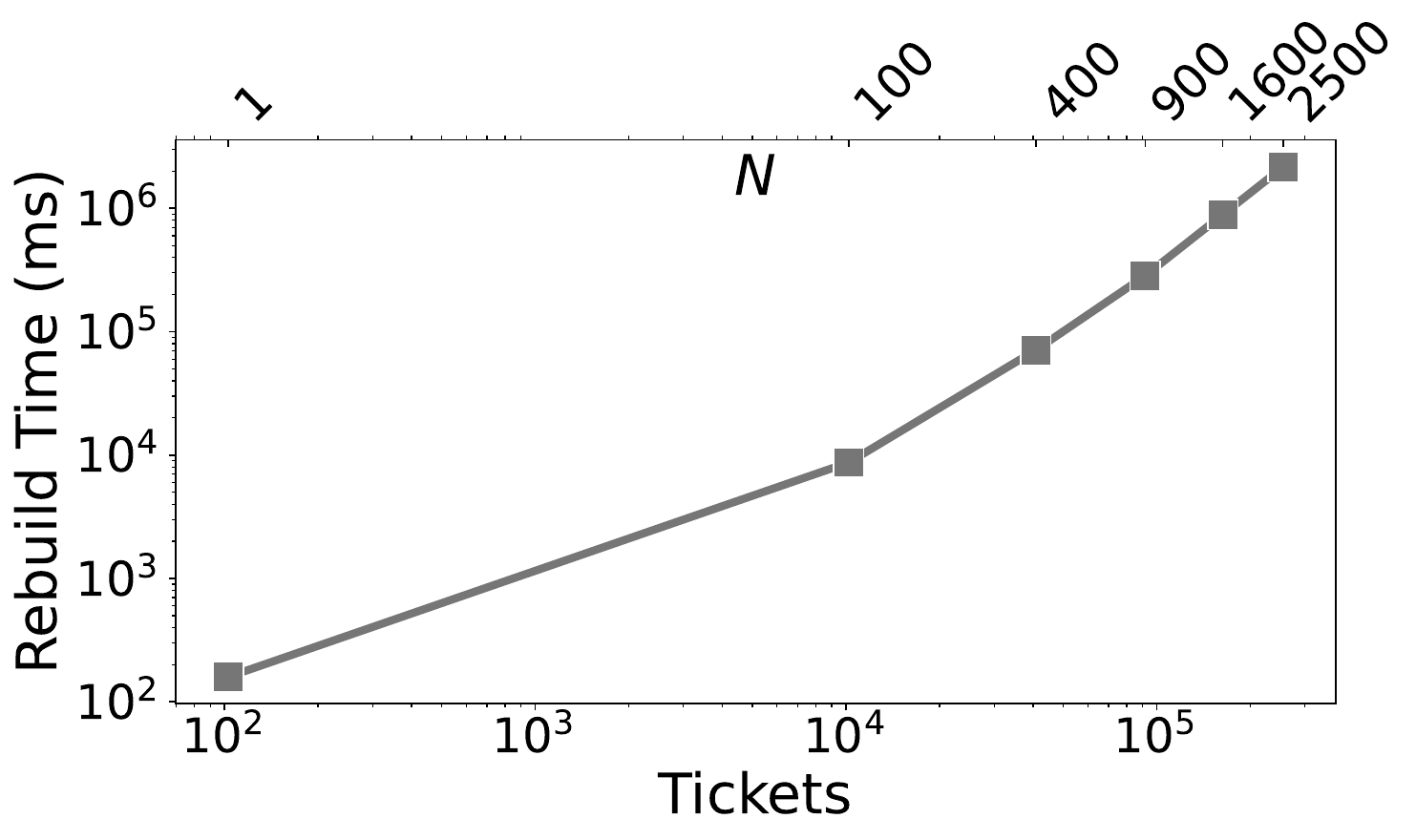}
        \caption{Rebuild time vs.\ $N$ ($M=10$).}
        \label{fig:rebuild_N}
    \end{subfigure}
    \begin{subfigure}[h]{0.48\textwidth}
        \centering
        \includegraphics[width=\textwidth]{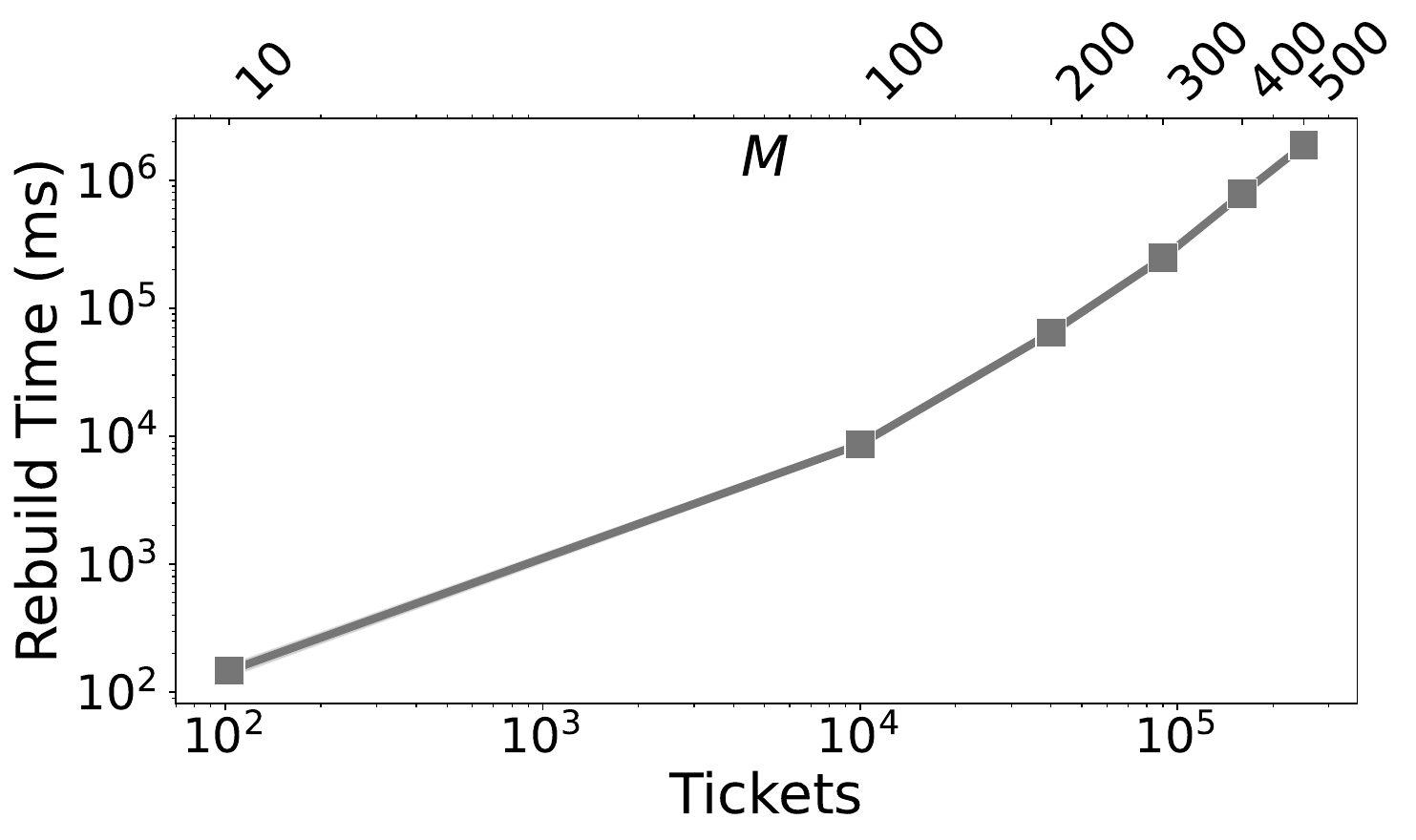}
        \caption{Rebuild time vs.\ $M$ ($N=1$).}
        \label{fig:rebuild_M}
    \end{subfigure}
    \caption{Recovery overhead for the four-task workflow (Figure~\ref{fig:recovery_workflow}) as the number of tasks $N$ and the per-task fanout $M$ vary. Both axes are log-scaled.}
    \label{fig:recovery}
\end{figure}

We evaluate the overhead of Operon's fault-tolerance primitives, \texttt{check} and \texttt{rebuild}, across varying workflow scales.
Figure~\ref{fig:recovery} measures the recovery overhead of \texttt{check} and \texttt{rebuild} on the four-task workflow in Figure~\ref{fig:recovery_workflow}, scaling the number of root tasks~$N$ (with $M{=}10$) and the per-task fanout~$M$ (with $N{=}1$) independently.
For \texttt{check} (Figures~\ref{fig:check_N} and~\ref{fig:check_M}), \texttt{trust-all} completes in constant time (${\sim}2$\,ms) regardless of scale, as it performs no verification.
The \texttt{metadata-only} and \texttt{quick} modes scale similarly, remaining under one second up to ${\sim}10^5$ tickets.
The \texttt{exhaustive} mode, which validates all persisted entities, is roughly an order of magnitude more expensive and reaches ${\sim}10^5$\,ms at the largest configurations.
For \texttt{rebuild} (Figures~\ref{fig:rebuild_N} and~\ref{fig:rebuild_M}), the cost grows roughly linearly with the total number of tickets in both the $N$ and $M$ dimensions, ranging from ${\sim}100$\,ms at $10^2$ tickets to ${\sim}10^6$\,ms at $5 \times 10^5$ tickets.

\subsection{Raw Measurement Data}
\begin{figure}[htbp]
    \centering
    \includegraphics[width=.6\textwidth]{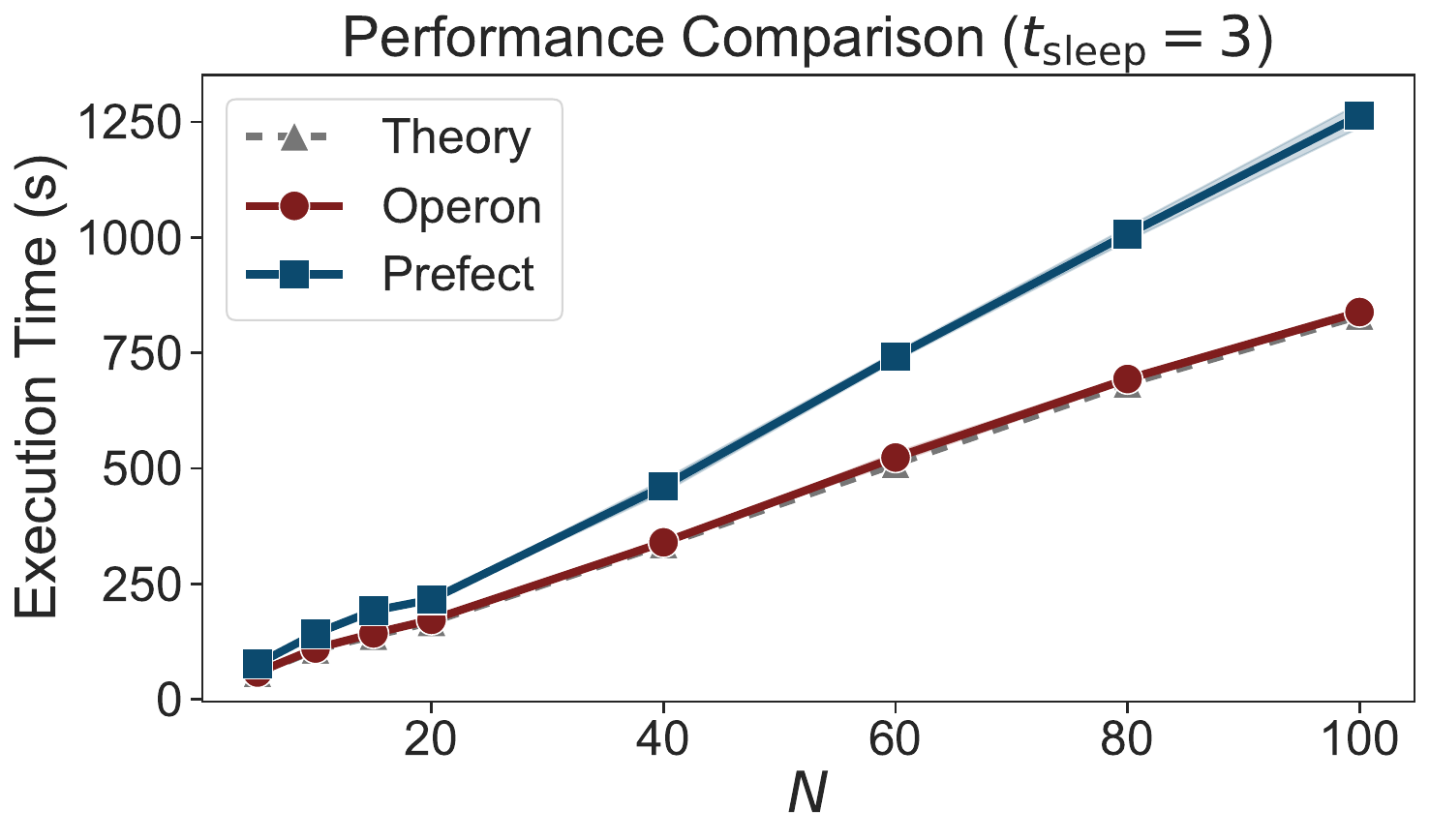}
    \caption{Total execution time of Operon (red) and Prefect (blue) versus the number of \texttt{PaperId}s $N$ at $t_{\text{sleep}} = 3$\,s. Companion to Figure~\ref{fig:experiment_sleep} in the body.
        The reference line ``Theory'' (gray) is the theoretical minimum $\left(\left\lceil\frac{\text{\# of vlm\_evaluate}}{64}\right\rceil + 1\right)\times t_\text{sleep}$.}
    \label{fig:experiment_N}
\end{figure}

Table~\ref{tab:performance-data} reports the per-trial execution times plotted in Figure~\ref{fig:experiment_N}, varying the number of papers $N$ at $t_{\text{sleep}} = 3$\,s.

\begin{table}[htbp]
\centering
\caption{Data of Figure~\ref{fig:experiment_N}}
\label{tab:performance-data}
\resizebox{\textwidth}{!}{%
\begin{tabular}{ll l rrrrrrrr}
\toprule
& & & \multicolumn{8}{c}{$t_{\text{sleep}} = 3$} \\
\cmidrule(lr){4-11}
& & & $N = 5$ & $N = 10$ & $N = 15$ & $N = 20$ & $N = 40$ & $N = 60$ & $N = 80$ & $N = 100$ \\
\midrule
\multirow{3}{*}{Operon}
& & Try 1 & 58.15 & 109.45 & 140.65 & 173.19 & 340.87 & 535.84 & 690.11 & 838.13 \\
& & Try 2 & 57.83 & 109.91 & 146.84 & 173.03 & 339.00 & 518.40 & 700.80 & 838.73 \\
& & Try 3 & 57.98 & 110.60 & 140.32 & 170.17 & 339.59 & 516.14 & 689.66 & 838.38 \\
\midrule
\multirow{3}{*}{Prefect}
& & Try 1 & 74.93 & 141.06 & 185.30 & 213.59 & 452.68 & 734.26 & 998.51 & 1290.88 \\
& & Try 2 & 78.00 & 139.22 & 199.10 & 217.88 & 453.80 & 752.55 & 1024.64 & 1238.67 \\
& & Try 3 & 76.97 & 146.49 & 191.00 & 216.71 & 477.41 & 739.78 & 998.43 & 1262.26 \\
\midrule
\multicolumn{3}{l}{Theory} & 57 & 108 & 138 & 168 & 336 & 510 & 684 & 831 \\
\bottomrule
\end{tabular}%
}
\end{table}

Table~\ref{tab:row-time-data} reports the per-trial execution times plotted in Figure~\ref{fig:experiment_sleep} (body), varying the sleep interval $t_{\text{sleep}}$ at $N = 20$.

\begin{table}[htbp]
\centering
\caption{Data of Figure~\ref{fig:experiment_sleep}}
\label{tab:row-time-data}
\resizebox{\textwidth}{!}{%
\begin{tabular}{ll l rrrrrr}
\toprule
& & & \multicolumn{6}{c}{$N = 20$} \\
\cmidrule(lr){4-9}
& & & $t_{\text{sleep}} = 0$ & $t_{\text{sleep}} = 1$ & $t_{\text{sleep}} = 2$ & $t_{\text{sleep}} = 3$ & $t_{\text{sleep}} = 4$ & $t_{\text{sleep}} = 5$ \\
\midrule
\multirow{3}{*}{Operon}
& & Try 1 & 8.32 & 57.64 & 113.75 & 169.98 & 226.22 & 282.10 \\
& & Try 2 & 8.49 & 57.70 & 113.97 & 169.84 & 227.54 & 281.93 \\
& & Try 3 & 8.52 & 57.70 & 113.84 & 169.93 & 225.94 & 282.00 \\
\midrule
\multirow{3}{*}{Prefect}
& & Try 1 & 126.87 & 144.07 & 175.22 & 218.61 & 269.51 & 314.82 \\
& & Try 2 & 125.55 & 138.22 & 181.05 & 217.82 & 272.58 & 312.72 \\
& & Try 3 & 126.06 & 137.74 & 179.65 & 218.55 & 264.00 & 316.78 \\
\midrule
\multicolumn{3}{l}{Theory} & 0 & 56 & 112 & 168 & 224 & 280 \\
\bottomrule
\end{tabular}%
}
\end{table}

Tables~\ref{tab:recovery-n1} and~\ref{tab:recovery-m10} report the per-trial recovery times plotted in Figure~\ref{fig:recovery}, sweeping the per-task fanout $M$ (at $N=1$) and the task count $N$ (at $M=10$), respectively.

\begin{table}[htbp]
\centering
\caption{Recovery time (ms) for $N = 1$.}
\label{tab:recovery-n1}
\resizebox{\textwidth}{!}{%
\begin{tabular}{ll l rrrrrr}
\toprule
& & & \multicolumn{6}{c}{$N = 1$} \\
\cmidrule(lr){4-9}
& & & $M = 10$ & $M = 100$ & $M = 200$ & $M = 300$ & $M = 400$ & $M = 500$ \\
& \multicolumn{2}{l}{Total Data $= NM^2$} & 100 & 10000 & 40000 & 90000 & 160000 & 250000 \\
\midrule
\multirow{12}{*}{check}
& \multirow{3}{*}{trust-all}
  & Try 1 & 1.9175 & 1.7510 & 1.6158 & 1.8316 & 2.0577 & 1.8906 \\
& & Try 2 & 1.8048 & 1.8620 & 1.9270 & 1.9247 & 1.5593 & 1.8480 \\
& & Try 3 & 1.7874 & 1.8889 & 1.9847 & 1.7197 & 2.4423 & 1.7151 \\
\cmidrule(lr){2-9}
& \multirow{3}{*}{metadata-only}
  & Try 1 & 15.517 & 23.335 & 50.750 & 69.403 & 96.157 & 147.44 \\
& & Try 2 & 16.483 & 27.875 & 53.292 & 76.580 & 91.854 & 138.24 \\
& & Try 3 & 12.805 & 24.433 & 46.541 & 72.911 & 110.47 & 119.28 \\
\cmidrule(lr){2-9}
& \multirow{3}{*}{quick}
  & Try 1 & 33.514 & 33.771 & 55.003 & 89.656 & 119.05 & 127.03 \\
& & Try 2 & 27.026 & 39.405 & 63.342 & 82.847 & 112.66 & 133.02 \\
& & Try 3 & 26.778 & 38.442 & 55.701 & 83.906 & 108.26 & 146.28 \\
\cmidrule(lr){2-9}
& \multirow{3}{*}{exhaustive}
  & Try 1 & 76.504 & 2834.0 & 10139 & 22706 & 44561 & 62874 \\
& & Try 2 & 87.630 & 2946.5 & 10160 & 22805 & 41585 & 63262 \\
& & Try 3 & 79.087 & 2814.3 & 10209 & 23219 & 40820 & 64013 \\
\midrule
\multirow{3}{*}{\shortstack[l]{rebuild\\(after check trust)}}
& & Try 1 & 129.82 & 8315.4 & 60001 & 2.5045e5 & 8.0363e5 & 1.8712e6 \\
& & Try 2 & 154.69 & 8484.0 & 64782 & 2.3873e5 & 7.8469e5 & 1.8869e6 \\
& & Try 3 & 157.94 & 9159.2 & 68318 & 2.5153e5 & 7.6430e5 & 1.8814e6 \\
\bottomrule
\end{tabular}%
}
\end{table}

\begin{table}[htbp]
\centering
\caption{Recovery time (ms) for $M = 10$.}
\label{tab:recovery-m10}
\resizebox{\textwidth}{!}{%
\begin{tabular}{ll l rrrrrr}
\toprule
& & & \multicolumn{6}{c}{$M = 10$} \\
\cmidrule(lr){4-9}
& & & $N = 1$ & $N = 100$ & $N = 400$ & $N = 900$ & $N = 1600$ & $N = 2500$ \\
& \multicolumn{2}{l}{Total Data $= NM^2$} & 100 & 10000 & 40000 & 90000 & 160000 & 250000 \\
\midrule
\multirow{12}{*}{check}
& \multirow{3}{*}{trust-all}
  & Try 1 & 1.9619 & 2.0045 & 1.8124 & 1.8395 & 2.1228 & 2.0901 \\
& & Try 2 & 2.0380 & 1.6079 & 1.6423 & 1.8790 & 1.7685 & 1.6900 \\
& & Try 3 & 1.8129 & 1.7913 & 1.8274 & 1.7463 & 1.7276 & 1.7248 \\
\cmidrule(lr){2-9}
& \multirow{3}{*}{metadata-only}
  & Try 1 & 14.226 & 80.622 & 199.13 & 382.79 & 656.93 & 1064.6 \\
& & Try 2 & 15.434 & 81.854 & 199.61 & 382.72 & 664.17 & 1095.2 \\
& & Try 3 & 14.355 & 82.257 & 196.34 & 396.04 & 714.94 & 1080.7 \\
\cmidrule(lr){2-9}
& \multirow{3}{*}{quick}
  & Try 1 & 28.247 & 90.127 & 201.47 & 392.87 & 677.33 & 1033.6 \\
& & Try 2 & 23.931 & 87.148 & 212.32 & 406.15 & 652.84 & 1139.2 \\
& & Try 3 & 26.689 & 90.399 & 210.52 & 400.97 & 660.13 & 1131.2 \\
\cmidrule(lr){2-9}
& \multirow{3}{*}{exhaustive}
  & Try 1 & 82.705 & 3436.6 & 12736 & 30505 & 48921 & 76264 \\
& & Try 2 & 75.182 & 3336.2 & 12727 & 29134 & 49362 & 76183 \\
& & Try 3 & 73.519 & 3344.8 & 12476 & 27487 & 48972 & 76084 \\
\midrule
\multirow{3}{*}{\shortstack[l]{rebuild\\(after check trust)}}
& & Try 1 & 157.31 & 8694.8 & 71560 & 2.8049e5 & 8.9604e5 & 2.1812e6 \\
& & Try 2 & 159.04 & 8724.1 & 68517 & 2.8268e5 & 9.1451e5 & 2.2142e6 \\
& & Try 3 & 162.21 & 8690.5 & 72714 & 2.8898e5 & 8.6015e5 & 2.0848e6 \\
\bottomrule
\end{tabular}%
}
\end{table}

\end{document}